\documentclass[a4paper,11pt]{JHEP3}
\usepackage{fancyheadings}
\usepackage[dvips]{graphicx}
\usepackage{ifthen}
\usepackage{amsmath}
\usepackage{epsfig}
\usepackage{youngtab}
\usepackage{multirow}
\usepackage[errorshow]{tracefnt}

\newcommand\nn{\nonumber}

\newcommand\benu{\begin{enumerate}}
\newcommand\eenu{\end{enumerate}}
\newcommand\bit{\begin{itemize}}
\newcommand\eit{\end{itemize}}

\newcommand{\bd}{\begin{displaymath}}
\newcommand{\ed}{\end{displaymath}}
\newcommand{\bq}{\begin{equation}}
\newcommand{\eq}{\end{equation}}

\newcommand\al{\alpha}
\newcommand\be{\beta}
\newcommand\ga{\gamma}
\newcommand\de{\delta}

\newcommand\eps{\epsilon}

\newcommand\Ga{\Gamma}
\newcommand\Om{\Omega}

\newcommand\tr{\mathrm{tr}}

\newcommand\ie{{\it i.e.}}
\newcommand\eg{{\it e.g.}}

\title{Supersymmetric Corrections to Eleven-Dimensional Supergravity}

\author{Martin Cederwall\thanks{martin.cederwall@fy.chalmers.se}\\
\hbox{Theor. Physics, G\"oteborg
  Univ. \&\ Chalmers Univ. of Techn., 41296 G\"oteborg, Sweden}}
\author{Ulf Gran\thanks{ugran@mth.kcl.ac.uk}\\ Dept. of Mathematics,
King's College, Strand, London WC2R 2LS, UK}
\author{Bengt E.W. Nilsson\thanks{tfebn@fy.chalmers.se}\\
\hbox{Theor. Physics, G\"oteborg
  Univ. \&\ Chalmers Univ. of Techn., SE-41296 G\"oteborg, Sweden}}
\author{Dimitrios Tsimpis\thanks{tsimpis@mppmu.mpg.de}\\
  Max-Planck-Institut f\"ur Physik, F\"ohringer Ring 6, D-80805
  M\"unchen, Germany}

\abstract{In this paper we study eleven-dimensional supergravity in
its most general form. This is done by implementing manifest
supersymmetry (and Lorentz invariance) through the use of the
geometric (torsion and curvature) superspace Bianchi identities.
These identities are solved to linear order in a deformation
parameter introduced via the dimension zero supertorsion given in
its most general form. The theory so obtained is referred to as the
deformed theory (to avoid the previously used term ``off-shell'').
An important by-product of this result is that any higher derivative
correction to ordinary supergravity of the same dimension as $R^4$,
but not necessarily containing it, derived \eg\ from M-theory, must
appear in a form compatible with
the equations obtained here. Unfortunately we have not yet much to
say about the explicit structure of these corrections in terms of
the fields in the massless supermultiplet. Our results are
potentially  powerful since if the dimension zero torsion could be
derived by other means, our reformulation of the Bianchi identities
as a number of algebraic relations implies that the full theory
would be known to first order
in the deformation, including the dynamics. We mention briefly
some methods to derive the information needed to obtain explicit
answers both in the context of supergravity and ten-dimensional
super-Yang--Mills where the situation is better understood. Other
relevant aspects like spinorial cohomology, the role of the 3- and
6-form potentials and the connection of these results to M2 and
M5 branes are also commented upon.}

\keywords{Supergravity Models, M-theory, Superspaces}
\preprint{hep-th/yymmddd
\\G\"oteborg ITP preprint
\\KCL-MTH-04-14
\\MPP-2004-113}
\begin{document}

\newcommand\ts{\textstyle}
\newcommand\punkt{\,\,.}
\newcommand\komma{\,\,,}
\newcommand\minus{\!-\!}
\newcommand\plus{\!+\!}
\newcommand{\msmall}[1]{{\hbox{$#1$}}}
\newcommand\half{\msmall{1\over2}}
\newcommand{\fraction}[1]{\msmall{1\over#1}}
\newcommand\fr{\fraction}
\newcommand{\Fraction}[2]{\msmall{#1\over#2}}
\newcommand\Fr{\Fraction}
\newcommand\Eg{{\tenit E.g.}}

\newcommand\nl{\hfill\break\indent}
\newcommand\nlni{\hfill\break}

\newcommand\e{\varepsilon}
\newcommand\g{\gamma}

\newcommand\G{\Gamma}

\newcommand\Z{{\Bbb Z}}
\newcommand\C{{\Bbb C}}
\newcommand\R{{\Bbb R}}

\newcommand\M{{\cal M}}
\newcommand\Ham{{\cal H}}

\newcommand\w{\wedge}
\newcommand\id{1\hskip-3.5pt 1}
\newcommand\Int{\int\limits}
\newcommand\bra{\,<\!\!}
\newcommand\ket{\!\!>\,}
\newcommand\cross{\!\times\!}
\newcommand\Tr{\hbox{Tr}\,}
\newcommand\Dslash{D\hskip-6.5pt/\hskip1.5pt}
\newcommand\pslash{\partial\hskip-5.5pt/\hskip.5pt}


\newcommand\m{\mu}

\newcommand\D{\Delta}

\newcommand\tS{\tilde S}
\newcommand\tZ{\tilde Z}
\newcommand\tZp{\tilde Z'}

\newcommand\sea{\searrow}

\newcommand\strike{\hskip0.5pt\lower1pt\hbox to9pt{\rm / \hfill}\hskip-9.5pt}



\section{Introduction}\label{Introduction}
Our understanding of M-theory is still very limited, mainly due to
the lack of powerful methods to probe
 it at the microscopic level.
 One approach to encoding information about M-theory is through
its low energy effective field theory. The short distance properties
are then built into terms appearing as higher-order corrections to
 the leading terms
given by the action \cite{Cremmer:1978}
\begin{align}
\begin{split}
S=&\frac{1}{2\kappa^2}\int
d^{11}x\sqrt{-g}\left(R-\frac{1}{2\cdot4!}H^{mnpq}H_{mnpq}\right)
+\frac{1}{12\kappa^2}\int C\wedge H\wedge H \\
&+\hbox{terms with fermions}\, , \label{Lagrangian}
\end{split}
\end{align}
which is of second order in $\#\hbox{(derivatives)}+$\half$
\#\hbox{(fermions)}$.
The ultimate goal is to be able  to derive the higher-derivative
corrections, \eg\ by means of  a microscopic version of M-theory.
Since this is not yet possible, our aim here is instead to
 solve the superspace Bianchi identities in order to
obtain the most general form such correction terms can take
restricted only by supersymmetry and Lorentz invariance in eleven
dimensions. To what extent such an approach can capture main
features of M-theory is an interesting question to which we have no
answer at this point.

The structure of these correction terms is in general extremely
complicated. Powers of the Riemann tensor of the kind $R^2$ and
$R^4$ are basic examples which have been extensively discussed in
the literature, primarily in the context of string theory and
ten-dimensional effective actions, but also in the
eleven-dimensional context relevant to M-theory. A recent overview is
given in ref. \cite{Howe:2004b}. The existence of
these terms can be inferred by a variety of means in string theory
(for a review see \cite{Peeters:2001ub}), while in M-theory one must rely on
anomaly cancellation arguments \cite{Vafa:1995,Duff:1995wd}, or
(superparticle) loop calculations
\cite{Green:1997,Russo:1997,Green:1998:2,Green:1999:3,Green:1999} in
conjunction  with results from  string theory uplifted to eleven
dimensions. Very recently, one-loop calculations were performed
directly in eleven dimensions \cite{Anguelova:2004}, using a
generalisation of Berkovits' pure spinor approach \cite{Berkovits:2000fe}.

The methods used so far to deduce the explicit form of
 such corrections in eleven
dimensions produce only isolated terms out of a large number of
terms making up the complete superinvariants they belong to. For a
discussion of superinvariants and a collection of references,
 see \eg\ \cite{Peeters:2000}. Since it would be useful to have a
better understanding of the possible superinvariants, there has
been a lot of work invested into the supersymmetrisation of some
isolated terms. In particular, the supersymmetrisation of $R^2$ and
$R^4$ terms in ten dimensions were considered already some time ago,
see ref.\ \cite{deRoo:1993} and references therein. More recently
also terms related to  $R^4$  in eleven dimensions have been
investigated \cite{Peeters:2000,Peeters:2000:2} including a detailed
study of superinvariants by lifting up results from string vertex
operator calculations to eleven dimensions.

Another approach would be to   develop methods based on superspace
in eleven dimensions \cite{Cremmer:1980, Brink:1980} that incorporate
supersymmetry in a manifest way. In ten dimensions $N$=1
supergravity has been constructed off-shell at the linear level in
terms of a superspace lagrangian \cite{Howe:1982}. Such a
formulation is useful when discussing  superinvariants
\cite{Kallosh:1998,Nilsson:1986}, and should in principle lend
itself to a complete analysis of possible superinvariants and
deduction of the corresponding higher-derivative terms in ordinary
component language. The situation in eleven-dimensional
supergravity, or M-theory, is, however, completely different due to
the fact that an off-shell lagrangian formulation with a finite
number of auxiliary fields is not known and may not even exist. From
a general counting argument by Siegel and Ro\v cek
\cite{Siegel:1981} we know that the latter is true for $N$=4
super-Yang--Mills in four dimensions (and consequently also in ten
dimensions) but that maximally supersymmetric supergravity passes
the test. Similar arguments
\cite{Rivelles:1983} suggest that, at the linearised level
and in the absence of central charges,
eleven-dimensional supergravity does not allow for an
off-shell lagrangian quadratic in the fields.
The analysis carried out in
the present paper will
in principle provide an independent check of that statement. In this
respect the approach advocated here is parallel to the discussion of
ten-dimensional super-Yang--Mills theory carried out in
refs.\ \cite{Nilsson:1981} and \cite{Nilsson:1986:3}, which proves
that an off-shell lagrangian based on these fields does not exist.

In order to implement eleven-dimensional supersymmetry and Lorentz
invariance, which must be part of  any M-theory effective action, in
a manifest way, we will here follow
refs.\ \cite{Nilsson:1998,Cederwall:2000ye}
and define the theory in superspace
by means of the superspace Bianchi identities (SSBI's). The latter are
integrability conditions when the theory is formulated in terms of
superspace field strengths. By imposing constraints on the
supertorsion components of dimension zero the identities turn
into non-trivial algebraic relations between certain tensor
superfields where some of the relations turn out to be equations of
motion.
 The outcome of the analysis of the SSBI's depends in a crucial way
on the choice of the dimensionless components of the supertorsion;
setting them equal to a Dirac matrix reproduces uniquely the standard
supergravity theory given above in eq. (1.1) as shown by Howe in
\cite{Howe:1997rf}. The goal in this paper is to complete the
analysis of the Bianchi identities started in \cite{Cederwall:2000ye}
based on the most general torsion constraints given in
\cite{Howe:1997wf,Nilsson:1998} and obtain the equations of motion of
the deformed theory. Our main result  is that we have
managed to reformulate the Bianchi identities to first order in a
general deformation
parameter, into a set of algebraic relations between
tensor superfields. These relations act as constraints which any
higher derivative correction must satisfy. Unfortunately, we cannot
as yet derive explicit expressions for the corrections in terms of the
massless physical fields since we have very little information about
how to express the dimension zero torsion in terms of these physical
fields.

Ultimately, however, we must express the dimension-0 components
of the supertorsion in terms of the physical fields. As already
mentioned, these components of the supertorsion are not arbitrary,
but satisfy certain constraints and may be subjected to certain
field redefinitions, as explained in more detail in section 3. The
problem of finding an explicit solution to these conditions is
equivalent to computing explicit representatives of a particular
spinorial cohomology group. This procedure was carried out
successfully in the case of $N=1$, $10d$ SYM in
\cite{Cederwall:2001bt,Cederwall:2001td,Cederwall:2001dx,Cederwall:2002df},
but the analogous problem in 11d supergravity, where the $R^4$ terms
enter at order $\ell_P^6$ (see sec. \ref{DeformedSupergravity} for
details), seems at present forbiddingly complicated to carry out by
brute force. Recently this analysis was carried out to order
$\ell_P^3$ by one of the authors in \cite{Tsimpis:2004rs}. At this
order there appears a superinvariant which turns out to be
topological in nature in that it can be redefined away by
appropriately shifting the flux quantisation condition of the
4-form.

Another approach to finding the explicit form of the torsion
constraints, advocated recently in \cite{Howe:2003cy}, is to use the
superspace Bianchi identities for the antisymmetric tensor field.
This approach was applied with some success to ten dimensions some
time ago \cite{Atick:1986de,Nilsson:1986md} in the context of $N=1$
supergravity coupled to SYM, using in particular the SSBI for the
3-form field strength $H$ with an $F^2$ topological term $dH=trF^2$.
In the work presented here the eleven-dimensional 3-form
potential emerges from the analysis of the geometric identities via
its own field equation and some (possibly anomalous) Bianchi
identity, and should not be introduced from the start. However, as
pointed out in \cite{Howe:2003cy} the superspace Bianchi identities
for the 4- and 7-form field strengths do in fact relate the
relevant dimension zero torsion components needed in our analysis to
perhaps even more basic negative dimension components of the
superspace 4-form field strength\footnote{ Further comments on
the relation between the SSBI for the superspace 4-form and those
for the geometric fields, can be found in section \ref{Fourform}.}.
The fact that setting all negative dimension components of the
4-form to zero implies via its own SSBI that no torsion
deformations are possible,
 was first noted in \cite{Cederwall:2000ye}.

It was argued in
 \cite{Howe:2003cy} that, under some plausible assumptions,
 the SSBI for the 7-form field strength
$dH_7=\frac{1}{2}H_4H_4+X_8$, where $X_8$ is the superform extension
of the anomaly term found in \cite{Duff:1995wd}, can be iteratively
solved without ambiguities in a similar manner to the
ten-dimensional case discussed in \cite{Nilsson:1986md}. This
approach was therefore proposed in \cite{Howe:2003cy} as a
systematic, albeit quite tedious, way to obtain information about
the zero-dimensional torsion components and, eventually, the
lagrangian of the deformed theory. In  contrast to the situation in
eleven dimensions, the problem simplifies enormously in ten
dimensions due to the fact that the equations can be solved without
relaxing the on-shell torsion constraints \cite{Bonora:1990mt}. Work
on solving the 7-form SSBI in eleven dimensions is in progress
\cite{Howe:2004}. Starting from the torsion SSBI (as in the present
paper) and demanding that the anomalous 7-form BI comes out of
the analysis at the level of equations of motion at dimension 2, is
expected to reflect on the structure of the zero-dimensional torsion
components \cite{Cederwall:2000ye}.

 A proper understanding of the superspace
 torsion components is also vital when proving
$\kappa$-invariance for M2 and M5 branes coupled to background
supergravity \cite{Bergshoeff:1987,Duff:1987,Cederwall:1998:2} and
M-theory corrected versions of it. In fact, one should compare to
the situation in IIA and IIB string theory and the coupling to
D-branes
\cite{Cederwall:1997:2,Aganagic:1997,Cederwall:1997:3,Bergshoeff:1997:2}.
 Here it has been established that there
are higher-derivative background field corrections also on the
world-volumes of the branes, see
\eg\ refs.\ \cite{Bachas:1999,Wyllard:2000}
and references therein.
World-volume corrections to the M2 brane
effective action at order $\ell_P^4$ were
recently computed directly in eleven dimensions in \cite{Howe:2003sa}.
The presence
of such terms complicates the issue of $\kappa$-invariance and it
becomes crucial to know the exact form of the supertorsion and to
understand its relation to the corrections both in target space and
on the brane.

Another aspect of the higher-derivative corrections is that it is to
a large extent unclear how supersymmetry organises the infinite set
of such terms into infinite subsets unrelated by supersymmetry. From
previous work \cite{Green:1998by} we know,
both in ten and eleven dimensions, that adding
one bosonic $R^2$ or $R^4$ term generates an infinite set of other
terms of progressively higher order in the number of derivatives. This
is clear in any on-shell theory, as discussed in detail in the type
IIB case in \eg\ ref.\ \cite{Green:1998by}. In the heterotic case in
ten dimensions an iterative procedure is needed also due to the fact that
there is an implicit dependence on
the 3-form field strength in the supercurvature that appears in
the SSBI, $dH=\tr(F^2-R^2)$, used to define the theory in superspace
\cite{Nilsson:1986md,Pesando:1992}. This situation resembles the one
for M-theory under discussion in this paper apart from the
fact that the corresponding SSBI for the 4-form field strength is
not added as a separate equation but will instead follow from the
geometric SSBI for the supertorsion
\cite{Candiello:1994,Howe:1997rf}.

It is expected that at higher orders there are terms which appear as
a result of iteration triggered by a lower-order term, as well as
terms which are part of genuinely new superinvariants. Of course, in
order to determine which series of terms do actually occur in
M-theory, one has to invoke some microscopic description of the
theory or rely on a comparison with string theory. In
super-Yang--Mills in ten dimensions recent results
\cite{Collinucci:2002ac,deRoo:2003xv} indicate a situation with new
superinvariants appearing at each higher order. Note that if one
chooses to truncate the theory to a certain number of fields, as is
the case when one considers linearised superinvariants, one
generally finds that there are more independent superinvariants (to
that order in the number of fields) than when supersymmetry is
required to all orders in the number of fields.

This paper is organised as follows. In section two we set up the
superspace formalism, review the standard undeformed theory and the
issue of the Weyl connection, and discuss how to obtain the torsion
constraints and in what sense they will produce the most general
deformed theory. The derivation of the deformed theory is then
summarised in 
section three, while the
actual details will be spelt out in full detail in appendix B. 
Section four contains some further comments and
conclusions.

\section{Superspace Formalism and
Undeformed Supergravity}\label{SuperSpace}In this section we will
review all the relevant formalism and the methods connected to
superspace geometry. This is meant to be a self-contained review of
superspace geometry with specific application to $11d$ supergravity.
We will give a systematic account of superspace, geometrical
variables (vielbein, spin connection, torsion, curvature) and
differential calculus. We will discuss in detail the issue of
torsion constraints, their classification in terms of conventional
and physical constraints, their implementation and significance, and
show how they affect the Bianchi identities in undeformed $11d$
supergravity. We find it necessary to include this background, since
much of the information, especially concerning conventional
constraints, is hard to extract from the existing literature and is
mostly conveyed as folklore. This will set the stage for deforming
the supergravity in the most general way allowed by supersymmetry,
which is the main subject of the paper.

\subsection{Superspace Geometry}\label{SuperGeometry}The superspace relevant
to eleven-dimensional supergravity \cite{Cremmer:1978,Cremmer:1980,Brink:1980}
has 11 bosonic and 32 fermionic directions.
The (super-)vielbeins\footnote{In the following, we will often omit the
  prefix ``super'', since it otherwise should be prepended to almost
  all terms related to superspace geometry.},
or frame 1-forms, are $E^A=dZ^ME_M{}^A$, where $M=(m,\m)$ are
coordinate basis (``curved'') tangent indices and $A=(a,\al)$ are
Lorentz frame (``inertial'', ``flat'') tangent indices. Bosonic
directions are denoted by Latin letters and fermionic by Greek.
$Z^M=(x^m,\theta^\m)$ are the superspace coordinates.

When dealing with differential forms and exterior derivatives on
superspace, we use standard superspace conventions. Since the
fermionic property of some components and differentials give signs
depending on ordering, this is a convenient way of handling these with
a minimum of extra signs. The expansion of a form in components is
always done with the differentials in front and in reverse order as
compared to the component field:
\begin{equation}
A_{(m)}={1\over m!}dZ^{M_m}\w\ldots\w dZ^{M_1}A_{M_1\ldots M_m}
={1\over m!}E^{A_m}\w\ldots\w E^{A_1}A_{A_1\ldots A_nm}\punkt
\end{equation}
Taking
care of the statistics of the building blocks then means that the
components of a wedge product $A_{(m)}\w B_{(n)}$ of two bosonic forms
come without signs as
\begin{equation}
(A_{(m)}\w B_{(n)})_{A_1\ldots A_{m+n}}={(m+n)!\over m!n!}
A_{[A_1\ldots A_m}B_{A_{m+1}\ldots A_{m+n})}
\end{equation}
(or the same expression
  for the components in coordinate basis),
where ``$[\ldots)$'' indicates graded symmetrisation. By
letting the exterior derivative act by wedge product from the right,
$dA=A\,\w\buildrel{\scriptstyle\leftarrow}\over d$, its component expression is made
to mimic the component expansion of the ordinary bosonic exterior derivative,
$(dA)_{MM_1\ldots M_n}=(n+1)\partial_{[M}A_{M_1\ldots M_n)}$, which
  facilitates a translation of identities for bosonic fields.

The superspace is equipped with a spin connection $\Om$, which is a
1-form taking values in the Lie algebra of the structure group.
Working with spin connections (in contrast to affine connections) is
a necessary aspect of supergeometry, since the concept of metric is
confined to the bosonic directions (for this reason, the term
vielbein is misleading; we will use it anyway). The choice of
structure group is of course an essential piece of input. We will
use the Lorentz group as structure group, with the 32 fermionic
components transforming as a spinor, so that
$\Om_{(a}{}^c\eta_{b)c}=0$,
$\Om_\al{}^\be={1\over4}(\G^a{}_b)_\al{}^\be\Om_a{}^b$ and
$\Om_a{}^\be=0=\Om_\al{}^b$. We will later comment on the
enlargement of the structure group with a scale transformation,
so-called ``Weyl superspace'' \cite{Howe:1997rf}. The choice of the
Lorentz group as (a factor in) the structure group is intuitively
clear---there must be some input telling the fermions that they are
supposed to behave as spinors. We are not aware of any attempt to
further modify the structure group, and this is a question to which
we hope to be able to devote a systematic investigation in the
future.

The spin connection is {\it a priori} completely unrelated to the
vielbein, so the amount of component fields (any field is of course
a superfield, depending on all superspace coordinates) is enormous.
To take it down to the physical field content of the supergravity
theory, one has to make certain choices. Most of those amount to
what goes under the name ``conventional constraints''. Among these
are some of a type familiar from the Cartan formulation of ordinary
gravity. Finally, a small set of choices, ``physical constraints'',
must be made that have physical significance and determine the exact
form of the equations of motion for the supergravity fields. The
systematics of the these different types of constraints are
explained in detail in the following subsection.

The amount of deviation of the vielbein from being covariantly
closed is the torsion, which is a 2-form with an inertial tangent
index, $T^A=DE^A=dE^A+E^B\w\Om_B{}^A$ (note that since derivatives
act from the right, so does the connection). Torsion is a crucial
object in superspace geometry and supergravity, and does not vanish
even in flat superspace. Many components will be set to zero by
constraints in the following subsection, and the remaining ones
constitute, together with curvature, the main tool of our
calculations. Curvature is defined as usual,
$R_A{}^B=d\Om_A{}^B+\Om_A{}^C\w\Om_C{}^B$. Torsion and curvature
play the r\^ole of field strengths in the theory, and obey the
Bianchi identities
\begin{align}
&DT^A=E^B\w R_B{}^A\komma\cr &DR_A{}^B=0\punkt\label{SSBI}
\end{align}
The first of these plays a central part in the calculations of this
paper, while the second need not be explicitly solved since
it is implied by the first one.
This last fact follows from a theorem by Dragon \cite{Dragon:1979nf} and
relies on the structure group being the Lorentz group.

For completeness, and partially for use in the following subsection,
we would like to exhibit the symmetries of the theory. Under the
local Lorentz symmetry (or, in general, the structure algebra) with
gauge parameter $\Lambda$, the connection transforms as
$\de_\Lambda\Om_A{}^B=D\Lambda_A{}^B$, while the vielbein, torsion
and curvature transform covariantly, $\de_\Lambda
E^A=-E^B\w\Lambda_B{}^A$, etc. Under diffeomorphisms generated by a
vector field $\xi=\xi^M\partial_M=\xi^AE_A$ any field is transformed
as $\D_\xi\phi={\cal L}_\xi\phi$, where ${\cal L}_\xi=i_\xi
d+di_\xi$ is the Lie derivative. In order to covariantise this under
the local structure algebra, this transformation is combined with a
structure transformation with parameter $-i_\xi\Om$, so that we
instead consider $\tilde\D_\xi={\cal L}_\xi+\de_{-i_\xi\Om}$. The
transformation rules of the geometric quantities under consideration
are then
\begin{align}
&\tilde\D_\xi E^A=D\xi^A+i_\xi T^A\komma\cr
&\tilde\D_\xi\Om_A{}^B=i_\xi R_A{}^B\komma\cr &\tilde\D_\xi
T^A=Di_\xi T^A+\xi^BR_B{}^A+E^B\w i_\xi R_B{}^A\komma\cr
&\tilde\D_\xi R_A{}^B=Di_\xi R_A{}^B\punkt
\end{align}
All calculations are made with components that carry inertial
indices, in order to have access to the structure group and its
invariant tensors (gamma matrices). This means that exterior
derivatives give rise to torsion,
\begin{equation}
(DA)_{AA_1\ldots A_n}=(n+1)D_{[A}A_{A_1\ldots A_n)}
+\Fr{n(n+1)}2\,T_{[AA_1}{}^BA_{|B|A_2\ldots A_n)}\punkt
\end{equation}

We end this section with a comment on the concept of dimension. It
is natural to assign to the superspace coordinates canonical
(inverse length) dimensions $(-1,-{1\over2})$ (for bosonic and
fermionic coordinates, respectively). This introduces a grading. All
components of our geometrical objects are conveniently labeled by
their canonical dimension. Since an ordinary bosonic vielbein, for
example, is dimensionless, the vielbein 1-forms carry dimension $-1$
($E^a$) and $-\half$ ($E^\al$). Their components have dimension 0
($E_m{}^a$, $E_\m{}^\al$), $\half$ ($E_m{}^\al$) and $-\half$
($E_\m{}^a$). The torsion has components with dimensions running
from 0 to $\Fr32$. For dimensional reasons, only the dimension-0
components ($T_{\al\be}{}^c$) can (and will) contain invariant
tensors of the structure group. Any calculation, like the main one
of this paper, can be made sequentially for increasing order of
dimension, since there are no operators involved that lower the
dimension of component fields.

\subsection{Conventional Constraints}\label{ConventionalConstraints}As
mentioned in the previous subsection, the constraints we will impose
are of different types. The property they have in common is that
they are effectuated by fixing some components of the torsion. This
ensures the gauge covariance of the constraints, and therefore of
the resulting physical system. In principle, some of the constraints
have the effect of eliminating certain superfluous components of the
vielbein, \ie\ components that after solving the SSBI's occur in
combinations such that they can be removed by field redefinitions
(as can be seen by not enforcing these constraints). However,
imposing them explicitly in terms of vielbeins would be unfortunate,
since such constraints would break diffeomorphism invariance. The
vielbeins carry one coordinate index and one inertial index, and the
coordinate index can not be converted into an inertial index (the
result would be the unit matrix). The torsion components, on the
other hand, carry an inertial index and in addition two lower
indices that can be taken in the inertial as well as in the
coordinate basis. All constraints are formulated in terms of the
torsion, and in terms of components with inertial indices only.
Since such components are scalars under diffeomorphisms, this is the
only covariant procedure to impose constraints. As long as they are
formulated in a way that respects the local structure symmetry, all
symmetries will be preserved.

Let us start by considering the {\it conventional constraints}
\cite{ Gates:1979wg, Gates:1979jv}. There are two kinds of
conventional constraints that can be associated with transformations
of the spin connection and the vielbein respectively, while the
other is held constant. These two transformations have the property
that they leave the torsion SSBI in (\ref{SSBI}) invariant and
therefore take a solution of the SSBI's into a new solution. This is
the reason why we can use these kinds of transformations in order to
find an as simple solution to the SSBI's as possible. The two kinds
of transformations clearly commute with each other.

The first kind shifts the spin connection by an arbitrary 1-form
(with values in the structure algebra) and leaves the vielbein
invariant:
\begin{equation}
\left.\begin{matrix}\hfill E^A & \rightarrow & E^A\hfill\\
\hfill\Om_A{}^B & \rightarrow & \Om_A{}^B+\D_A{}^B\hfill\\\end{matrix}\right\}
\quad\Longrightarrow\quad T^A\rightarrow T^A+E^B\w\D_B{}^A   \punkt
\label{FirstTransformation}
\end{equation}
This kind of redefinition serves to remove the independent degrees
of freedom in $\Om$, which can be achieved by constraints on $T$ as
long as there are no irreducible representations of the structure
group residing in $\Om$ that do not occur in $T$ (all structure
groups under consideration fulfill this requirement, as will be seen
later). This shift is often expressed as the torsion being absorbed
in the spin connection. The canonical example is ordinary bosonic
geometry, where one gets $T_{ab}{}^c\rightarrow
T_{ab}{}^c+2\D_{[ab]}{}^c$, where $\D$ is antisymmetric in the last
two indices, meaning that the transformation can be used to set the
torsion identically to zero, leaving the vielbeins as the only
independent variables. In supergravity the analysis is more subtle.
Only certain representations in the torsion can be brought to zero.

The second kind of transformation consists of a change of tangent
bundle, while the connection is left invariant:
\begin{equation}
\left.\begin{matrix}\hfill E^A&\rightarrow&E^BM_B{}^A\hfill\cr
\hfill\Om_A{}^B&\rightarrow&\Om_A{}^B\hfill\cr
\end{matrix}\right\}
\quad\Longrightarrow\quad T^A\rightarrow T^BM_B{}^A+E^B\w DM_B{}^A   \punkt
\label{SecondTransformation}
\end{equation}
Again, it is essential that one implements the constraints on the
torsion. This will mean that not all components in $M$ can be used.
In fact, the remaining degrees of freedom will all reside in the
component $E_\m{}^a$ of negative dimension, as will become clear in
section \ref{ImplementationOfConventionalConstraints}. The form of
the transformation of $T$ will in practice mean that the
transformations have to be implemented sequentially in increasing
dimension, in order for the second term not to interfere with
constraints obtained by using the first term. We will do this in
detail for $11d$ supergravity below. This second kind of
transformation has no relevance in purely bosonic geometry---there
$M$ has dimension 0, and can not be used to algebraically eliminate
torsion components of dimension 1 (which are taken care of by the
first kind of transformation, anyway). It should also be noted that
not all matrices $M$ are relevant. If $M$ is an element in the
structure group, the transformations in eq.
\eqref{SecondTransformation} can be supplemented by a transformation
of the first kind from eq. \eqref{FirstTransformation} with suitable
parameter ($\D=M^{-1}dM+M^{-1}\Om M-\Om$) so that the total
transformation is a gauge transformation.

\subsection{Implementation of the
  Conventional Constraints}\label{ImplementationOfConventionalConstraints}
Having discussed the general aspects of conventional constraints and
their associated transformations, we would now like to go through
the details for $11d$ supergravity.

The transformations \eqref{FirstTransformation} and \eqref{SecondTransformation}
act in a highly non-linear way on torsion components with inertial
indices. This is because the inertial components even of an invariant
differential form change when the frame field is transformed. For
example, the first term in the torsion transformation of
\eqref{SecondTransformation} reads
\begin{equation}
T_{AB}{}^C\rightarrow
(M^{-1})_A{}^{A'}(M^{-1})_B{}^{B'}T_{A'B'}{}^{C'}M_{C'}{}^C+\ldots\punkt
\end{equation}
Instead of considering large transformations, bringing the torsion
components in different irreducible representations to their
constrained values, we find it much simpler to treat infinitesimal
transformations. Then we just have to check that any transformation
corresponding to a conventional constraint acts by taking us out of
the ``constraint surface''; if this is the case, the conventional
constraint constitutes a valid choice.

We start by displaying a table of torsion components and transformation
parameters ($\D$ and $M$), classified according to dimension and
further divided into irreducible representations of the Lorentz
group\footnote{Representations of the Lorentz group Spin(1,10) are
  specified with standard Dynkin labels, where (10000) is the vector
  and (00001) the spinor. Note that only the representations relevant
  for the conventional constraints are explicitly displayed.}.
$$
\begin{matrix}\hbox{\underbar{Dim.}}
          &\hbox{\underbar{Torsion}}\hfill
          &\hbox{\underbar{$\D$}}\hfill
          &\hbox{\underbar{$M$}}\hfill\cr
&&&\cr
\hfill-\half\,\,&&&M_\al{}^b\hfill\cr
&&&\cr
\hfill0\,\,
&T_{\al\be}{}^c\,\hfill\scriptstyle(00000)\oplus(01000)\oplus(20000)&
          &M_a{}^b\,\hfill\scriptstyle(00000)\oplus(01000)\oplus(20000)\cr
     &\hfill\scriptstyle\oplus(10000)\oplus(00100)\oplus(11000)
        &&M_\al{}^\be\,\hfill\scriptstyle(00000)\oplus(10000)\oplus(01000)\cr
&\hfill\scriptstyle\oplus(00010)\oplus(10002)
           &&\hfill\scriptstyle\oplus(00100)\oplus(00010)\oplus(00002)\cr
&&&\cr
\hfill\half\,\,
   &T_{\al b}{}^c\hfill\,\scriptstyle(20001)\oplus2(10001)\oplus(01001)
   &\D_{\al b}{}^c\hfill\,\scriptstyle(01001)\oplus(10001)\oplus(00001)
           &M_a{}^\be\hfill\scriptstyle(10001)\oplus(00001)\cr
&\hfill\scriptstyle\oplus2(00001)&&\cr
&T_{\al\be}{}^\g\,\hfill\scriptstyle(00003)\oplus(00011)\oplus(00101)&&\cr
&\hfill\scriptstyle\oplus2(01001)\oplus3(10001)\oplus3(00001)&&\cr
&&&\cr
\hfill1\,\,
   &T_{ab}{}^c\hfill\,\scriptstyle(11000)\oplus(00100)\oplus(10000)
   &\D_{ab}{}^c\hfill\,\scriptstyle(11000)\oplus(00100)\oplus(10000)&\cr
        &T_{a\be}{}^\g\hfill&&\cr
&&&\cr \hfill\Fr32\,\, &T_{ab}{}^\g\hfill&&
\end{matrix}
$$

Using a transformation parameter at a certain dimension affects the
torsion components at that dimension and higher, so we may implement
the conventional constraints sequentially in increasing dimension
without the risk of subsequent transformations interfering with
conventional constraints already imposed.

At dimension $-\half$, we have no torsion. Therefore, the
transformation with $M_\al{}^b$ is not used. This means that we do not
remove the degrees of freedom in $E_\m{}^a$. Note that we want to
avoid using a transformation to eliminate degrees of freedom at a
higher dimension; this is of course possible in principle, but would
not lead to the algebraic elimination of entire superfields.

At dimension 0, it is clear that the torsion components in (11000) and
(10002) cannot be algebraically removed, as they do not occur in $M$. We
also note that the transformation
$T_{\al\be}{}^c\rightarrow(M^{-1})_\al{}^{\al'}(M^{-1})_\be{}^{\be'}
   T_{\al'\be'}{}^{c'}M_{c'}{}^c$ is linear in the dimension-0 torsion,
   so it will not be possible to set it to zero. Starting from the
   ordinary
term $2\G^c_{\al\be}$, it is easily seen that all representations
except $(11000)\oplus(10002)$ are generated by a transformation. Out
of the representations in the transformation parameter, some are still
unused, namely $(00000)\oplus(01000)\oplus(00002)$. The (00000) will
be interpreted later as corresponding to a local Weyl (scale) transformation
(when supplemented with the suitable transformation of the
connection). It is the combination $M_a{}^b=e^\sigma\de_a{}^b$,
$M_\al{}^\be=e^{\sigma/2}\de_\al{}^\be$ that leaves $\G^a_{\al\be}$ invariant.
The (01000) is the combination (infinitesimally)
$M_a{}^b=\de_a{}^b+\e j_a{}^b$,
$M_\al{}^\be=\de_\al{}^\be+\e\fr4(\G^a{}_b)_\al{}^\be j_a{}^b$ corresponding
to a local Lorentz transformation. As argued in the previous
subsection, such transformations, lying in the structure group, are irrelevant.
In conclusion, the general torsion at dimension 0 is
\begin{equation}
T_{\al\be}{}^c=2\left({{\G}_{\al\be}}^{c}
+\fr2{{\G}_{\al\be}}^{d_1d_2}{X_{d_1d_2,}}^c+
\fr{5!}{{\G}_{\al\be}}^{d_1\ldots d_5}{Y_{d_1\ldots
d_5,}}^c\right)\komma \label{GeneralDimZeroTorsion}
\end{equation}
where $X$ and $Y$ are in the representations (11000) and (10002) of
the Lorentz group, respectively, \ie, $X_{[a_1a_2,a]}=0$,
$X_{ab,}{}^b=0$, $Y_{[a_1\ldots a_5,a]}=0$, $Y_{a_1\ldots
a_4b,}{}^b=0$.

At dimension $\half$, there is an overlap between the irreducible
representations in $\D$ and $M$, and one has to check that the
corresponding transformations act on $T$ in a non-degenerate way. The
choice of which representations to eliminate,
among the ones multiply occurring in $T$, is
not unique. Our choice is to eliminate one (10001) and
one (00001) representation
in each of the $T_{\al b}{}^c$, $T_{\al\be}{}^\g$.

At dimension 1, finally, the conventional constraints are, as usual,
$T_{ab}{}^c=0$. This part is identical to the elimination of $\Om$ in
bosonic gravity.

Once the conventional constraints have been fixed, using the
transformations discussed above, certain torsion components are
constrained to vanish or to take certain values. The torsion Bianchi
identities, which are automatically satisfied when torsion is
defined in terms of vielbein and spin connection, then cease to be
identities. In $11d$ supergravity, as in other maximally
supersymmetric theories lacking an off-shell supersymmetric
formulation, the Bianchi identities imply the field equations. The
main philosophy of this paper is to turn this property into an
advantage. The set of physically distinct theories differ by the
choice of non-conventional constraints, as explained in the
following subsection. Keeping the torsion components connected to
this last choice general does not take the theory off-shell, but
gives all allowed forms of the field equations. These components
contain fields in a stress tensor multiplet occurring in the field
equations.

In conclusion, by using conventional constraints (for the case that
the structure group is the Lorentz group), the torsion is
brought to the form
\vfill\eject

\begin{equation}
\begin{matrix}
\hbox{dim 0:}\hfill
&\hfill T_{\al\be}{}^{c}&=&
2\Bigl(\G_{\al\be}{}^{c}\hfill&\cr
&&&+\fr2{{\G}_{\al\be}}^{d_1d_2}{X_{d_1d_2}}^c
        \hfill&\hfill(11000)\cr
&&&+\fr{5!}\G_{\al\be}{}^{d_1\ldots d_5}Y_{d_1\ldots d_5}{}^c\Bigr)
        \hfill&\hfill(10002)\cr
&&&&\cr
\hbox{dim $\half$:}\hfill
&\hfill T_{\al b}{}^{c}&=&\tS_b{}^c{}_\al
        \hbox{\vbox to10pt{}}\hfill&\hfill(20001)\cr
&&&+2(\G_{(b}\tS_{d)})_\al\eta^{cd}
        \hbox{\vbox to10pt{}}\hfill&\hfill(10001)\cr
&&&+\de_b^c\tS_\al\hfill&\hfill(00001)\cr
&&&&\cr
&\hfill T_{\al\be}{}^{\ga}&=&
\fr{120}\G^{d_1\ldots d_5}_{\al\be}\tZ_{d_1\ldots d_5}{}^\ga
        \hbox{\vbox to10pt{}}\hfill&\hfill(00003)\cr
&&&+\fr{24}\G^{d_1\ldots d_5}_{\al\be}(\G_{d_1}\tZ_{d_2\ldots d_5})^\ga
        \hbox{\vbox to10pt{}}\hfill&\hfill(00011)\cr
&&&+\fr{12}\G^{d_1\ldots d_5}_{\al\be}(\G_{d_1d_2}\tZ_{d_3d_4d_5})^\ga
        \hbox{\vbox to10pt{}}\hfill&\hfill(00101)\cr
&&&+\fr{12}\G^{d_1\ldots d_5}_{\al\be}(\G_{d_1d_2d_3}\tZ_{d_4d_5})^\ga
+\fr2\G^{d_1d_2}_{\al\be}\tZp_{d_1d_2}{}^c
        \hbox{\vbox to10pt{}}\hfill&\hfill2(01001)\cr
&&&+\fr{24}\G^{d_1\ldots d_5}_{\al\be}(\G_{d_1\ldots d_4}\tZ_{d_5})^\ga
+\G^{d_1d_2}_{\al\be}(\G_{d_1}\tZp_{d_2})^\ga
        \hbox{\vbox to10pt{}}\hfill&\hfill2(10001)\cr
&&&+\fr{120}\G^{d_1\ldots d_5}_{\al\be}(\G_{d_1\ldots d_5}\tZ)^\ga
+\fr2\G^{d_1d_2}_{\al\be}(\G_{d_1d_2}\tZp)^\ga
        \hbox{\vbox to10pt{}}\hfill&\hfill2(00001)\cr
&&&&\cr
\hbox{dim 1:}\hfill
&\hfill T_{ab}{}^{c}&=&0\hfill&\hfill\cr
&&&&\cr
&\hfill T_{a\be}{}^{\ga}&=&
\fr{24}(\G^{d_1\ldots d_4})_\be{}^\ga A_{d_1\ldots d_4a}
+\fr{120}(\G_a{}^{d_1\ldots d_5})_\be{}^\ga A'_{d_1\ldots d_5}
        \hbox{\vbox to10pt{}}\hfill&\hfill2(00002)\cr
&&&+\fr6(\G^{d_1d_2d_3})_\be{}^\ga A_{d_1d_2d_3a}
+\fr{24}(\G_a{}^{d_1\ldots d_4})_\be{}^\ga A'_{d_1\ldots d_4}
        \hbox{\vbox to10pt{}}\hfill&\hfill2(00010)\cr
&&&+\fr2(\G^{d_1d_2})_\be{}^\ga A_{d_1d_2a}
+\fr6(\G_a{}^{d_1d_2d_3})_\be{}^\ga A'_{d_1d_2d_3}
        \hbox{\vbox to10pt{}}\hfill&\hfill2(00100)\cr
&&&+(\G^d)_\be{}^\ga A_{da}
+\fr2(\G_a{}^{d_1d_2})_\be{}^\ga A'_{d_1d_2}
        \hbox{\vbox to10pt{}}\hfill&\hfill2(01000)\cr
&&&+(\G_a{}^d)_\be{}^\ga A_d+A'_a\de_\be{}^\g
        \hbox{\vbox to10pt{}}\hfill&\hfill2(10000)\cr
&&&+(\G_a)_\be{}^\ga A
        \hbox{\vbox to10pt{}}\hfill&\hfill(00000)\cr
&&&+\fr{120}(\G^{d_1\ldots d_5})_\be{}^\ga B_{d_1\ldots d_5,a}
        \hbox{\vbox to10pt{}}\hfill&\hfill(10002)\cr
&&&+\fr{24}(\G^{d_1\ldots d_4})_\be{}^\ga B_{d_1\ldots d_4,a}
        \hbox{\vbox to10pt{}}\hfill&\hfill(10010)\cr
&&&+\fr6(\G^{d_1d_2d_3})_\be{}^\ga B_{d_1d_2d_3,a}
        \hbox{\vbox to10pt{}}\hfill&\hfill(10100)\cr
&&&+{1\over2}(\G^{d_1d_2})_\be{}^\ga B_{d_1d_2,a}
        \hbox{\vbox to10pt{}}\hfill&\hfill(11000)\cr
&&&+(\G^d)_\be{}^\ga B_{d,a}
        \hbox{\vbox to10pt{}}\hfill&\hfill(20000)\cr
&&&&\cr
\hbox{dim $3\over2$:}\hfill
&\hfill T_{ab}{}^{\ga}&=&
\tilde t_{ab}{}^{\ga}
        \hbox{\vbox to10pt{}}\hfill&\hfill(01001)\cr
&&&+2(\G_{[a}\tilde t_{b]})^\ga
        \hbox{\vbox to10pt{}}\hfill&\hfill(10001)\cr
&&&+(\G_{ab}\tilde t)^\ga
        \hbox{\vbox to10pt{}}\hfill&\hfill(00001)\cr
\end{matrix}
\label{ConventionalTorsion}
\end{equation}
\vfill\eject

\subsection{Physical (Non-Conventional)
Constraints and Spinorial
Cohomology}\label{PhysicalConstraintsAndCohomology}The form of
torsion \eqref{ConventionalTorsion} arrived at in the previous
subsection is actually the starting point for the calculation of
this paper, as it is presented in section
\ref{DeformedSupergravity}. It is general enough to contain any
``deformation'' allowed by supersymmetry, \ie, when substituted in
the torsion Bianchi identities it will contain components
corresponding to the most general stress tensor multiplet.

In order to arrive at a specific version of $11d$ supergravity, one
has to make a few more choices. It was shown in ref.\
\cite{Howe:1997rf} that taking $T_{\al\be}{}^c=2\G_{\al\be}^c$ at
dimension zero gives the superspace formulation of ordinary
``undeformed'' supergravity. In that paper, the structure group was
enlarged to include a Weyl (scale) transformation. As a byproduct of
our analysis, we will find the same result for the Lorentz group
below.

There exists a very helpful method for determining exactly which
torsion components contain information of the deformation, \ie,
which torsion components have to be subjected to physical, or
non-conventional, constraints, in order to put the theory on-shell
expressed in terms of the physical fields. This is the theory of
spinorial cohomology, put forward in the context of $10d$
super-Yang--Mills in ref.\ \cite{Cederwall:2001bt}, and further
generalised in \cite{Cederwall:2001dx,Cederwall:2001xk}. A purely
tensorial definition, \ie, not relying on particular
representations, was given in \cite{Howe:2003cy}. We will not give a
detailed account of the theory here. Its validity is general and not
confined to the supergravity considered in this paper. The statement
obtained for $11d$ supergravity is that the gauge transformation
(diffeomorphism) parameter $\xi^a$ in (10000), the vielbein
$E_\al{}^a$ in (10001), the torsion $T_{\al\be}{}^a$ in
$(11000)\oplus(10002)$ and the Bianchi identities in
$(11001)\oplus(10003)$ are part of a complex
\begin{equation}
\vtop{\baselineskip0pt\lineskip0pt
\ialign{
$\hfill#\hfill$&$\,\hfill#\hfill\,$&$\hfill#\hfill$&$\,\hfill#\hfill\,$&$\hfill#\hfill$&$\,\hfill#\hfill\,$&$\hfill#\hfill$&$\,\hfill#\hfill\,$&$\hfill#\hfill$&$\hfill#\hfill$\cr
\xi&\buildrel\D\over\rightarrow&E&\buildrel\D\over\rightarrow &T&\buildrel\D\over\rightarrow&\hbox{BI}&\buildrel\D\over\rightarrow &&\ldots\cr
\phantom{(10000)}&&&&&&&&&\cr
(10000)&\rightarrow&(10001)&\rightarrow &(10002)&\rightarrow &(10003)&\rightarrow &(10004)&\ldots\cr
       &   &       &\sea&       &\sea&       &\sea&       &      \cr
       &   &       &    &(11000)&\rightarrow &(11001)&\rightarrow &(11002)&\ldots\cr
       &   &       &    &       &    &       &\sea&       &      \cr
       &   &       &    &       &    &       &    &(12000)&\ldots\cr
       &   &       &    &       &    &       &    &       &\phantom{\rightarrow}\cr
}}\label{ElevenSGComplex}
\end{equation}
The operator $\D$ is a nilpotent fermionic ``exterior derivative''
 given by the action of the covariant fermionic
derivative together with a projection onto the relevant
representations. Its cohomology (seen as bosonic/fermionic
components of superfields) describes gauge transformations, physical
fields and the stress tensor multiplet, respectively (the meaning of
cohomology at the level of Bianchi identities and higher has not
been understood). The full cohomology of $11d$ supergravity is
summarised in the following table, where the entries are denoted by
the irreducible representations of the respective component fields,
$n$ denotes the horizontal level in the complex
\eqref{ElevenSGComplex} and the dimensions of the fields are given
in the vertical axis. \vfill\break \eject

\begin{equation}
\hskip-1cm
\vtop{\baselineskip25pt\lineskip0pt
\ialign{
$\hfill#\quad$&$\scriptstyle\,\hfill#\hfill\,$&$\scriptstyle\,\hfill#\hfill\,$
&$\scriptstyle\,\hfill#\hfill\,$&$\scriptstyle\,\hfill#\hfill\,$&$\scriptstyle\,\hfill#\hfill$
&$\scriptstyle\,\hfill#\hfill$&$\scriptstyle\,\hfill#\hfill$
                &\quad#\cr
            &\ts n=0&\ts n=1&\ts n=2&\ts n=3&\ts n=4&\ts n=5&\ts n=6&\cr
\hbox{dim}=-1&\quad(10000)\quad
                &\quad\phantom{(00000)}\quad
                &\quad\phantom{(00000)}\quad
                &\quad\phantom{(00000)}\quad
                &\quad\phantom{(00000)}\quad
                &\quad\phantom{(00000)}\quad
                &\quad\phantom{(00000)}\quad
                &\cr
        -\fr2&(00001)&\bullet&               &&&       &\cr
           0&\bullet&(20000)&\bullet&       &&&       &\cr
       \Fr12&\bullet&\raise3pt\vtop{\baselineskip6pt\ialign{
                                        \hfill$#$\hfill\cr
                                        \scriptstyle(00001)\cr
                                        \scriptstyle(10001)\cr}}
                        &\bullet&\bullet&&&&\cr
           1&\bullet&\raise3pt\vtop{\baselineskip6pt\ialign{
                                        \hfill$#$\hfill\cr
                                        \scriptstyle(00010)\cr
                                        \scriptstyle(10000)\cr}}
                        &\bullet&\bullet&\bullet&&&\cr
       \Fr32&\bullet&\bullet&\raise3pt\vtop{\baselineskip6pt\ialign{
                                        \hfill$#$\hfill\cr
                                        \scriptstyle(00001)\cr
                                        \scriptstyle(10001)\cr}}
                        &\bullet&\bullet&\bullet&&\cr
           2&\bullet&\bullet&\raise6pt\vtop{\baselineskip6pt\ialign{
                                        \hfill$#$\hfill\cr
                                        \scriptstyle(00000)(00002)\cr
                                        \scriptstyle(00100)(01000)\cr
                                        \scriptstyle(10000)(20000)\cr}}
                                &\bullet&\bullet&\bullet&\bullet&\cr
       \Fr52&\bullet&\bullet&\bullet&\bullet&\bullet&\bullet&\bullet&
\hfill\cr
           3&\bullet&\bullet&\bullet&\raise6pt\vtop{\baselineskip6pt\ialign{
                                        \hfill$#$\hfill\cr
                                        \scriptstyle(00000)(00002)\cr
                                        \scriptstyle(00100)(01000)\cr
                                        \scriptstyle(10000)(20000)\cr}}
                                &\bullet&\bullet&\bullet\cr
       \Fr72&\bullet&\bullet&\bullet&\raise3pt\vtop{\baselineskip6pt\ialign{
                                        \hfill$#$\hfill\cr
                                        \scriptstyle(00001)\cr
                                        \scriptstyle(10001)\cr}}
                        &\bullet&\bullet&\bullet&\cr
           4&\bullet&\bullet&\bullet&\bullet
                                &\raise3pt\vtop{\baselineskip6pt\ialign{
                                        \hfill$#$\hfill\cr
                                        \scriptstyle(00010)\cr
                                        \scriptstyle(10000)\cr}}
                        &\bullet&\bullet&\cr
       \Fr92&\bullet&\bullet&\bullet&\bullet
                                &\raise3pt\vtop{\baselineskip6pt\ialign{
                                        \hfill$#$\hfill\cr
                                        \scriptstyle(00001)\cr
                                        \scriptstyle(10001)\cr}}
                        &\bullet&\bullet&\cr
           5&\bullet&\bullet&\bullet&\bullet&(20000)&\bullet&\bullet&\cr
    \Fr{11}2&\bullet&\bullet&\bullet&\bullet&\bullet&(00001)&\bullet&\cr
           6&\bullet&\bullet&\bullet&\bullet&\bullet&(10000)&\bullet&\cr
       \Fr{13}2&\bullet&\bullet&\bullet&\bullet&\bullet&\bullet&\bullet&\cr
}}
\label{gorgias}
\end{equation}
All information is thus contained in the
lowest-dimensional superfield of each type. The stress tensor fields,
\ie, the deformations, are contained in the torsion representations
$(11000)\oplus(10002)$ at dimension 0. These are the ones encoding
the exact form of the interactions and,
therefore, these are the ones that should be
subjected to physical constraints.
The undeformed supergravity is thus obtained by imposing the physical
constraint
\begin{equation}
T_{\al\be}{}^c=2\G_{\al\be}^c\punkt\
\end{equation}

\subsection{Bianchi Identities and Undeformed
  Supergravity}\label{Undeformed}
Eleven-dimensional
supergravity contains, in addition
  to the metric and the gravitino, a 3-form potential $C$ with
  field strength $H=dC$ and field equation $d\star H=\half H\w
  H$. These fields can be read off the table of spinorial cohomologies
  at $n=1$, where the tensor field enters via its field strength $H$,
  due to gauge invariance. We also find a spinor at dimension $\half$ and a
  vector at dimension 1, that will be interpreted as the Weyl
  connection.
 Remember that spinorial cohomology is not
  {\it a priori} supersymmetric, in that it only encodes objects of
  lowest dimensionality. Higher-dimensional Bianchi identities will
  restrict the fields occurring.
$H$ may be promoted to a 4-form in superspace, but this is
  not necessary: like all supergravity fields it is found in
  the geometric superspace variables.
This subsection contains a brief review of the Bianchi
  identity calculation in the undeformed case, with the purpose of
  illustrating how the supergravity degrees of freedom arise, how the
  Bianchi identities lead to the equations of motion, and to
  what extent the result is unique. Some relevant equations for
  undeformed supergravity are collected in appendix A.

The torsion Bianchi identity is $DT^A=E^B\w R_B{}^A$, which in
inertial components reads
\begin{equation}
3R_{[ABC)}{}^D=3D_{[A}T_{BC)}{}^D+3T_{[AB}{}^ET_{|E|C)}{}^D
\punkt\label{ComponentTorsionBI}
\end{equation}
The procedure for solving the Bianchi identities is to consider this
equation, starting from the lowest dimension and moving upwards,
decomposing in all occurring irreducible representations of the
Lorentz group. If a curvature is allowed to carry a certain
representation, the information contained in eq.
\eqref{ComponentTorsionBI} is the value of this curvature component.
The only conditions on torsion components come from situations where
the curvature is constrained by the structure algebra. The r\^ole of
the structure group is double in this sense: a larger structure
group serves on one hand to eliminate torsion components via
conventional constraints of the first kind, on the other hand it
gives fewer restrictions on the torsion through the Bianchi
identities.

The complete set of torsion Bianchi identities
is
\begin{equation}
\begin{matrix}
\hfill\hbox{\rm dim.}&\scriptstyle{1\over2}:\hfill&\hfill\scriptstyle
3(\strike R_{(\al\be}){}_{\g)}{}^d
         &=&\scriptstyle 3D_{(\al}T_{\be\g)}{}^d
         +3T_{(\al\be}{}^eT_{|e|\g)}{}^d+3T_{(\al\be}{}^\e T_{|\e|\g)}{}^d\hfill\cr
&&&&\mathstrut\cr
\hfill\hbox{\rm dim.}&\scriptstyle1:\hfill&\hfill\scriptstyle
3(R_{(\al\be}){}_{\g)}{}^\de&=&\scriptstyle 3D_{(\al}T_{\be\g)}{}^\de
         +3T_{(\al\be}{}^eT_{|e|\g)}{}^\de+3T_{(\al\be}{}^\e
         T_{|\e|\g)}{}^\de\hfill\cr
&&\hfill\scriptstyle\updownarrow\hskip16pt&&\cr
&&\hfill\scriptstyle 2(\strike R_{c(\al}){}_{\be)}{}^d+(R_{\al\be}){}_c{}^d
       &=&\scriptstyle 2D_{(\al}T_{\be)c}{}^d+D_cT_{\al\be}{}^d+
       T_{\al\be}{}^e \strike T_{ec}{}^d+T_{\al\be}{}^\e T_{\e c}{}^d\hfill\cr
&&&&\hfill\scriptstyle+2T_{c(\al}{}^e T_{|e|\be)}{}^d+2T_{c(\al}{}^\e T_{|\e|\be)}{}^d\cr
&&&&\mathstrut\cr
\hfill\hbox{\rm dim.}&\scriptstyle{3\over2}:\hfill&\hfill\scriptstyle
(\strike R_{\al\be}){}_c{}^\de+2(R_{c(\al}){}_{\be)}{}^\de
       &=&\scriptstyle 2D_{(\al}T_{\be)c}{}^\de+D_cT_{\al\be}{}^\de+
       T_{\al\be}{}^e T_{ec}{}^\de+T_{\al\be}{}^\e T_{\e c}{}^\de\hfill\cr
&&&&\hfill\scriptstyle+
       2T_{c(\al}{}^e T_{|e|\be)}{}^\de+2T_{c(\al}{}^\e T_{|\e|\be)}{}^\de
\cr
&&\hfill\scriptstyle\updownarrow\hskip16pt&&\cr
&&\hfill\scriptstyle (\strike R_{bc}){}_\al{}^d+2(R_{\al[b}){}_{c]}{}^d&=&\scriptstyle
       D_\al \strike T_{bc}{}^d+2D_{[b}T_{c]\al}{}^d
       +2T_{\al[b}{}^e \strike T_{|e|c]}{}^d+\strike T_{bc}{}^e T_{e\al}{}^d
       \hfill\cr
&&&&\hfill\scriptstyle+2T_{\al[b}{}^\e T_{|\e|c]}{}^d+T_{bc}{}^\e T_{\e\al}{}^d\cr
&&&&\mathstrut\cr
\hfill\hbox{\rm dim.}&\scriptstyle2:\hfill&\hfill\scriptstyle
2(\strike R_{\al[b}){}_{c]}{}^\de+(R_{bc}){}_\al{}^\de&=&\scriptstyle
       D_\al T_{bc}{}^\de+2D_{[b}T_{c]\al}{}^\de
       +2T_{\al[b}{}^e T_{|e|c]}{}^\de+\strike T_{bc}{}^e T_{e\al}{}^\de
       \hfill\cr
&&&&\hfill\scriptstyle+2T_{\al[b}{}^\e T_{|\e|c]}{}^\de+T_{bc}{}^\e T_{\e\al}{}^\de\cr
&&\hfill\scriptstyle\updownarrow\hskip16pt&&\cr
&&\hfill\scriptstyle 3(R_{[ab}){}_{c]}{}^d&=&\scriptstyle 3D_{[a}\strike T_{bc]}{}^d+
        3\strike T_{[ab}{}^e T_{|e|c]}{}^d
       +3T_{[ab}{}^\e T_{|\e|c]}{}^d
   \hfill\cr
&&&&\mathstrut\cr
\hfill\hbox{\rm dim.}&\scriptstyle{5\over2}:\hfill&\hfill\scriptstyle
3(\strike R_{[ab}){}_{c]}{}^\de&=&\scriptstyle 3D_{[a}T_{bc]}{}^\de
       +3\strike T_{[ab}{}^e T_{|e|c]}{}^\de
       +3T_{[ab}{}^\e T_{|\e|c]}{}^\de
\hfill\cr
\label{LorentzComponentBI}
\end{matrix}
\end{equation}
Here, we have striked out curvature components that vanish due to
the bosonic property of the structure group and torsions that have
been set to zero using the ordinary bosonic form of the first kind
of conventional constraint, and indicated with arrows curvature
components that are related to each other due to the Lorentz
condition. Of course both the structure group and the vanishing of
certain torsion components has a finer structure than can be taken
care of by dividing into bosonic and fermionic indices; it has to be
accounted for by performing a full decomposition into irreducible
representations. Note that only (linear combinations of) equations
without curvature contain information.

According to the previous subsection, the only physical constraint
that has to be imposed on the conventionally constrained torsion of
eq. \eqref{ConventionalTorsion} is
$$
T_{\al\be}{}^c|_{(11000)\oplus(10002)}=0.
$$
The Bianchi identity at
dimension $\half$ therefore reads
\begin{equation}
0=\G_{(\al\be}^eT_{|e|\g)}{}^d+T_{(\al\be}{}^\e \G_{|\e|\g)}^d\punkt
\end{equation}
Let us compare the content of irreducible representations in
this equation, obtained as
$(10000)\otimes(00001)^{\otimes_s3}$,
 to the one in the torsion according to
eq. \eqref{ConventionalTorsion}.
\newskip\repskip
\repskip=-8pt
\begin{equation}
\begin{matrix}\tabskip=100pt
\hbox{\rm equation: }&\scriptstyle(10003)\oplus(11001)\oplus(20001)\oplus(00003)
     \oplus(00011)\oplus(00101)\oplus2(01001)\oplus3(10001)\oplus2(00001)\cr
\hbox{\rm torsion: }\hfill&\hfill\scriptstyle(20001)\oplus(00003)
     \oplus(00011)\oplus(00101)\oplus2(01001)\oplus3(10001)\oplus3(00001)\cr
\end{matrix}
\end{equation}
From this comparison it follows that the Bianchi identities {\it
may} set the entire dimension-$\half$ torsion to 0, except for a
spinor. Note that it does not prove that this actually happens; in
principle, and this will be the case for Bianchi identities at
higher dimension, there can be a linear dependence between equations
in the same representation when expressed in terms of the torsion
components, leading to more solutions for the torsion than would be
guessed by counting representations. At dimension $\half$, however,
there is no degeneracy, and everything except for a single spinor is
set to zero. A detailed calculation shows that all dimension-$\half$
components of the torsion in eq. \eqref{ConventionalTorsion} vanish
except for the spinors $\tS$, $\tZ$ and $\tZp$, and that
$\tZ=\Fr3{88}\tS$, $\tZp=-\fr{44}\tS$.

The procedure at dimension $\half$ illustrates the general method. At
dimension 1, decomposing in irreducible representations and taking
into account the Lorentz condition, gives the non-vanishing torsion
components $A$, $A_a$, $A'_a$, $A_{abcd}$ and $A'_{abcd}$,
together with the
relations (only relations where curvature components are eliminated
are displayed)
\begin{align}
D\tS&=-4224A\komma\cr D\G_a\tS&=64A_a=-64A'_a\komma\cr
D\G_{ab}\tS&=0\komma\cr D\G_{abc}\tS&=0\komma\cr
D\G_{abcd}\tS&=-1408(A_{abcd}+2A'_{abcd})\komma\cr
D\G_{abcde}\tS&=0\punkt
\end{align}
Note that this implies $D_{(\al}\tS_{\be)}=2\G^a_{\al\be}A_a$.

In the following subsection, we demonstrate how the spinor $\tS$ and
the vector $A_a$ are identified as spinor and vector components of a
Weyl (scaling) connection, and how they can be brought to zero by a
conventional constraint. The rest of the discussion in the present
subsection is based on this being done.

The remaining calculation for the undeformed $11d$ supergravity is
well known, and we will not relate all details leading to equations
of motion etc. In the absence of $\tS$, one finds the only
non-vanishing torsion components at dimension 1 to be $A_{abcd}$ and
$A'_{abcd}$, with the relation $A_{abcd}+2A'_{abcd}=0$. This field
is identified as proportional to the 4-form field strength $H$ of
$11d$ supergravity (see appendix A), \ie, $T_{a\be}{}^\ga\sim
(\G^{d_1d_2d_3})_\be{}^\ga H_{d_1d_2d_3a} -\fr{8}(\G_a{}^{d_1\ldots
d_4})_\be{}^\ga H_{d_1\ldots d_4}$. At dimension $3\over2$, the torsion
$T_{ab}{}^\ga$ is the gravitino field strength, and its gamma traces
are set to zero as equations of motion, $\tilde t_a{}^\ga=0$,
$\tilde t^\ga=0$. In addition, one gets from the Bianchi identities
information about how the gravitino field strength sits inside the
superfield $H$: $D_\al H_{abcd}\sim(\G_{[ab}\tilde t_{cd]})^\al$. At
dimension 2, the Weyl tensor appears in the representation (02000)
and is expressible as (schematically) $D\tilde t+H^2$. The Bianchi
identities at this dimension imply the Bianchi identities as well as
the field equations for $H$ together with the Einstein equations.

\subsection{The Weyl Connection}\label{Weyl}Apart from the ordinary
  supergravity fields, the only freedom allowed by the torsion Bianchi
  identities resides in the spinor superfield $\tS$
at dimension $\half$. From the dimension-$\half$ Bianchi identities it
  follows that it is constrained to obey the
  equation $D_{(\al}\tS_{\be)}-2\G^c_{\al\be}A_c=0$. Letting
  $V_\al=\tS_\al$,
$V_a=-\half A_a$, and $V=dZ^AV_A$, this equation is the dimension-1
  component of $dV=0$. Indeed, without going into details, it is
  confirmed that the Bianchi identities at dimension $3\over2$ and 2 imply
  that the 1-form $V$ is closed. Modulo topologically
  non-trivial configurations, $V$ is exact, $V=d\phi$, where $\phi$ is
  a scalar superfield of dimension 0.

We now recall that there was a scalar transformation among the ones
connected to conventional constraints that was never used. This
``Weyl transformation'' can be used to shift $\phi$ to zero. In the
present situation, where we have already chosen conventional
constraints at dimension $\half$ and higher, this has to be done
carefully. The reason why we always perform conventional
transformations by increasing dimension was that they affect torsion
components also at higher dimension. Shifting $\phi$ affects the
torsion constraints already fixed, and has to be accompanied by new
conventional transformations in order to restore the
constraints\footnote{There exists a choice of conventional
constraints adapted to Weyl transformations, where this is not
needed. Unfortunately, this is not the convention adopted in this
paper.}. For example, at dimension $\half$, all torsion is eliminated by
a Weyl rescaling with $\sigma=-{67\over66}\phi$, followed by a
conventional transformation of the first kind with $\D_{\al
b}^c=-{1\over66}e^{-{\sigma\over2}}(\G_b{}^cD\phi)_\al$ and one of
the second kind with $M_a{}^\be={1\over
132}e^{-{\sigma\over2}}(\G_aD\phi)^\be$.

If $\phi$ is a non-trivial flat connection, it can no longer be set
to zero, but it can be shifted to a representative in its cohomology
class. Such non-trivial Weyl connections have been used to construct
massive supergravity in lower dimensions
\cite{Howe:1998qt,Lavrinenko:1998qa}. Even though the formulation
due to Howe \cite{Howe:1997rf} with a structure group enlarged to
encompass scalings is more geometrical, the exact same statements
hold true for the Lorentz group formulation: the two versions are
completely equivalent (recall the fact that a conventional
transformation of the second kind with values in the structure group
can be traded for a gauge transformation).

An interesting question is whether the flatness of the Weyl connection
remains in the deformed theory. This question can
equally well be addressed with or without a Weyl component of the
structure group. As we will see in section \ref{DeformedSupergravity},
the answer is negative.

\subsection{The 4-form}\label{Fourform}The 4-form field strength $H$
occurs as a component of the torsion at dimension 1 in the geometric
approach to $11d$ supergravity pursued here. As is well known, it
can also be promoted to a 4-form in {\it superspace}, which we
denote by the same letter. Its components have dimensions ranging
from $-1$ ($H_{\al\be\ga\de}$) to 1 ($H_{abcd}$). $H$ can be
expressed as the exterior derivative of a superspace 3-form
potential $C$, $H=dC$, so its Bianchi identity reads $dH=0$.
Conventional constraints corresponding to redefinitions of the
potential may be imposed, analogous to the ones redefining the
vielbein in section (\ref{ConventionalConstraints}), whereupon the
Bianchi identities cease to be automatically satisfied.

The gauge transformations, field content and deformations are now
related to cohomologies of another complex, namely that containing
$\G$-traceless parts of $n$ symmetrised spinors:
\begin{equation}
\vtop{\baselineskip0pt\lineskip0pt \ialign{
$\hfill#\hfill$&$\,\hfill#\hfill\,$&$\hfill#\hfill$&$\,\hfill#\hfill\,$&$\hfill#\hfill$&$\,\hfill#\hfill\,$&$\hfill#\hfill$&$\,\hfill#\hfill\,$&$\hfill#\hfill$&$\,\hfill#\hfill\,$&$\hfill#\hfill$&$\hfill#\hfill$\cr
&&\ldots&\buildrel\D\over\rightarrow&\Lambda&\buildrel\D\over\rightarrow
  &C&\buildrel\D\over\rightarrow&H&\buildrel\D\over\rightarrow
  &\hbox{BI}&\ldots\cr
\phantom{(00000)}&&&&&&&&&\cr
(00000)&\rightarrow&(00001)&\rightarrow &(00002)&\rightarrow
&(00003)&\rightarrow &(00004)&\rightarrow &(00005)&
                                \ldots\cr
       &   &       &\sea&       &\sea&       &\sea&       &\sea&&\cr
       &   &       &    &(01000)&\rightarrow &(01001)&\rightarrow &(01002)&\rightarrow &(01003)&
                                \ldots\cr
       &   &       &    &       &    &       &\sea&       &\sea&&      \cr
       &   &       &    &       &    &       &    &(02000)&\rightarrow &(02001)&
                                \ldots\cr
       &   &       &    &       &    &       &    &       &&&\phantom{\rightarrow}\cr
}}\label{ElevenComplex}
\end{equation}

The non-conventional constraint that has to be imposed in order to
obtain the undeformed supergravity is the vanishing of the dimension
-1 components of $H$ in the representations
$(02000)\oplus(01002)\oplus(00004)$. The Bianchi identities then
imply the equations of motion for the fields.

The cohomology of the complex (\ref{ElevenComplex}) is
\cite{Cederwall:2001dx}

\begin{equation}
\hskip-2cm \vtop{\baselineskip25pt\lineskip0pt \ialign{
$\hfill#\quad$&$\scriptstyle\,\hfill#\hfill\,$&$\scriptstyle\,\hfill#\hfill\,$
&$\scriptstyle\,\hfill#\hfill\,$&$\scriptstyle\,\hfill#\hfill\,$&$\scriptstyle\,\hfill#\hfill$
&$\scriptstyle\,\hfill#\hfill$&$\scriptstyle\,\hfill#\hfill$
&$\scriptstyle\,\hfill#\hfill$&$\scriptstyle\,\hfill#\hfill$&\quad#\cr
&\ts n=0&\ts n=1&\ts n=2&\ts n=3&\ts n=4&\ts n=5&\ts n=6&\ts n=7&\ts
n=8&\cr \hbox{dim}=-3&\quad(00000)\quad
        &\phantom{\quad(00000)\quad}&\phantom{\quad(00000)\quad}
        &\phantom{\quad(00000)\quad}&\phantom{\quad(00000)\quad}
        &\phantom{\quad(00000)\quad}&\phantom{\quad(00000)\quad}
        &\phantom{\quad(00000)\quad}&\phantom{\quad(00000)\quad}&\cr
        -\Fr52&\bullet&\bullet&               &&&       &\cr
           -2&\bullet&(10000)&\bullet&       &&&       &\cr
       -\Fr32&\bullet&\bullet&\bullet&\bullet&&&       &\cr
           -1&\bullet&\bullet&\raise3pt\vtop{\baselineskip6pt\ialign{
                    \hfill$#$\hfill\cr
                    \scriptstyle(01000)\cr
                    \scriptstyle(10000)\cr}}
            &\bullet&\bullet&&\cr
       -\Fr12&\bullet&\bullet&(00001)
                &\bullet&\bullet&\bullet&&\cr
           0&\bullet&\bullet&\bullet&\raise6pt\vtop{\baselineskip6pt\ialign{
                    \hfill$#$\hfill\cr
                    \scriptstyle(00000)\cr
                    \scriptstyle(00100)\cr
                    \scriptstyle(20000)\cr}}
                &\bullet&\bullet&\bullet&&\cr
       \Fr12&\bullet&\bullet&\bullet&\raise3pt\vtop{\baselineskip6pt\ialign{
                    \hfill$#$\hfill\cr
                    \scriptstyle(00001)\cr
                    \scriptstyle(10001)\cr}}
                &\bullet&\bullet&\bullet&\bullet&\cr
           1&\bullet&\bullet&\bullet&\bullet&\bullet&\bullet
            &\bullet&\bullet&\bullet&\cr
       \Fr32&\bullet&\bullet&\bullet&\bullet
                &\raise3pt\vtop{\baselineskip6pt\ialign{
                    \hfill$#$\hfill\cr
                    \scriptstyle(00001)\cr
                    \scriptstyle(10001)\cr}}
                &\bullet&\bullet&\bullet&\bullet\cr
           2&\bullet&\bullet&\bullet&\bullet
                &\raise6pt\vtop{\baselineskip6pt\ialign{
                    \hfill$#$\hfill\cr
                    \scriptstyle(00000)\cr
                    \scriptstyle(00100)\cr
                    \scriptstyle(20000)\cr}}
                &\bullet&\bullet&\bullet&\bullet&\cr
       \Fr52&\bullet&\bullet&\bullet&\bullet&\bullet&(00001)&\bullet
            &\bullet&\bullet&\cr
           3&\bullet&\bullet&\bullet&\bullet&\bullet
                &\raise3pt\vtop{\baselineskip6pt\ialign{
                    \hfill$#$\hfill\cr
                    \scriptstyle(01000)\cr
                    \scriptstyle(10000)\cr}}
                &\bullet&\bullet&\bullet&\cr
       \Fr72&\bullet&\bullet&\bullet&\bullet&\bullet&\bullet
                &\bullet&\bullet&\bullet&\cr
           4&\bullet&\bullet&\bullet&\bullet&\bullet&\bullet&(10000)
            &\bullet&\bullet&\cr
       \Fr92&\bullet&\bullet&\bullet&\bullet&\bullet&\bullet&\bullet
            &\bullet&\bullet&\cr
           5&\bullet&\bullet&\bullet&\bullet&\bullet&\bullet&\bullet
                &(00000)&\bullet&\cr
    \Fr{11}2&\bullet&\bullet&\bullet&\bullet&\bullet&\bullet&\bullet
            &\bullet&\bullet&\cr
}}\label{ElevenCohomology}
\end{equation}
\vskip\parskip

It is interesting to compare the cohomologies here, referred to
below as ``$H$-cohomology'' to the ones obtained for the geometric
quantities stated in eq. (\ref{gorgias}) (``geometric
cohomology''). Starting with the gauge transformations, we see that
they, in addition to the spinor and vector parameters of superspace
diffeomorphisms, contain a 2-form of dimension -1, which is
expected. At the level of fields, the 4-form field strength in the
geometric cohomology (that can only contain quantities invariant
under 2-form gauge transformations) is replaced by the 3-form
potential. In addition, there is a scalar at dimension 0. The spinor
at dimension $\half$ is still present, but the vector is absent. At
the level of the field equations, we find representations fitting
the Einstein equations as well as the gravitino equations both in
the $H$-cohomology and in the geometric one. The representation
corresponding to the equation of motion for $C$ is present in both,
and the Bianchi identity in (00002) has gone away, which is
consistent with the formulation being based on the potential instead
of the field strength.

In short, the differences between the two cohomologies are in part
attributed to the replacement of the field strength by its
potential, in part to a difference concerning the Weyl connections.

We should mention that although all fields are contained in the
cohomology of the 3-form $C$, there is no existing formalism based
solely on this field, without reference to superspace geometry. One
should therefore not {\it a priori} interpret components of the
$H$-cohomology not present in the geometric cohomology to constitute
independent fields or deformations. Similarly, a field or
deformation occurring in the geometric cohomology but not in the
$H$-cohomology should not be ruled out by inspection only, since it
may be present without explicitly occurring in \eg\ the $H$ field.

In the undeformed supergravity, the components in
$(02000)\oplus(01002)\oplus(00004)$ at dimension $-1$ are taken to
vanish. This is the non-conventional constraint. The scalar at
dimension 0 occurs because $H_{ab\ga\de}$ is not invariant under
Weyl transformations. It is not possible to set it equal to
$2(\G_{ab}){}_{\ga\de}$ by a conventional transformation related to
redefinitions of $C$. A conventional Weyl rescaling of the vielbein
is needed for this.

It is clear that the cohomology in
$(02000)\oplus(01002)\oplus(00004)$ in $H_{\al\be\ga\de}$ is
sufficient to encode modifications to the equations of motion for
all fields in $11d$ supergravity. A detailed analysis of the
superspace Bianchi identities for $H$ up to dimension 0
 has been performed in ref. \cite{Howe:2003cy}.

It has been widely assumed, mostly for \ae sthetical reasons, that
formulations with or without explicit use of $H$ should be
equivalent. This is certainly the case for undeformed supergravity.
It is not obvious, however, that this statement remains true in the
deformed case. As we will see in the following section, the purely
geometrical approach of this paper allows for non-vanishing Weyl
curvature, which is expressible in terms of the torsion components
in $(11000)\oplus(10002)$ at dimension 0. On the other hand, the
$H$-cohomology does not contain the vector component of the Weyl
connection. The geometric cohomology contains the Bianchi identity
for the $H$-field, and we are not guaranteed that a deformation will
allow for the identification of a globally closed 4-form, although
this is of course not excluded.

It was shown in \cite{Howe:2003cy} that the system including
the $H$ field implies the
Bianchi identities in the geometric picture, up to dimension
$\half$. Provided no new constraints arise at dimensions higher than
$\half$ (this is indeed the case at dimension 1 as we will see in
the following sections),
this shows that the $H$ field formulation
implies the geometric formulation.

For the two formulations to be equivalent
the converse should hold as well, and one would expect to find
integrability conditions on the $X$ and $Y$ tensors in
$(11000)\oplus(10002)$ stating their integrability to the tensors in
$(02000)\oplus(01002)\oplus(00004)$. As explained in detail in
the following section, so far we have not found any candidates up to
dimension 1 for such conditions, other than the constraints in
$(11001)\oplus(10003)$. However, it is not at all clear that
the latter can play this r\^ole.

A conclusion concerning the equivalence of our geometric approach
with the one containing the superspace 4-form has to await further
results at the level of the equations of motion.

\section{Deformed Supergravity}\label{DeformedSupergravity}

In this section we solve the SSBI's by using the most general form of
the torsion components, subject to the conventional constraints
analysed previously. In particular, this implies that the
zero-dimension component of the torsion includes the $X$ and $Y$
tensors introduced in equation \eqref{GeneralDimZeroTorsion}. Recall
that $X$ and $Y$ are set to zero in the case of ordinary 11d
supergravity and as a consequence most of the torsion components are
set to zero by the SSBI's. As we have seen in section
\ref{Undeformed}, the only components of the torsion that are not
set to zero by the SSBI's correspond to the 4-form field strength
$H:=A_{(4)}=-2A^\prime_{(4)}$ and the gravitino field strength
$\tilde{t}$. The curvature tensor $R$ appears at dimension 2.

In the deformed case, this is no longer the case: the SSBI's will now
solve for the previously vanishing components of the torsion in
terms of (derivatives of) $X$ and $Y$. It is by substituting $X$ and
$Y$ into the SSBI's and solving up to dimension 2, that one arrives
at the deformed equations of motion and, eventually, the lagrangian,
after specifying $X$ and $Y$ in terms of the physical fields.
Clearly, in this approach the deformation is parametrised by $X$,
$Y$.

Eleven-dimensional supergravity has no coupling constant, since
there is no scalar in 11d whose VEV could play this role. There is,
however, the possibility of a low-energy (curvature) expansion in
the Planck length $\ell_P$. It is believed that the first such
correction occurs at order $\ell_P^6$, corresponding to the still
undetermined $R^4$ superinvariant\footnote{Subject to some plausible
assumptions, it was argued in \cite{Howe:2003cy} that there is a
unique $R^4$ superinvariant consistent with the $C\wedge X_8$
Chern--Simons term.}. As has recently been shown in
\cite{Tsimpis:2004rs}, at order $\ell_P^3$ there appears a
superinvariant which turns out to be topological in nature in that
it can be removed by appropriately shifting the flux quantisation
condition of the gauge field. More generally, let us introduce a
deformation parameter $\beta$ and consider the tensors $X$ and $Y$
to be of order $\beta$. The reader may want to think of $\beta$ as
being proportional to $\ell_P^6$, but our analysis is valid
irrespectively of the actual value of $\beta$. We treat the problem
of solving the deformed Bianchi identities perturbatively, to first
order in $\beta$. This means, in particular, that we ignore terms
quadratic or higher in  $X$, $Y$ \footnote{Note that this an
improvement on the analysis of
\cite{Cederwall:2000ye,Cederwall:2000:2} where non-linear terms of
the form $H Y$ and bosonic derivatives of $Y$ were ignored. In
addition, in those references $X$ was set to zero.}. Furthermore the
Weyl spinor $\tilde{S}$ is also of order $\beta$, since it is set to
zero by the SSBI's in the undeformed case. However, as noted in
\cite{Howe:1997rf}, this is only true for a simply connected
space-time manifold. We will henceforth assume this to be the case.

Let us now turn to the actual procedure of solving the deformed
SSBI's. Just as in the undeformed case, we need to project onto each
irreducible representation. This is most conveniently done by
appropriately contracting with gamma matrices. The computation is
straightforward and conceptually the same as in the undeformed case.
It is however much more tedious and we have found GAMMA
\cite{Gran:2001:2,Gran:2004} to be an extremely useful tool.

The reader can find all the details of the calculation in app. B.
Here we summarise a few salient points:

$\bullet$ At dimension $\half$,
the BI's impose constraints on the tensors  $X$, $Y$. Explicitly these
read
\begin{align}
Y^1_{a_1\dots a_5,b}&=0\komma\cr Y^1_{a_1a_2,b}&={1\over
7}X^1_{a_1a_2,b}\komma\label{hera}
\end{align}
where the superscript refers to the $\theta$-level of the
corresponding superfield; our notation is further explained in app.
B. These constraints restrict the possible deformations, \ie, the
possible admissible expressions of $X$, $Y$ in terms of physical
fields. The problem of finding the explicit form of $X$, $Y$ which
satisfy \eqref{hera} and are not removable by field redefinitions of
the form
\begin{equation}
T^a_{\alpha\beta}\rightarrow T^a_{\alpha\beta}+D_{(\alpha}
\delta E_{\beta)}{}^a~,
\end{equation}
is equivalent to solving the spinorial cohomology problem for the
theory \cite{Cederwall:2001dx,Cederwall:2001xk}. The case referred
to here, \ie, when $X$, $Y$ are functions of the physical fields of
the theory, was dubbed in \cite{Howe:2003cy} `spinorial cohomology
with physical coefficients'. This is to be contrasted with `spinorial
cohomology with unrestricted coefficients', in which case $X$, $Y$
are freely given superfields. The latter cohomology is summarised,
for the case of 11d supergravity, in table (\ref{gorgias}) of section
\ref{PhysicalConstraintsAndCohomology}. Spinorial cohomology with
unrestricted coefficients is isomorphic to pure spinor cohomology
\cite{Nilsson:1986:3,Howe:1991},  which has recently found
application in the covariant quantisation of the superstring
\cite{Berkovits:2000fe}.

$\bullet$ All the components of the dimension-$\half$ torsion
are solved for in terms of (spinor derivatives of) $X$, $Y$,
except for the Weyl spinor $\tilde{S}$. However, this does not imply that
$\tilde{S}$ is an extra degree of freedom,
because its derivative $D_{(\alpha}\tilde{S}_{\beta)}$
(which is part of the Weyl curvature)
is completely determined in terms $X$, $Y$,
by the dimension-1 BI. Explicitly
\begin{align}
D\Gamma_a\tilde{S}&=64A_a\nn\komma\\
\frac{11}{8}D\Ga_{ab}\tilde{S}&=4D^fX_{ab,f} +A^{i_1\dots
i_4}Y_{i_1\dots
i_4[a,b]}+ 16A_{ab}+72A_{ab}^{\prime}\nn\komma\\
\frac{11}{8}D\Ga_{a_1\dots a_5}\tilde{S}& =-4D^eY_{a_1\dots
a_5,e}-120A_{[a_1a_2a_3}{}^iX_{a_4a_5],i}
+\frac{1}{3}\epsilon_{[a_1a_2a_3|i_1\dots i_8} A^{i_1i_2
i_3}{}_{|a_4|}Y^{i_4\dots i_8}{}_{,|a_5]}\punkt
\end{align}
Note that once the deformation is turned on, \ie, for $X$,
$Y\neq 0$, the Weyl curvature ceases to be flat.


$\bullet$ At dimension $1$ the SSBI's impose a number of equations which
appear to be new constraints on $X$, $Y$. Explicitly
\begin{alignat}{3}
A\circ X_{a_1a_2,b} &=\frac{11}{15300}A^\prime
\circ Y_{a_1a_2,b}+\frac{120}{17}X^2_{a_1a_2,b}
-\frac{80}{17}X^{2\prime}_{a_1a_2,b}&\nn\\
&+\frac{280}{17}Y^2_{a_1a_2,b}
-\frac{3360}{17}Y^{2\prime}_{a_1a_2,b}&(11000)\nn\komma\\
A\circ Y^{(2)}_{a_1a_2a_3,b}&=\frac{351}{259}A\circ
Y^{(1)}_{a_1a_2a_3,b} -\frac{405}{37}DX_{a_1a_2a_3,b}
-\frac{1080}{37}X^{2}_{a_1a_2a_3,b}&\nn\\
&-\frac{1080}{37}X^{2\prime}_{a_1a_2a_3,b}
-\frac{1080}{37}Y^{2}_{a_1a_2a_3,b}
-\frac{540}{37}Y^{2\prime}_{a_1a_2a_3,b}&\nn\\
&-\frac{17280}{37}Y^{2\prime\prime}_{a_1a_2a_3,b}&(10100)\nn\komma\\
A\circ X^{(1)}_{a_1\dots a_4,b}&=\frac{81}{37}A\circ
X^{(2)}_{a_1\dots a_4,b} +\frac{119}{88800}A\circ Y^{(1)}_{a_1\dots
a_4,b} -\frac{217}{66600}A\circ Y^{(2)}_{a_1\dots a_4,b}&\nn\\
&-\frac{7}{111}DY_{a_1\dots a_4,b}
+\frac{60}{37}X^{2}_{a_1\dots a_4,b}
+\frac{60}{37}X^{2\prime}_{a_1\dots a_4,b}&\nn\\
&-\frac{30}{37}Y^{2}_{a_1\dots a_4,b}
-\frac{60}{37}Y^{2\prime}_{a_1\dots a_4,b}
+\frac{20}{37}Y^{2\prime\prime}_{a_1\dots a_4,b}&\nn\\
&-\frac{80}{37}Y^{2\prime\prime\prime}_{a_1\dots a_4,b}&(10010)\nn\komma\\
A\circ Y^{(2)}_{a_1\dots a_5,b}&=-\frac{11}{2016}A\circ X_{a_1\dots
a5,b} -\frac{11}{42}A\circ Y^{(1)}_{a_1\dots a_5,b} +X^{2}_{a_1\dots
a_5,b}&\nn\\
&-X^{2\prime}_{a_1\dots a_5,b}
-\frac{5}{2}Y^{2}_{a_1\dots a_5,b} -3Y^{2\prime}_{a_1\dots a_5,b}&\nn\\
&+5Y^{2\prime\prime}_{a_1\dots a_5,b}
+4Y^{2\prime\prime\prime}_{a_1\dots a_5,b}
+\frac{10}{3}Y^{2\prime\prime\prime\prime}_{a_1\dots a_5,b}&(10002)\komma\nn\\
DY_{a_1\dots a_5,b}&=\frac{55}{12}A\circ X_{a_1\dots a5,b} +3220
A\circ Y^{(1)}_{a_1\dots a_5,b} -840 X^{2}_{a_1\dots a_5,b}&\nn\\
&+840 X^{2\prime}_{a_1\dots a_5,b}
-7980 Y^{2}_{a_1\dots a_5,b} -7560 Y^{2\prime}_{a_1\dots a_5,b}&\nn\\
&+5880 Y^{2\prime\prime}_{a_1\dots a_5,b} +6720
Y^{2\prime\prime\prime}_{a_1\dots a_5,b} +5600
Y^{2\prime\prime\prime\prime}_{a_1\dots a_5,b}&(10002)\komma \label{taube}
\end{alignat}
where the quantities involved are defined in app. B. However, all
these should follow from the dimension-$\half$ constraints
\eqref{hera}. This is expected merely on the grounds of
representation theory. Namely, taking the tensor product of a spinor
and the irreducible representations occurring in the dimension-$\half$
constraints, leads to a number of irreducible representations
occurring in the dimension-1 SSBI. These are exactly the ones we
find above. Explicitly
\begin{align}
(00001)\otimes (11001)&=(11000)\oplus (10100)\oplus
(10010)\oplus(10002) \oplus \dots \komma\cr (00001)\otimes (10003)&=
(10002) \oplus\dots\punkt\cr
\end{align}
In conclusion, no new constraints on $X$ and $Y$ occur at dimension
1.

$\bullet$ The fact that we find no new constraints on $X$ and $Y$ at
dimension $1$ is a strong indication that the computation we have
done is correct\footnote{The results in the previous publications
\cite{Cederwall:2000ye,Cederwall:2000:2} are not entirely correct.},
since any computational error generically introduces extra
constraints. This also implies that there are no bugs in GAMMA
\cite{Gran:2001:2,Gran:2004}.

$\bullet$ Apart from the purely geometrical description of 11d
supergravity in terms of the torsion, the system admits an
alternative formulation in terms of a closed 4-form in
superspace. The SSBI's for the 4-form were analysed in detail in
\cite{Howe:2003cy} and constraints analogous to \eqref{hera} were
derived. It was also shown how to make contact with the supertorsion
formulation, by deriving the expressions of $X$, $Y$ in terms of the
lowest, purely spinorial, component of the 4-form. Furthermore it
was shown that these expressions should automatically satisfy the
supertorsion constraints \eqref{hera}. In other words, the 4-form
formulation implies the geometric one.

The converse is less straightforward: If
the geometric and the 4-form formulations
turn out to be equivalent, the constraints \eqref{hera}
will be the
integrability conditions for the system to be equivalent
to a closed 4-form in superspace. It is far from clear, however, that
this will turn out to be the case.

$\bullet$ The problem of computing explicit representatives of
spinorial cohomology with physical coefficients is extremely
complicated in general, even at order $\ell_P^6$. It was argued in
\cite{Howe:2003cy} that it would be advantageous to tackle this
issue within the context of the 4-form (or the `dual' 7-form)
formulation of supergravity. In order to arrive at the deformed
equations of motion and eventually at the lagrangian, one would
still need to make contact with the geometric formulation of the
present paper. In this sense the two approaches are complementary.

$\bullet$ It was argued in \cite{Howe:2003cy} that no new
constraints appear at dimensions higher than $\half$. We have seen
that this is indeed the case at dimension $1$. We expect this result
to hold at higher dimensions as well. This means in particular that
at dimension $3\over2$ the SSBI's simply solve for the corresponding
components of the torsion. In practice, instead of continuing our
analysis of the SSBI's at dimension $3\over2$ or higher, in order to
arrive at the equations of motion it is more convenient to simply
substitute the explicit expressions of $X$, $Y$ in terms of the
physical fields directly into the BI, along the lines of
\cite{Cederwall:2001td}.

Let us briefly review the procedure. As we can see from
\eqref{LorentzComponentBI} the dimension-$3\over2$ torsion
component is given by the spinor derivative of the dimension-$1$
torsion which is, in its turn, given by two spinor derivatives on
$X$, $Y$. Schematically:
\begin{equation}
T_{3\over2}\sim D_\alpha T_{1}\sim D_{\alpha}^3X+D_{\alpha}^3Y\punkt
\end{equation}
The relevant objects to compute are then $D_{\alpha}X$
and $D_{\alpha}Y$. To first order in $\beta$, this can
readily be done as follows. Recall that $X$, $Y$ are assumed to be
functions of the physical field strengths of the theory
$H,\tilde{t},R$.
The action of the spinor derivative on the latter
is known (from the
undeformed theory) to lowest order in $\beta$. Schematically
\begin{align}
D_{\alpha}H&=\tilde{t}+{\cal O}(\beta)\komma\cr D_{\alpha}\tilde{t}
&=R+\partial
H+H^2+{\cal O}(\beta)\komma\cr D_{\alpha}R&=\partial \tilde{t}
+H \tilde{t}+{\cal
O}(\beta)\punkt\label{PRIP}
\end{align}
As noted before, the tensors $X$, $Y$ are of order ${\cal
O}(\beta)$. Therefore the ${\cal O}(\beta)$ terms in \eqref{PRIP}
can be ignored as they would give rise to ${\cal O}(\beta^2)$ terms
in $D_{\alpha}X$, $D_{\alpha}Y$.

In appendix B, we use the described method to compute some of the
relevant SSBI's at dimension $3\over2$ and 2, leading to equations of
motion.

\section{Summary and conclusions}

M-theory has, as far as we know, no coupling constants in which to
do perturbation theory, and is therefore often viewed as a
non-perturbative second quantised theory without well defined
one-particle states. As a consequence, in order to avoid discussing
the full theory
 we must
rely on some kind of low energy approximation. At low energies the theory
has eleven-dimensional supersymmetry and local Lorentz covariance, and one may ask which generalizations of
ordinary eleven-dimensional supergravity are compatible with imposing only these symmetries.
This may or may not yield a more general structure than a low energy
approximation of M-theory.

In this paper we implement these symmetries by the use of
superspace. From the supervielbein one defines in a standard fashion
the supertorsion and super-Riemann tensors, and derives their
respective super-Bianchi identities which we refer to as the
geometric SSBI's. This step has in fact introduced all three fields
in low energy eleven-dimensional supergravity; the elfbein, the spin $3\over2$
field and the three-index antisymmetric tensor gauge field. This can
be seen by setting  the zero dimension torsion tensor equal to a
gamma matrix which turns the SSBI's into the lowest order dynamical
supergravity equations corresponding to all the three fields
\cite{Howe:1997rf}.

However, as explained in sect. \ref{ConventionalConstraints},
 by using only the freedom of performing field redefinitions on the supervielbein and
spin connection one finds that the zero dimension torsion component is in the most general
situation actually expressed
in terms of two unspecified tensor superfields, X and Y,
 in certain representations of the Lorentz structure group.

In section \ref{DeformedSupergravity} we have taken a step towards
solving the SSBI's in terms of these two tensors by presenting the
solution to linear order in X and Y of all SSBI's of dimension $\half$
and 1. This solution is then used in order to obtain deformed
equations of motion at dimension $3\over2$ and 2. 
The problem of finding explicit forms of the equations of
motion is then shifted to finding out the structure of the tensors X
and Y in terms of the physical gauge covariant fields, \ie, the
Riemann tensor etc. This is a very difficult problem, much more
complex than the corresponding problem in SYM, simply because the
number of independent combinations of fields in the appropriate
representations is large, but once the structure of X and Y are
known the full theory is obtainable (as can be seen from the
formulas in app. B). The analysis of X and Y is a kind of spinorial
cohomology problem, discussed previously in the simpler case of
super-Yang--Mills theory in ref.
\cite{Cederwall:2001bt,Cederwall:2001td,Cederwall:2002df}. In the
case of supergravity some results were recently derived in ref.
\cite{Tsimpis:2004rs}, where the cohomology was solved to order
$\ell_P^3$. At this order the first
 possible non-trivial term appeared which turns out to be purely topological in nature and
related to the 4-form quantisation conditions discussed by Witten
in ref. \cite{Witten:1996md}. The often discussed $R^4$ terms are
expected at order $\ell_P^6$ and will require a substantial amount
of work to analyse in full generality. Even the task of just writing
down an Ansatz for $X$ and $Y$ in terms of the physical fields
(obeying the undeformed field equations) looks formidable, since the
independent combinations in the representations of $X$ and $Y$ at
this dimension are counted in thousands. We would like to return to
this in a future publication, but think that input of some other
kind is needed to avoid that type of brute force calculation.

A different approach to finding the form of X and Y is to introduce
also the superspace BI's for the gauge fields, either the one  for
the 4-form field strength only or in combination with the SSBI's
for the dual 6-form potential. In the latter case the anomaly
related term $C_3X_8$ in the lagrangian can be most naturally
introduced in superspace via the generalised SSBI
$dH_7={1\over2}H_4\wedge H_4+X_8$. Once this is done the central role played
by the dimension zero torsion is taken over by the lowest dimension
component of the the 4-form $H_4$, namely $H_{\alpha \beta \gamma
\delta}$ as discussed in detail in \cite{Howe:2003cy}. Restricting
this field affects the structure of the theory, \eg, setting it to
zero leaves the theory in the lowest second order form
\cite{Cederwall:2000ye}. More important, however, is that the
deformation in the geometric sector can probably more easily be
derived by relating it to the deformation in the gauge sector, as
emphasised in \cite{Howe:2003cy,Howe:2004}.

In fact, the geometric sector of the superspace version of the
theory may be viewed as secondary to the gauge sector. That is,
since the supertorsion tensor appears explicitly in the component
equations of the gauge SSBI's one can obtain the geometric
deformation in terms of the deformation in the gauge sector and
consistency will probably also require the geometric SSBI's to be
satisfied. The cohomology tables presented in sect. \ref{SuperSpace}
have a bearing on this issue. As one can see by comparing the
tables, the geometric and gauge systems do not seem to be in a one
to one correspondence, a fact that is not yet understood. The
differences were discussed in sect. \ref{Fourform}, where some of
them where explained. Some differences remain obscure, however,
among them the question of the existence of a closed 4-form in
the geometric formulation. 
Perhaps the most
efficient way to proceed will in the end turn out to be to use all
the SSBI's simultaneously, provided the discrepancies between the
cohomology tables are not a symptom of any deeper structural
differences between the two systems.

\section*{Acknowledgments}

We would like to thank P.~Howe and P.~Vanhove for useful discussions.
This work is supported in part by the
Swedish Research Council.

\appendix
\section{Undeformed 11d supergravity in superspace}

The nonzero components of the supertorsion and supercurvature 
of undeformed 11d supergravity are given by
\begin{align} 
T_{\al\beta}{}^c&=2(\Gamma^c)_{\al\beta}\nn\\
T_{a\beta}{}^{\ga}&= -{1\over36}\left((\Gamma^{bcd})_{\beta}{}^{\ga}
H_{abcd} +{1\over8} (\Gamma_{abcde})_{\beta}{}^{\ga} H^{abcd}\right)
\end{align}
(the field strength $H$ is related to the torsion component $A$ used
in this paper by $H=-6A$)
and
\begin{align}
R_{\al\beta ab} &=-{1\over3}\left((\Gamma^{cd})_{\al\beta} H_{abcd} +
{1\over24}(\Gamma_{abcdef})_{\al\beta} H^{cdef}\right)\nn\\ 
R_{\al bcd}&=-(\Gamma_bT_{cd})_\al-(\Gamma_cT_{bd})_\al
+ (\Gamma_dT_{bc})_\al ~.
\label{tara}
\end{align}
Note that the Lorentz condition implies
\begin{equation}
R_{AB\al}{}^\beta=\frac{1}{4}R_{ABcd}(\Gamma^{cd})_\al{}^\beta~.
\end{equation}
The action of the spinorial derivative on the physical
field strengths and their equations of motion are given by
\begin{align}
D_{\al}H_{abcd}&=-12(\Gamma_{[ab}T_{cd]})_\al\nn\\
D_{\al}T_{ab}{}^\beta&=\frac{1}{4}R_{abcd}(\Gamma^{cd})_{\al}{}^\beta
-2D_{[a}T_{b]\al}{}^\beta-2T_{[a|\al}{}^\epsilon T_{|b]\epsilon}{}^\beta\nn\\
D_{\al}R_{abcd}&=2D_{[a|}R_{\al|b]cd}-T_{ab}{}^\epsilon R_{\epsilon \al cd}
+2T_{[a|\al}{}^\epsilon R_{\epsilon|b]cd}
\end{align}
\label{derivatives}
and
\begin{align}
D_{[a}H_{bcde]}&=0\nn\\
D^fH_{fabc}&=-\frac{1}{2(4!)^2}
\varepsilon_{abcd_1\dots d_8}H^{d_1\dots d_4}H^{d_5\dots d_8} \nn\\
(\Gamma^aT_{ab})_\al&=0\nn\\
R_{ab}-\frac{1}{2}\eta_{ab}R&=-\frac{1}{12}\left(H_{adfg}H_b{}^{dfg}
-\frac{1}{8}\eta_{ab}H_{dfge}H^{dfge}  \right) ~.
\end{align}
The above equations can be integrated to an action whose
bosonic part is given by eq. (\ref{Lagrangian}). 
%
%

\section{Solution of the SSBI's}

In this section we present the full details of our solution to
the SSBI's at dimension $\half$ and dimension $1$, and partial results
(deformations of the equations of motion) at dimension $3\over2$ and 2.
The procedure was explained in the main body of the paper:
the torsion components are
expanded in irreducible representations as in
(\ref{ConventionalTorsion}) of section \ref{Undeformed},
and substituted into the SSBI's given in equation (\ref{LorentzComponentBI}).
By appropriately contracting with gamma matrices one then projects
onto each irreducible representation.
The fields
$H_{abcd}:=A_{abcd}=-2A^{\prime}_{abcd},\, \tilde{t}_{ab}{}^\ga, \,
R_{abcd}$,
are the only ones that are non-zero in the ordinary
(undeformed) eleven-dimensional supergravity.
All other superfields are of linear order in
the deformation parameter $\beta$, as explained
in section \ref{DeformedSupergravity}.
In the following we discard terms of order ${\cal O}(\beta^2)$.
This is the {\it only} approximation we use.

A note on notation: A tilde denotes a spinor superfield
(\eg\ $\tilde{S}$).
A numerical superscript $n$ on the superfield
$A$ denotes a superfield $A^n$ sitting at
$\theta^n$-level in $A$.

\subsection{The dimension-$\half$ SSBI's}

The SSBI at dimension $\half$ reads
\begin{equation}
0={R_{(\al\be\ga)}}^d=D_{(\al}{T_{\be\ga)}}^d
+{T_{(\al\be}}^E{T_{|E|\ga)}}^d.
\label{dimhalfssbi}
\end{equation}
It decomposes in irreducible
representations as
\begin{align}
(00001)^{{\otimes}_s 3}&\otimes(10000)= \nn\\
&2\times(00001)\oplus(00003)
\oplus(00011)\oplus(00101)\oplus \nn\\
&2\times(01001)\oplus3\times(10001)
\oplus(10003)\oplus(11001)\oplus(20001)
\end{align}
Since the SSBI involves the fields at $\theta^1$-level
in $X$ and $Y$,
we also need to expand $D_{\al}X_{a_1a_2,b}$ and
$D_{\al}Y_{a_1\ldots a_5,b}$ into
irreducible tensors. Explicitly:
\begin{align}
DY_{a_1\dots a_5,b}&=
5\left(\Ga_{[a_1}{}Y^1_{a_2\dots a_5]b}
+\Ga_b{}Y^1_{a_1\dots a_5}\right) \nn\\
&+10\left( \Ga_{[a_1a_2}{}Y^1_{a_3a_4a_5]b}
+\Gamma_{b[a_1}{}Y^1_{a_2\dots a_5]}
+{6\over 7}{}Y^1_{[a_1\dots a_4}\eta_{a_5]b} \right)\nn\\
&+10\left( \Gamma_{[a_1a_2a_3}{}Y^1_{a_4a_5]b}
+\Gamma_{b[a_1a_2}{}Y^1_{a_3a_4a_5]}
+{12\over 7}\eta_{b[a_1}\Gamma_{a_2}{}Y^1_{a_3a_4a_5]} \right)\nn\\
&+5\left( \Gamma_{[a_1\dots a_4}{}Y^1_{a_5]b}
+\Gamma_{b[a_1a_2a_3}{}Y^1_{a_4a_5]}
+{18\over 7}\eta_{b[a_1}\Gamma_{a_2a_3}{}Y^1_{a_4a_5]} \right)\nn\\
&+\left( \Gamma_{a_1\dots a_5}{}Y^1_{b}
+\Gamma_{b[a_1\dots a_4}{}Y^1_{a_5]}
+{24\over 7}\eta_{b[a_1}\Gamma_{a_2a_3a_4}{}Y^1_{a_5]} \right)\nn\\
&+{}Y^1_{a_1\dots a_5,b}+5\Gamma_{[a_1}{}Y^1_{a_2\dots a_5],b}
+10\Gamma_{[a_1a_2}{}Y^1_{a_3a_4a_5],b}\nn\\
&+10\Gamma_{[a_1a_2a_3}{}Y^1_{a_4a_5],b}
+5\Gamma_{[a_1\dots a_4}{}Y^1_{a_5],b}
\end{align}
and
\begin{align}
DX_{a_1a_2,b}&=
2\left( \Ga_b{}X^1_{a_1a_2} -\Ga_{[a_1}{}X^1_{a_2]b} \right)
-\left( \Ga_{a_1a_2}{}X^1_{b} -\Ga_{b[a_1}{}X^1_{a_2]}
+{3\over 10}{}X^1_{[a_1}\eta_{a_2]b} \right)\nn\\
&+2\Ga_{[a_1}{}X^1_{a_2],b}+{}X^1_{a_1a_2,b}\punkt
\end{align}
The inversions which we will need later are
\begin{align}
{X}^1_a&={10\over 1287}\Ga^{ij}DX_{ij,a}\komma\nn\\
{X}^1_{a_1a_2}&={4\over 117}(\Ga^iDX_{a_1a_2,i}
-{1\over 8}\Ga_{[a_1|}{}^{ij}DX_{ij,|a_2]})\komma\nn\\
{X}^1_{a_1 , a_2}&={10\over 117}\big(\Ga^i DX_{i (a_1,a_2)}
+{3\over 20}\Ga_{(a_1|}{}^{ij}DX_{ij|,a_2)}\big)\komma\nn\\
&\dots \komma
\end{align}

\begin{align}
{Y}^1_{a_1\dots a_5}&={1\over 195}
(\Ga^iDY_{a_1\dots a_5,i}-\Ga_{[a_1}{}^{ij}DY_{a_2\dots a_5]i,j}
-{1\over 2}\Ga_{[a_1a_2}{}^{ijk}DY_{a_3a_4a_5]ij,k}\nn\\
&+{1\over 6}\Ga_{[a_1a_2a_3}{}^{ijkl}DY_{a_4a_5]ijk,l}
+{1\over 24}\Ga_{[a_1\dots a_4}{}^{ijklm}DY_{a_5]ijkl,m} )\komma\nn\\
{Y}^1_{a_1\dots a_4}&={1\over 312}
(\Ga^{ij}DY_{a_1\dots a_4i,j}+{4\over 5}\Ga_{[a_1}{}^{ijk}DY_{a_2a_3a_4]ij,k}
-{3\over 10}\Ga_{[a_1a_2}{}^{ijkl}DY_{a_3a_4]ijk,l}\nn\\
&-{1\over 15}\Ga_{[a_1a_2a_3}{}^{ijklm}DY_{a_4]ijkl,m})\komma\nn\\
{Y}^1_{a_1a_2a_3}&=-{1\over 936}
(\Ga^{ijk}DY_{a_1a_2a_3ij,k}-{3\over 5}\Ga_{[a_1}{}^{ijkl}DY_{a_2a_3]ijk,l}
-{3\over 20}\Ga_{[a_1a_2}{}^{ijklm}DY_{a_3]ijkl,m})\komma\nn\\
{Y}^1_{a_1a_2}&=-{1\over 3510}
(\Ga^{ijkl}DY_{a_1a_2ijk,l}+{2\over 5}
\Ga_{[a_1}{}^{ijklm}DY_{a_2]ijkl,m})\komma\nn\\
{Y}^1_{a}&={1\over 61776}
\Ga^{ijklm}DY_{ijklm,a}\komma\nn\\
{Y}^1_{a_1,a_2}&={1\over 5616}
(\Ga^{ijkl}DY_{ijkl(a_1,a_2)}-{6\over 35}
\Ga_{(a_1|}{}^{ijklm}DY_{ijklm,|a_2)})\komma\nn\\
&\dots \komma
\end{align}
where the ellipses stand for the irreducible representations
which drop out of the SSBI's and will  not be needed
in the following. Plugging the above and the explicit
expressions for the torsion components (\ref{ConventionalTorsion})
into the SSBI, we get
\begin{align}
\tilde{Z}_{a1\dots a5}&=-5Y^1_{a_1\dots a_5},\\
\tilde{Z}_{a1\dots a4}&=-{13\over 7}Y^1_{a_1\dots a_4}\\
\tilde{Z}_{a1\dots a3}&=-{39\over 14}Y^1_{a_1\dots a_3},\\
\tilde{Z}_{ab}&={1\over 8}X^1_{ab}
-{255\over 112}Y^1_{ab},\\
\tilde{Z^{\prime}}_{ab}&=-{17\over 8}X^1_{ab}
+{25\over 16}Y^1_{ab},\\
\tilde{S}_{a}&=\frac{693}{460}X^1_a
-\frac{198}{115}Y^{1}_{a}, \\
\tilde{Z}_a&=\frac{221}{920}X^1_a
-\frac{741}{1610}Y^{1}_{a} ,\\
\tilde{Z}^\prime_a&=-\frac{923}{460}X^1_a
+\frac{221}{115}Y^{1}_{a} ,\\
\tilde{Z}&=\frac{3}{88}\tilde{S},\\
\tilde{Z^\prime}&=-\frac{1}{44}\tilde{S},\\
Y^1_{a_1...a_5,b}&=0,\\
Y^1_{a1a2,b}&={1\over 7}X^1_{a_1a_2,b},\\
\tilde{S}_{a,b}&=X^1_{a,b} +14Y^1_{a,b}.
\end{align}

\subsection{The dimension-$1$ SSBI's}

We now turn to the SSBI's at dimension 1. There are two such
equations, namely
\begin{align}
&R_{\al\be c}{}^d=2D_{(\al}T_{\be)c}{}^d+D_cT_{\al\be}{}^d
    +T_{\al\be}{}^ET_{Ec}{}^d+2T_{c(\al}{}^ET_{|E|\be)}{}^d\komma\\
&R_{(\al\be\ga)}{}^\de=D_{(\al}T_{\be\ga)}{}^\de
+T_{(\al\be}{}^ET_{|E|\ga)}{}^\de\punkt
\label{DimOneBI}
\end{align}
These decompose as
\begin{align}
(00001)^{{\otimes}_s 2}\otimes (10000)^{\otimes 2}&=
(00000)\oplus 3\times (10000)\oplus 3\times(01000)\oplus\nn\\
&2\times(00100)\oplus 2\times(00010)\oplus 3\times(00002)\oplus\nn\\
&2\times(10002)\oplus 2\times(10010)\oplus 2\times(10100)\oplus\nn\\
&2\times(11000)\oplus 2\times(20000)\oplus\dots
\end{align}
and
\begin{align}
(00001)^{{\otimes}_s 3}\otimes(00001)&=
(00000)\oplus 2\times(10000)\oplus3\times(01000)\oplus\nn\\
&3\times(00100)\oplus3\times(00010) \oplus4\times(00002)
\oplus \nn\\
&3\times(10002)\oplus 2\times(10010)\oplus 2\times(10100)\oplus\nn\\
&2\times(11000)\oplus (20000)\oplus\dots ~,
\end{align}
respectively. We will need the $\theta^2$-level expansion of
$X_{ab,c},\, Y_{abcde,f}$. We have
\begin{align}
(00001)^{{\otimes}_a 2}\otimes(10002)&=
(01000)\oplus
2\times(00100)\oplus2\times(00010) \oplus2\times(00002)\nn\\
&5\times(10002)\oplus 4\times(10010)\oplus 3\times(10100)\oplus\nn\\
&2\times(11000)\oplus \times(20000)
\oplus \dots
\end{align}
and
\begin{align}
(00001)^{{\otimes}_a 2}\otimes(11000)&=
(01000)\oplus
(00100)\oplus(00010) \oplus(00002)\nn\\
&2\times(10002)\oplus 2\times(10010)\oplus 2\times(10100)\oplus\nn\\
&2\times(11000)\oplus \times(20000)
\oplus \dots ~.
\end{align}
Explicitly we expand
\begin{align}
\frac{1}{10}&D_{[\al}D_{\be]}Y_{a_1\ldots a_5,b}                \nn\\
&=\Ga_{[a_1a_2}{}^eY^2_{a_3a_4a_5]be}
    +\Ga_{b[a_1}{}^eY^2_{a_2\dots a_5]e}
    +\frac{6}{7}\eta_{b[a_1}\Ga_{a_2}{}^{e_1e_2}Y^2_{a_3a_4a_5]e_1e_2}  \nn\\
&+\frac{1}{6}\big(\Ga_{[a_1\ldots a_4}{}^{e_1e_2e_3}Y^{2\prime}_{a_5]be_1e_2e_3}
    +\Ga_{b[a_1a_2 a_3}{}^{e_1e_2e_3}Y^{2\prime}_{a_4a_5]e_1e_2e_3}
    +\frac{6}{7}\eta_{b[a_1}
        \Ga_{a_2a_3 a_4}{}^{e_1\ldots e_4}Y^{2\prime}_{a_5]e_1\ldots e_4}\big)   \nn\\
&+\Ga_{[a_1a_2a_3}{}^eY^2_{a_4a_5]be}
    +\Ga_{b[a_1a_2}{}^eY^2_{a_3a_4a_5]e}
    -\frac{6}{7}\eta_{b[a_1}\Ga_{a_2a_3}{}^{e_1e_2}Y^2_{a_4a_5]e_1e_2}  \nn\\
&+\frac{1}{6}\big(\Ga_{a_1\ldots a_5}{}^{e_1e_2e_3}Y^{2\prime}_{be_1e_2e_3}
    +\Ga_{b[a_1\ldots a_4}{}^{e_1e_2e_3}Y^{2\prime}_{a_5]e_1e_2e_3}
    -\frac{6}{7}\eta_{b[a_1}
        \Ga_{a_2\ldots a_5]}{}^{e_1\ldots e_4}Y^{2\prime}_{e_1\ldots e_4}
\big)   \nn\\
&+\Ga_{[a_1a_2a_3}Y^2_{a_4a_5]b}
    +\Ga_{b[a_1a_2}Y^2_{a_3a_4a_5]}
    -\frac{6}{7}\eta_{b[a_1}\Ga_{a_2a_3}{}^eY^2_{a_4a_5]e}          \nn\\
&+\frac{1}{2}\big(\Ga_{a_1\ldots a_5}{}^{e_1e_2}Y^{2\prime}_{be_1e_2}
    +\Ga_{b[a_1\ldots a_4}{}^{e_1e_2}Y^{2\prime}_{a_5]e_1e_2}
    -\frac{6}{7}\eta_{b[a_1}\Ga_{a_2\ldots a_5]}{}^{e_1e_2e_3}Y^{2\prime}_{e_1e_2e_3}\big)\nn\\
&+\Ga_{[a_1\dots a_4}Y^2_{a_5]b}
    +\Ga_{b[a_1a_2a_3}Y^2_{a_4a_5]}
    +\frac{6}{7}\eta_{b[a_1}\Ga_{a_2a_3a_4}{}^eY^2_{a_5]e}          \nn\\
&+\frac{1}{2}\big(\Ga_{[a_1a_2|}{}^{e_1e_2}Y^{2}_{e_1e_2|a_3a_4a_5],b}
    +\Ga_{b[a_1|}{}^{e_1e_2}Y^{2}_{e_1e_2|a_2\dots ,a_5]}\big) \nn\\
&+\frac{1}{2}\big(\Ga_{[a_1a_2|}{}^{e_1e_2}Y^{2\prime}_{e_1e_2|a_3a_4a_5],b}
    +\Ga_{[a_1a_2|}{}^{e_1e_2}Y^{2\prime}_{e_1e_2b|a_3a_4 ,a_5]}
  +\frac{4}{7}\eta_{b[a_1}
        \Ga_{a_2|}{}^{e_1e_2e_3}Y^{2\prime}_{e_1e_2 e_3|a_3a_4 ,a_5]}\big)\nn\\
&+\frac{1}{24}\big(\Ga_{[a_1\dots a_4|}{}^{e_1\dots e_4}
Y^{2\prime\prime}_{e_1\dots e_4|a_5],b}
    +\Ga_{b[a_1a_2a_3|}{}^{e_1\dots e_4}Y^{2\prime\prime}_{e_1\dots e_4|a_4 ,a_5]}
\big)\nn\\
&+\frac{1}{24}\big(\Ga_{[a_1\dots a_4|}{}^{e_1\dots e_4}
Y^{2\prime\prime\prime}_{e_1\dots e_4|a_5],b}
    +\Ga_{[a_1\dots a_4|}{}^{e_1\dots e_4}
Y^{2\prime\prime\prime}_{e_1\dots e_4b,|a_5]}\nn\\
  &~~~~~~~~~~~~~~~~~~~~~+\frac{24}{35}\eta_{b[a_1}
        \Ga_{a_2a_3 a_4|}{}^{e_1\dots e_5}
Y^{2\prime\prime\prime}_{e_1\dots e_5,|a_5]}\big)\nn\\
&+
Y^{2\prime\prime\prime\prime}_{a_1\dots a_5,b}\nn\\
&+\Ga_{[a_1a_2|}{}^{e}Y^{2}_{e|a_3a_4a_5],b}
    +\Ga_{b[a_1|}{}^{e}Y^{2}_{e|a_2\dots ,a_5]}
  -\frac{3}{14}\eta_{b[a_1}
        \Ga_{a_2|}{}^{e_1e_2}Y^{2}_{e_1e_2|a_3a_4 ,a_5]}\nn\\
&+\Ga_{[a_1a_2|}{}^{e}Y^{2\prime}_{e|a_3a_4a_5],b}
    +\Ga_{[a_1a_2|}{}^{e}Y^{2\prime}_{eb|a_3a_4 ,a_5]}
  -\frac{5}{7}\eta_{b[a_1}
        \Ga_{a_2|}{}^{e_1e_2}Y^{2\prime}_{e_1e_2|a_3a_4 ,a_5]}\nn\\
&+\frac{1}{6}\big(
      \Ga_{[a_1\dots a_4|}{}^{e_1e_2e_3}
      Y^{2\prime\prime}_{e_1e_2e_3|a_5],b}
     +\Ga_{b[a_1a_2a_3|}{}^{e_1e_2e_3}
     Y^{2\prime\prime}_{e_1e_2e_3|a_4 ,a_5]}\nn\\
     &~~~~~~~~~~~~~~~~~~~~~-\frac{1}{28}\eta_{b[a_1}
        \Ga_{a_2a_3 a_4|}{}^{e_1\dots e_4}
      Y^{2\prime\prime}_{e_1\dots e_4,|a_5]}
       \big)\nn\\
&+\frac{1}{6}\big(
      \Ga_{[a_1\dots a_4|}{}^{e_1e_2e_3}
      Y^{2\prime\prime\prime}_{e_1e_2e_3|a_5],b}
      +\Ga_{[a_1\dots a_4|}{}^{e_1e_2e_3}
      Y^{2\prime\prime\prime}_{e_1e_2e_3b ,|a_5]}
      -\frac{5}{7}\eta_{b[a_1}
        \Ga_{a_2a_3 a_4|}{}^{e_1\dots e_4}
      Y^{2\prime\prime\prime}_{e_1\dots e_4,|a_5]}
      \big)\nn\\
&+\Ga_{[a_1a_2a_3|}{}^{e}Y^{2}_{e|a_4a_5],b}
    +\Ga_{b[a_1a_2|}{}^{e}Y^{2}_{e|a_3a_4,a_5]}
  -\frac{2}{7}\eta_{b[a_1}
        \Ga_{a_2a_3|}{}^{e_1e_2}Y^{2}_{e_1e_2|a_4 ,a_5]}\nn\\
&+\Ga_{[a_1a_2a_3|}{}^{e}Y^{2\prime}_{e|a_4a_5],b}
    -\Ga_{[a_1a_2a_3|}{}^{e}Y^{2\prime}_{eb|a_4,a_5]}
  -\frac{6}{7}\eta_{b[a_1}
        \Ga_{a_2a_3|}{}^{e_1e_2}Y^{2\prime}_{e_1e_2|a_4 ,a_5]}\nn\\
&+\frac{1}{6}
   \big(\Ga_{a_1\dots a_5}{}^{e_1e_2e_3}Y^{2\prime\prime}_{e_1e_2e_3,b}
    +\Ga_{b[a_1\dots a_4|}{}^{e_1e_2e_3}
    Y^{2\prime\prime}_{e_1e_2e_3,|a_5]} \big)\nn\\
&+\Ga_{[a_1a_2a_3}Y^{2}_{a_4a_5],b}
  +\frac{6}{7}\eta_{b[a_1}
        \Ga_{a_2a_3|}{}^{e}Y^{2}_{e|a_4 ,a_5]}\nn\\
&+\frac{1}{2}
   \big(\Ga_{a_1\dots a_5}{}^{e_1e_2}Y^{2\prime}_{e_1e_2,b}
    +\Ga_{b[a_1\dots a_4|}{}^{e_1e_2}
    Y^{2\prime}_{e_1e_2,|a_5]} \big)\nn\\
&+\Ga_{[a_1a_2a_3a_4}Y^{2}_{a_5],b}
+\frac{4}{7}\eta_{b[a_1}\Ga_{a_2a_3a_4|}{}^eY^{2}_{e,|a_5]}\nn\\
&+\ldots
\label{DfiveX}
\end{align}
and
\begin{align}
\frac{1}{2}&D_{[\al}D_{\be]}X_{a_1a_2,b}                \nn\\
&=\frac{1}{6}\big(\Ga_{[a_1}{}^{e_1e_2e_3}X^2_{a_2]be_1e_2e_3}
    -\Ga_{b}{}^{e_1e_2e_3}X^2_{a_1a_2e_1e_2e_3}
    -\frac{3}{10}\eta_{b[a_1}\Ga_{}{}^{e_1\dots e_4}X^2_{a_2]e_1\dots e_4}\big)  \nn\\
&+\frac{1}{2}\big(\Ga_{[a_1}{}^{e_1e_2}X^2_{a_2]be_1e_2}
    -\Ga_{b}{}^{e_1e_2}X^2_{a_1a_2e_1e_2}
    -\frac{3}{10}\eta_{b[a_1}\Ga_{}{}^{e_1e_2 e_3}X^2_{a_2]e_1e_2e_3}\big)   \nn\\
&+\frac{1}{2}\big(\Ga_{a_1a_2}{}^{e_1e_2}X^2_{be_1e_2}
    -\Ga_{b[a_1}{}^{e_1e_2}X^2_{a_2]e_1e_2}
    +\frac{3}{10}\eta_{b[a_1}\Ga_{a_2]}{}^{e_1e_2 e_3}X^2_{e_1e_2e_3}\big)   \nn\\
&+\Ga_{a_1a_2}{}^{e}X^2_{be}
    -\Ga_{b[a_1}{}^{e}X^2_{a_2]e}
    +\frac{3}{10}\eta_{b[a_1}\Ga_{a_2]}{}^{e_1e_2}X^2_{e_1e_2}  \nn\\
&+\frac{1}{6}\big(\Ga^{e_1e_2e_3}X^2_{e_1e_2e_3a_1a_2,b}
    -\Ga^{e_1e_2e_3}X^2_{e_1e_2e_3b[a_1,a_2]}\big)   \nn\\
&+\frac{1}{120}\big(\Ga_{a_1a_2}{}^{e_1\dots e_5}X^{2\prime}_{e_1\dots e_5,b}
    -\Ga_{b[a_1}{}^{e_1\dots e_5}X^{2\prime}_{e_1\dots e_5,|a_2]}\big)   \nn\\
&+\frac{1}{6}\big(\Ga_{[a_1|}{}^{e_1e_2e_3}X^2_{e_1e_2e_3|a_2],b}
    -\Ga_{b}{}^{e_1e_2e_3}X^2_{e_1e_2e_3[a_1,a_2]}
    -\frac{1}{20}\eta_{b[a_1|}\Ga^{e_1 \dots e_4}
    X^2_{e_1 \dots e_4,|a_2]}\big) \nn\\
&+\frac{1}{6}\big(\Ga_{[a_1|}{}^{e_1e_2e_3}X^{2\prime}_{e_1e_2e_3|a_2],b}
    +\Ga_{[a_1|}{}^{e_1e_2e_3}X^{2\prime}_{e_1e_2e_3b,|a_2]}
    +\frac{1}{8}\eta_{b[a_1|}\Ga^{e_1 \dots e_4}
    X^{2\prime}_{e_1 \dots e_4,|a_2]}\big) \nn\\
&+\frac{1}{2}\big(\Ga_{[a_1|}{}^{e_1e_2}X^2_{e_1e_2|a_2],b}
    -\Ga_{b}{}^{e_1e_2}X^2_{e_1e_2[a_1,a_2]}
    +\frac{1}{30}\eta_{b[a_1|}\Ga^{e_1e_2e_3}
    X^2_{e_1e_2e_3,|a_2]}\big) \nn\\
&+\frac{1}{2}\big(\Ga_{[a_1|}{}^{e_1e_2}X^{2\prime}_{e_1e_2|a_2],b}
    +\Ga_{[a_1|}{}^{e_1e_2}X^{2\prime}_{e_1e_2b,|a_2]}
     -\frac{2}{15}\eta_{b[a_1|}\Ga^{e_1e_2e_3}
    X^{2\prime}_{e_1e_2e_3,|a_2]}\big) \nn\\
&+\frac{1}{2}\big(\Ga_{a_1a_2}{}^{e_1e_2}X^2_{e_1e_2,b}
    -\Ga_{b[a_1|}{}^{e_1e_2}X^2_{e_1e_2,|a_2]}\big)\nn\\
&+X^{2\prime}_{a_1a_2,b} \nn\\
&+\Ga_{a_1a_2}{}^{e}X^2_{e,b}
    -\Ga_{b[a_1|}{}^{e}X^2_{e,|a_2]} \nn\\
&+\ldots ~.
\label{DtwoX}
\end{align}
Let us also note that
\begin{align}
2D_{(\al}D_{\be)}X_{a_1a_2,b}
=&2\left[  2( \Ga^{ij})_{\al\be}A_{ij[a_1|}{}^c
-{1\over 6}( \Ga_{[a_1|}{}^{cijkl})_{\al\be}A^{\prime}_{ijkl}\right]
X_{c|a_2],b}\nn\\
&+\left[  2( \Ga^{ij})_{\al\be}A_{ijb}{}^c
-{1\over 6}( \Ga_{b}{}^{cijkl})_{\al\be}A^{\prime}_{ijkl}\right]
X_{a_1a_2,c}\nn\\
&-2( \Ga^{f})_{\al\be}D_fX_{a_1a_2,b}
\end{align}
and
\begin{align}
2D_{(\al}D_{\be)}Y_{a_1\dots a_5,b}
=&5\left[  2( \Ga^{ij})_{\al\be}A_{ij[a_1|}{}^c
-{1\over 6}( \Ga_{[a_1|}{}^{cijkl})_{\al\be}A^{\prime}_{ijkl}\right]
Y_{c|a_2\dots a_5],b}\nn\\
&+\left[  2( \Ga^{ij})_{\al\be}A_{ijb}{}^c
-{1\over 6}( \Ga_{b}{}^{cijkl})_{\al\be}A^{\prime}_{ijkl}\right]
Y_{a_1\dots a_5,c}\nn\\
&-2( \Ga^{f})_{\al\be}D_fY_{a_1\dots a_5,b} ~.
\end{align}
We are now ready to project onto each irreducible representation.

\bigskip

\noindent{\it The singlet}
\newline
\noindent From the 1st SSBI we get
\begin{equation}
(\Ga^{bc})^{\al\be}R_{\al\be bc}=-14080A ~.
\label{(00000)a}
\end{equation}
From the 2nd SSBI we get
\begin{align}
3(\Ga_{e})^{\al\be}(\Ga^e)_{\de}{}^{\ga}R_{(\al\be \ga)}{}^{\de}
&={7\over 2}(\Ga^{bc})^{\al\be}R_{\al\be bc}\nn\\
&=9856A+14D^{\al}S_{\al} ~.
\label{(00000)b}
\end{align}
Eqs. (\ref{(00000)a}, \ref{(00000)b}) give,
\begin{align}
D\tilde{S}&=-4224A~.
\label{(00000)}
\end{align}

\noindent{\it The vectors}

\noindent From the 1st SSBI we get
\begin{equation}
(\Ga^{b})^{\al\be}R_{\al\be ba}=2D\Ga_a\tilde{S}+1280A_a+128A^\prime_a\komma
\label{(10000)a}
\end{equation}
\begin{equation}
(\Ga^{c})^{\al\be}R_{\al\be ac}=2D\Ga_a\tilde{S}+1408A^\prime_a\komma
\end{equation}
\begin{equation}
0=\de^{bc}(\Ga_a)^{\al\be}R_{\al\be bc}=22D\Ga_a\tilde{S}
-1280A_a+128A^\prime_a\punkt
\end{equation}
From the 2nd SSBI we get
\begin{align}
3(\Ga^{e})^{\al\be}(\Ga_{e a})_{\de}{}^{\ga}R_{(\al\be \ga)}{}^{\de}
&=8(\Ga^{b})^{\al\be}R_{\al\be ab}\nn\\
&=20D\Ga_a\tilde{S}-10240A_a+1280A^\prime_a\komma
\end{align}
\begin{align}
3(\Ga_{a})^{\al\be}\de_{\de}{}^{\ga}R_{(\al\be \ga)}{}^{\de}
&=(\Ga^{b})^{\al\be}R_{\al\be ab}\nn\\
&=34D\Ga_a\tilde{S}-1280A_a+2176A^\prime_a~.
\label{(10000)e}
\end{align}
From eqs. (\ref{(10000)a}-\ref{(10000)e}) we get
\begin{align}
D\Ga_a \tilde{S}&=64A_a ~,\nn\\
A^{\prime}_a&=-A_a~.
\end{align}

\noindent{\it The 2-forms}

\noindent From the 1st SSBI we get
\begin{align}
(\Ga_{[a_1|}{}^{b})^{\al\be}R_{\al\be b|a_2]}&= 2D\Ga_{a_1a_2}\tilde{S}
-{960\over 23}D^eX_{a_1a_2,e}
\nn\\
&-{209664\over 115}X^2_{a_1a_2}
+{29952\over 23}Y^2_{a_1a_2} \nn\\
&-{64\over 3}A^{i_1\dots i_4}Y_{i_1\dots i_4[a_1,a_2]}
+{2288\over 69}A^{\prime i_1\dots i_4}Y_{i_1\dots i_4[a_1,a_2]}\nn\\
&-128A_{a_1a_2}+1152A^\prime_{a_1a_2}
\komma
\label{(01000)a}
\end{align}
\begin{align}
(\Ga_{[a_1|}{}^{c})^{\al\be}R_{\al\be| a_2]c}&=2D\Ga_{a_1a_2}\tilde{S}
-{224\over 23}D^eX_{a_1a_2,e}\nn\\
&-{209664\over 115}X^2_{a_1a_2}+{29952\over 23}Y^2_{a_1a_2}\nn\\
&-{96\over 23}A^{\prime i_1\dots i_4}Y_{i_1\dots i_4[a_1,a_2]}
-1280A_{a_1a_2}\komma
\end{align}
\begin{align}
0=\de^{bc}(\Ga_{a_1a_2})^{\al\be}R_{\al\be bc}&=22D\Ga_{a1 a2}\tilde{S}
-64D^eX_{a_1a_2,e} \nn\\
&-{32\over 3}A^{i_1\dots i_4}Y_{i_1\dots i_4[a_1,a_2]}
+{32\over 3}A^{\prime i_1\dots i_4}Y_{i_1\dots i_4[a_1,a_2]}\nn\\
&-256A_{a_1a_2}-1152A^\prime_{a_1a_2}\punkt
\end{align}

\noindent From the 2nd SSBI we get
\begin{align}
3(\Ga_{a_1a_2})^{\al\be}\de_{\de}{}^{\ga}R_{(\al\be \ga)}{}^{\de}
&=-2(\Ga_{[a_1|}{}^{b})^{\al\be}R_{\al\be b|a_2]}\nn\\
&={382\over 11}D\Ga_{a_1a_2}\tilde{S} -{349696\over 3289}D^eX_{a_1a_2,e}
\nn\\
&-{5359104\over 1265}X^2_{a_1a_2}
+{1552896\over 253}Y^2_{a_1a_2} \nn\\
&-{32\over 3}A^{i_1\dots i_4}Y_{i_1\dots i_4[a_1,a_2]}
+{118688\over 9867}A^{\prime i_1\dots i_4}Y_{i_1\dots i_4[a_1,a_2]}\nn\\
&-256A_{a_1a_2}-1152A^\prime_{a_1a_2}
\komma
\end{align}
\begin{align}
3(\Ga_{[a_1})^{\al\be}(\Ga_{a_2]})_{\de}{}^{\ga}R_{(\al\be \ga)}{}^{\de}
&=0\nn\\
&={14\over 11}D\Ga_{a_1a_2}\tilde{S} +{3072\over 253}D^eX_{a_1a_2,e}
\nn\\
&+{2875392\over 1265}X^2_{a_1a_2}
-{1198080\over 253}Y^2_{a_1a_2} \nn\\
&+{32\over 3}A^{i_1\dots i_4}Y_{i_1\dots i_4[a_1,a_2]}
-{3200\over 759}A^{\prime i_1\dots i_4}Y_{i_1\dots i_4[a_1,a_2]}\nn\\
&-2048A_{a_1a_2}-1152A^\prime_{a_1a_2}
\komma
\end{align}
\begin{align}
3(\Ga^{e})^{\al\be}(\Ga_{ea_1a_2})_{\de}{}^{\ga}R_{(\al\be \ga)}{}^{\de}
&=14(\Ga_{[a_1|}{}^{b})^{\al\be}R_{\al\be b|a_2]}\nn\\
&={126\over 11}D\Ga_{a_1a_2}\tilde{S} +{398848\over 3289}D^eX_{a_1a_2,e}
\nn\\
&+{10639872\over 253}X^2_{a_1a_2}
-{14685696\over 253}Y^2_{a_1a_2} \nn\\
&+{32}A^{i_1\dots i_4}Y_{i_1\dots i_4[a_1,a_2]}
+{13536\over 3289}A^{\prime i_1\dots i_4}Y_{i_1\dots i_4[a_1,a_2]}\nn\\
&-2304A_{a_1a_2}-10368A^\prime_{a_1a_2}~.
\label{(01000)f}
\end{align}
From eqs. (\ref{(01000)a}-\ref{(01000)f}) we get
\begin{align}
\frac{11}{8}D\Ga_{ab}\tilde{S}&=4D^fX_{ab,f} +A^{i_1\dots
i_4}Y_{i_1\dots i_4[a,b]}+
16A_{ab}+72A_{ab}^{\prime}\nn\komma\\
A_{ab}&=-\frac{1}{320528}(1636D^fX_{ab,f} +1289A^{i_1\dots i_4}Y_{i_1\dots
i_4[a,b]})\nn\\
&-\frac{18}{7705}(87X^2_{ab}+365Y^2_{ab})\nn\komma\\
A_{ab}^{\prime}&=\frac{11}{480792}(1056D^fX_{ab,f} +881A^{i_1\dots
i_4}Y_{i_1\dots i_4[a,b]})\nn\\
&+\frac{2}{1541}(1907X^2_{ab}-2131Y^2_{ab})\nn\komma\\
A^{i_1\dots i_4}Y_{i_1\dots i_4[a,b]}&= -2A^{\prime i_1\dots i_4}Y_{i_1\dots
i_4[a,b]}~.
\end{align}
Note that the last line is redundant, as it is implied by the zeroth order
equation $A_{i_1\dots i_4}=-2A^{\prime}_{i_1\dots i_4}$.

\noindent{\it The 3-forms}

\noindent From the 1st SSBI we get
\begin{align}
(\Ga_{a_1a_2a_3}{}^{bc})^{\al\be}R_{\al\be bc}&= -{2\over
5}\epsilon_{[a_1a_2|i_1\dots i_9}A^{i_1\dots i_4} Y^{i_5\dots i_9}{}_{,|a_3]}
+192A^{ ij}{}_{[a_1a_2|}X_{ij,|a_3]}\nn\\
&+{8\over 15}\epsilon_{[a_1a_2|i_1\dots i_9}A^{\prime i_1\dots i_4}
Y^{i_5\dots i_9}{}_{,|a_3]} -1152A^{\prime ij}{}_{[a_1a_2|}X_{ij,|a_3]}\nn\\
&-7168A_{a_1a_2a_3}\komma
\label{(00100)a}
\end{align}
\begin{align}
(\Ga_{[a_1|})^{\al\be}R_{\al\be |a_2a_3]}&= -{2\over
45}\epsilon_{[a_1a_2|i_1\dots i_9}A^{\prime i_1\dots i_4} Y^{i_5\dots
i_9}{}_{,|a_3]}
-64A^{ ij}{}_{[a_1a_2|}X_{ij,|a_3]}\nn\\
&-128A_{a_1a_2a_3}\punkt
\end{align}
\noindent From the 2nd SSBI we get
\begin{align}
3(\Ga_{[a_1})^{\al\be}(\Ga_{a_2 a_3]})_{\de}{}^{\ga}R_{(\al\be
\ga)}{}^{\de}&=-15(\Ga_{[a_1|})^{\al\be}R_{\al\be |a_2a_3]}+\frac{1}{2}(\Ga_{a_1a_2a_3}{}^{bc})^{\al\be}R_{\al\be bc}\nn\\
&=2D\Ga_{a_1 a_2 a_3}\tilde S-\frac{139776}{23}X^2_{a_1 a_2
a_3}-\frac{146432}{23}Y^2_{a_1 a_2 a_3}\nn\\
&-\frac{1597440}{23}Y^{2 \prime}_{a_1 a_2 a_3}-1920 A_{a_1 a_2 a_3}+1024A^{\prime}_{a_1 a_2 a_3}\nn\\
&+\frac{2}{15}\epsilon_{[a_1a_2|i_1\dots i_9}A^{i_1\dots i_4} Y^{i_5\dots
i_9}{}_{,|a_3]}
+\frac{576}{23}A^{ ij}{}_{[a_1a_2|}X_{ij,|a_3]}\nn\\
&+\frac{2}{69}\epsilon_{[a_1a_2|i_1\dots i_9}A^{\prime i_1\dots i_4}
Y^{i_5\dots i_9}{}_{,|a_3]} +192A^{\prime ij}{}_{[a_1a_2|}X_{ij,|a_3]}\komma
\end{align}
\begin{align}
3(\Ga_{[a_1 a_2})^{\al\be}(\Ga_{a_3]})_{\de}{}^{\ga}R_{(\al\be
\ga)}{}^{\de}&=(\Ga_{[a_1|})^{\al\be}R_{\al\be |a_2a_3]}+\frac{1}{2}(\Ga_{a_1a_2a_3}{}^{bc})^{\al\be}R_{\al\be bc}\nn\\
&=\frac{6}{11}D\Ga_{a_1 a_2 a_3}\tilde S+\frac{6404608}{1265}X^2_{a_1 a_2
a_3}-\frac{611328}{253}Y^2_{a_1 a_2 a_3}\nn\\
&+\frac{3522560}{253}Y^{2 \prime}_{a_1 a_2 a_3}+128 A_{a_1 a_2 a_3}-1024A^{\prime}_{a_1 a_2 a_3}\nn\\
&-\frac{2}{15}\epsilon_{[a_1a_2|i_1\dots i_9}A^{i_1\dots i_4} Y^{i_5\dots
i_9}{}_{,|a_3]}
+\frac{797504}{3289}A^{ ij}{}_{[a_1a_2|}X_{ij,|a_3]}\nn\\
&+\frac{21634}{148005}\epsilon_{[a_1a_2|i_1\dots i_9}A^{\prime i_1\dots i_4}
Y^{i_5\dots i_9}{}_{,|a_3]} +192A^{\prime ij}{}_{[a_1a_2|}X_{ij,|a_3]}\komma
\end{align}
\begin{align}
3(\Ga^e)^{\al\be}(\Ga_{e a_1 a_2 a_3})_{\de}{}^{\ga}R_{(\al\be
\ga)}{}^{\de}&=-24(\Ga_{[a_1|})^{\al\be}R_{\al\be |a_2a_3]}+2(\Ga_{a_1a_2a_3}{}^{bc})^{\al\be}R_{\al\be bc}\nn\\
&=\frac{80}{11}D\Ga_{a_1 a_2 a_3}\tilde S+\frac{23052288}{1265}X^2_{a_1 a_2
a_3}-\frac{6889472}{253}Y^2_{a_1 a_2 a_3}\nn\\
&-\frac{14008320}{253}Y^{2 \prime}_{a_1 a_2 a_3}-3072 A_{a_1 a_2 a_3}+10240A^{\prime}_{a_1 a_2 a_3}\nn\\
&+\frac{4}{15}\epsilon_{[a_1a_2|i_1\dots i_9}A^{i_1\dots i_4} Y^{i_5\dots
i_9}{}_{,|a_3]}
-\frac{3891072}{3289}A^{ ij}{}_{[a_1a_2|}X_{ij,|a_3]}\nn\\
&-\frac{6496}{49335}\epsilon_{[a_1a_2|i_1\dots i_9}A^{\prime i_1\dots i_4}
Y^{i_5\dots i_9}{}_{,|a_3]} -768A^{\prime ij}{}_{[a_1a_2|}X_{ij,|a_3]}\punkt
\label{(00100)e}
\end{align}
A useful identity is
\begin{equation}
\epsilon_{a_1a_2a_3i_1\dots i_8}A^{i_1i_2 i_3j} Y_j{}^{i_4\dots i_7,i_8}
=-{3\over 20}\epsilon_{[a_1a_2|i_1\dots i_9}A^{i_1\dots i_4} Y^{i_5\dots
i_9}{}_{,|a_3]}\punkt
\end{equation}
From eqs. (\ref{(00100)a}-\ref{(00100)e}) we get,
\begin{align}
A^{\prime}_{a_1a_2a_3}&=A_{a_1a_2a_3}\komma\nn\\
A_{a_1a_2a_3}&=-\frac{5457}{66976}A^{
ij}{}_{[a_1a_2|}X_{ij,|a_3]}+\frac{193}{4592640}\epsilon_{[a_1a_2|i_1\dots
i_9}A^{i_1\dots i_4} Y^{i_5\dots
i_9}{}_{,|a_3]}\nn\\
&-\frac{1}{230}(593X^2_{a_1 a_2
a_3}-60Y^2_{a_1 a_2 a_3}+2900Y^{2 \prime}_{a_1 a_2 a_3})\komma\nn\\
D\Ga_{a_1 a_2 a_3}\tilde S&=\frac{1546236}{2093}A^{
ij}{}_{[a_1a_2|}X_{ij,|a_3]}-\frac{14671}{35880}\epsilon_{[a_1a_2|i_1\dots
i_9}A^{i_1\dots i_4} Y^{i_5\dots
i_9}{}_{,|a_3]}\nn\\
&+\frac{64}{115}(7239X^2_{a_1 a_2 a_3}+5540Y^2_{a_1 a_2 a_3}+71100Y^{2
\prime}_{a_1 a_2 a_3})~,
\end{align}
by imposing the conditions
\begin{align}
A^{ij}{}_{[a_1a_2|}X_{ij,|a_3]}&=
-2A^{\prime ij}{}_{[a_1a_2|}X_{ij,|a_3]}\nn\\
\epsilon_{[a_1a_2|i_1\dots
i_9}A^{i_1\dots i_4} Y^{i_5\dots i_9}{}_{,|a_3]}&=
-2\epsilon_{[a_1a_2|i_1\dots
i_9}A^{\prime i_1\dots i_4} Y^{i_5\dots i_9}{}_{,|a_3]}\nn ~,
\end{align}
which are implied by the zeroth-order equation
$A_{a_1\dots a_4}=-2A^\prime_{a_1\dots a_4}$.

\noindent{\it The 4-forms}

\noindent From the 1st SSBI we get
\begin{align}
(\Ga_{a_1a_2a_3a_4}{}^{bc})^{\al\be}R_{\al\be bc}&=
-512A^{ijk}{}_{[a_1|}
Y_{ijk|a_2a_3,a_4]}+1024A^{\prime ijk}{}_{[a_1|}
Y_{ijk|a_2a_3,a_4]}\nn\\
&-5376A^\prime_{a_1a_2a_3a_4}\komma
\label{(00010)a}
\end{align}
\begin{align}
(\Ga_{[a_1a_2|})^{\al\be}R_{\al\be |a_3a_4]}&={64\over 3}A^{ijk}{}_{[a_1|}
Y_{ijk|a_2a_3,a_4]}-{128\over 3}A^{\prime ijk}{}_{[a_1|}
Y_{ijk|a_2a_3,a_4]}\nn\\
&+128A_{a_1a_2a_3a_4}\punkt
\end{align}
\noindent From the 2nd SSBI we get
\begin{align}
3(\Ga_{[a_1})^{\al\be}(\Ga_{a_2a_3a_4]})_\de{}^\ga R_{(\al\be\ga)}{}^\de&=
\frac{1}{2}(\Ga_{a_1a_2a_3a_4}{}^{b c})^{\al\be}R_{\al\be b c}\nn\\
&=\frac{2080512}{1265}X^2_{a_1a_2a_3a_4}
+\frac{1400320}{253}Y^2_{a_1a_2a_3a_4}\nn\\
&+\frac{6266880}{253}Y^2{}'_{a_1a_2a_3a_4}
+\frac{774016}{3289}A^{i_1i_2i_3}{}_{[a_1|}Y_{i_1i_2i_3|a_2a_3,a_4]}\nn\\
&-\frac{660480}{3289}A'^{i_1i_2i_3}{}_{[a_1|}Y_{i_1i_2i_3|a_2a_3,a_4]}
+\frac{14}{11}D\Ga_{a_1a_2a_3a_4}\tilde{S}\nn\\
&+1792A_{a_1a_2a_3a_4}+896A'_{a_1a_2a_3a_4}\komma
\end{align}
\begin{align}
3(\Ga_{[a_1a_2})^{\al\be}(\Ga_{a_3a_4]})_\de{}^\ga R_{(\al\be\ga)}{}^\de&=-14(\Ga_{[a_1a_2|})^{\al\be}R_{\al\be |a_3a_4]}+\frac{1}{2}(\Ga_{a_1a_2a_3a_4}{}^{b c})^{\al\be}R_{\al\be b c}\nn\\
&=\frac{2080512}{1265}X^2_{a_1a_2a_3a_4}+\frac{1400320}{253}Y^2_{a_1a_2a_3a_4}\nn\\
&+\frac{6266880}{253}Y^2{}'_{a_1a_2a_3a_4}
-\frac{488960}{3289}A^{i_1i_2i_3}{}_{[a_1|}Y_{i_1i_2i_3|a_2a_3,a_4]}\nn\\
&+\frac{2228480}{9867}A'^{i_1i_2i_3}{}_{[a_1|}Y_{i_1i_2i_3|a_2a_3,a_4]}
+\frac{14}{11}D\Ga_{a_1a_2a_3a_4}\tilde S\nn\\
&+896A'_{a_1a_2a_3a_4}\komma
\end{align}
\begin{align}
3(\Ga_{e})^{\al\be}(\Ga^e{}_{a_1a_2a_3a_4})_\de{}^\ga R_{(\al\be\ga)}{}^\de&=-42(\Ga_{[a_1a_2|})^{\al\be}R_{\al\be |a_3a_4]}+\frac{3}{2}(\Ga_{a_1a_2a_3a_4}{}^{b c})^{\al\be}R_{\al\be b c}\nn\\
&=\frac{2080512}{253}X^2_{a_1a_2a_3a_4}
+\frac{7001600}{253}Y^2_{a_1a_2a_3a_4}\nn\\
&+\frac{31334400}{253}Y^2{}'_{a_1a_2a_3a_4}
+\frac{3449088}{3289}A^{i_1i_2i_3}{}_{[a_1|}Y_{i_1i_2i_3|a_2a_3,a_4]}\nn\\
&+\frac{907520}{3289}A'^{i_1i_2i_3}{}_{[a_1|}Y_{i_1i_2i_3|a_2a_3,a_4]}
+\frac{70}{11}D\Ga_{a_1a_2a_3a_4}\tilde{S}\nn\\
&+3584A_{a_1a_2a_3a_4}+9856A'_{a_1a_2a_3a_4} \label{(00010)e}\punkt
\end{align}
From eqs. (\ref{(00010)a}-\ref{(00010)e}) we get
\begin{align}
A_{a_1a_2a_3a_4}&=-2A'_{a_1a_2a_3a_4}-\frac{1}{1408}D\Ga_{a_1a_2a_3a_4}\tilde
S\nn\\
&-\frac{1161}{1265}X^2_{a_1a_2a_3a_4}
-\frac{5470}{1771}Y^2_{a_1a_2a_3a_4}\nn\\
&-\frac{24480}{1771}Y^2{}'_{a_1a_2a_3a_4}-\frac{21783}{46046}A^{i_1i_2i_3}{}_{[a_1|}Y_{i_1i_2i_3|a_2a_3,a_4]}\komma\nn\\
A^{i_1i_2i_3}{}_{[a_1|}Y_{i_1i_2i_3|a_2a_3,a_4]}&=-2A'^{i_1i_2i_3}{}_{[a_1|}Y_{i_1i_2i_3|a_2a_3,a_4]}\punkt
\end{align}
Note that the second equation is implied by the first one.

\noindent{\it The 5-forms}\\
\noindent From the 1st SSBI we get
\begin{align}
(\Ga_{[a_1a_2a_3a_4|}{}^{b})^{\al\be}R_{\al\be b|a_5]}&= 2D\Ga_{a_1\dots
a_5}\tilde{S}-{69888\over 115}X^2_{a_1\dots a_5}
+{49920\over 161}Y^2_{a_1\dots a_5}\nn\\
&-{99840\over 161}Y^{2\prime}_{a_1\dots a_5}
+{1536\over 115}D^eY_{a_1\dots a_5,e}\nn\\
&+128A_{a_1\dots a_5}-{2272\over 1035}\epsilon_{[a_1a_2a_3|i_1\dots i_8}
A^{i_1i_2 i_3}{}_{|a_4|}Y^{i_4\dots i_8}{}_{,|a_5]}\nn\\
&+256A_{[a_1a_2a_3}{}^iX_{a_4a_5],i}\nn\\
&+768A^{\prime}_{a_1\dots a_5} +{2656\over 1035}\epsilon_{[a_1a_2a_3|i_1\dots
i_8}
A^{\prime i_1i_2 i_3}{}_{|a_4|}Y^{i_4\dots i_8}{}_{,|a_5]}\nn\\
&-{44288\over 23}A^{\prime}_{[a_1a_2a_3}{}^iX_{a_4a_5],i}\komma
\label{(00002)a}
\end{align}
\begin{align}
(\Ga_{[a_1a_2a_3a_4|}{}^{c})^{\al\be}R_{\al\be |a_5]c}&= 2D\Ga_{a_1\dots
a_5}\tilde{S}- {69888\over 115}X^2_{a_1\dots a_5}
+{49920\over 161}Y^2_{a_1\dots a_5}\nn\\
&-{99840\over 161}Y^{2\prime}_{a_1\dots a_5} +{64\over 115}D^eY_{a_1\dots
a_5,e}
\nn\\
&+896A_{a_1\dots a_5}+{224\over 345}\epsilon_{[a_1a_2a_3|i_1\dots i_8}
A^{i_1i_2 i_3}{}_{|a_4|}Y^{i_4\dots i_8}{}_{,|a_5]}\nn\\
&-768A_{[a_1a_2a_3}{}^iX_{a_4a_5],i}\nn\\
&-{832\over 345}\epsilon_{[a_1a_2a_3|i_1\dots i_8} A^{\prime i_1i_2
i_3}{}_{|a_4|}Y^{i_4\dots i_8}{}_{,|a_5]} -{3072\over
23}A^{\prime}_{[a_1a_2a_3}{}^iX_{a_4a_5],i}\komma
\end{align}
\begin{align}
0=\de^{bc}(\Ga_{a_1\dots a_5})^{\al\be}R_{\al\be bc}&
=22D\Ga_{a_1\dots a_5}\tilde{S}+64D^eY_{a_1\dots a_5,e}\nn\\
&+640A_{a_1\dots a_5}-{32\over 9}\epsilon_{[a_1a_2a_3|i_1\dots i_8}
A^{i_1i_2 i_3}{}_{|a_4|}Y^{i_4\dots i_8}{}_{,|a_5]}\nn\\
&+1280A_{[a_1a_2a_3}{}^iX_{a_4a_5],i}\nn\\
&-768A^{\prime}_{a_1\dots a_5}+{32\over 9}\epsilon_{[a_1a_2a_3|i_1\dots i_8}
A^{\prime i_1i_2 i_3}{}_{|a_4|}Y^{i_4\dots i_8}{}_{,|a_5]}\nn\\
&-1280A^{\prime}_{[a_1a_2a_3}{}^iX_{a_4a_5],i}\punkt
\end{align}
\noindent From the 2nd SSBI we get
\begin{align}
3(\Ga_{[a_1})^{\al\be}(\Ga_{a_2a_3a_4a_5]})_\de{}^\ga
R_{(\al\be\ga)}{}^\de&=-3(\Ga_{[a_1a_2a_3a_4|}{}^{c})^{\al\be}R_{\al\be |a_5]c}\nn\\
&= \frac{10}{11}D\Ga_{a_1\dots a_5}\tilde{S}+{466176\over 1265}X^2_{a_1\dots
a_5}
-{9795840\over 1771}Y^2_{a_1\dots a_5}\nn\\
&-{9553920\over 1771}Y^{2\prime}_{a_1\dots a_5}
-{179456\over 16445}D^eY_{a_1\dots a_5,e}\nn\\
&+1664A_{a_1\dots a_5}+{430432\over 148005}\epsilon_{[a_1a_2a_3|i_1\dots i_8}
A^{i_1i_2 i_3}{}_{|a_4|}Y^{i_4\dots i_8}{}_{,|a_5]}\nn\\
&+256A_{[a_1a_2a_3}{}^iX_{a_4a_5],i}\nn\\
&-768A^{\prime}_{a_1\dots a_5} +{135584\over
148005}\epsilon_{[a_1a_2a_3|i_1\dots i_8}
A^{\prime i_1i_2 i_3}{}_{|a_4|}Y^{i_4\dots i_8}{}_{,|a_5]}\nn\\
&+{6217472\over 3289}A^{\prime}_{[a_1a_2a_3}{}^iX_{a_4a_5],i}\komma
\end{align}
\begin{align}
3(\Ga_{[a_1a_2})^{\al\be}(\Ga_{a_3a_4a_5]})_\de{}^\ga
R_{(\al\be\ga)}{}^\de&=-(\Ga_{[a_1a_2a_3a_4|}{}^{c})^{\al\be}R_{\al\be |a_5]c}\nn\\
&= 2D\Ga_{a_1\dots a_5}\tilde{S}-{69888\over 115}X^2_{a_1\dots a_5}
+{49920\over 161}Y^2_{a_1\dots a_5}\nn\\
&-{99840\over 161}Y^{2\prime}_{a_1\dots a_5}
+{1536\over 115}D^eY_{a_1\dots a_5,e}\nn\\
&+128A_{a_1\dots a_5}-{2272\over 1035}\epsilon_{[a_1a_2a_3|i_1\dots i_8}
A^{i_1i_2 i_3}{}_{|a_4|}Y^{i_4\dots i_8}{}_{,|a_5]}\nn\\
&+1792A_{[a_1a_2a_3}{}^iX_{a_4a_5],i}\nn\\
&+768A^{\prime}_{a_1\dots a_5} +{2656\over 1035}\epsilon_{[a_1a_2a_3|i_1\dots
i_8}
A^{\prime i_1i_2 i_3}{}_{|a_4|}Y^{i_4\dots i_8}{}_{,|a_5]}\nn\\
&+{26368\over 23}A^{\prime}_{[a_1a_2a_3}{}^iX_{a_4a_5],i}\komma
\end{align}
\begin{align}
3(\Ga^e)^{\al\be}(\Ga_{ea_1a_2a_3a_4a_5})_\de{}^\ga
R_{(\al\be\ga)}{}^\de&=-20(\Ga_{[a_1a_2a_3a_4|}{}^{c})^{\al\be}R_{\al\be |a_5]c}\nn\\
&= \frac{60}{11}D\Ga_{a_1\dots a_5}\tilde{S}+{2380800\over 253}X^2_{a_1\dots
a_5}
-{65280000\over 1771}Y^2_{a_1\dots a_5}\nn\\
&-{44313600\over 1771}Y^{2\prime}_{a_1\dots a_5}
-{217600\over 3289}D^eY_{a_1\dots a_5,e}\nn\\
&+3840A_{a_1\dots a_5}+{99520\over 9867}\epsilon_{[a_1a_2a_3|i_1\dots i_8}
A^{i_1i_2 i_3}{}_{|a_4|}Y^{i_4\dots i_8}{}_{,|a_5]}\nn\\
&-7680A_{[a_1a_2a_3}{}^iX_{a_4a_5],i}\nn\\
&-9216A^{\prime}_{a_1\dots a_5} -{20736\over
3289}\epsilon_{[a_1a_2a_3|i_1\dots i_8}
A^{\prime i_1i_2 i_3}{}_{|a_4|}Y^{i_4\dots i_8}{}_{,|a_5]}\nn\\
&-{29383680\over 3289}A^{\prime}_{[a_1a_2a_3}{}^iX_{a_4a_5],i}\komma
\end{align}
\begin{align}
3(\Ga_{a_1a_2a_3a_4a_5})^{\al\be}\de_\de{}^\ga
R_{(\al\be\ga)}{}^\de&=5(\Ga_{[a_1a_2a_3a_4|}{}^{c})^{\al\be}R_{\al\be |a_5]c}\nn\\
&= \frac{386}{11}D\Ga_{a_1\dots a_5}\tilde{S}-{258816\over 253}X^2_{a_1\dots
a_5}
+{10387200\over 1771}Y^2_{a_1\dots a_5}\nn\\
&+{8371200\over 1771}Y^{2\prime}_{a_1\dots a_5}
+{342272\over 3289}D^eY_{a_1\dots a_5,e}\nn\\
&+640A_{a_1\dots a_5}-{123680\over 29601}\epsilon_{[a_1a_2a_3|i_1\dots i_8}
A^{i_1i_2 i_3}{}_{|a_4|}Y^{i_4\dots i_8}{}_{,|a_5]}\nn\\
&+1280A_{[a_1a_2a_3}{}^iX_{a_4a_5],i}\nn\\
&-768A^{\prime}_{a_1\dots a_5} +{119456\over
29601}\epsilon_{[a_1a_2a_3|i_1\dots i_8}
A^{\prime i_1i_2 i_3}{}_{|a_4|}Y^{i_4\dots i_8}{}_{,|a_5]}\nn\\
&-{4747520\over 3289}A^{\prime}_{[a_1a_2a_3}{}^iX_{a_4a_5],i}~.
\label{(00002)g}
\end{align}
The following identities are useful
\begin{align}
\epsilon_{i_1\dots i_8 [a_1a_2a_3}A^{i_1i_2i_3}{}_{a_4}
Y^{i_4\dots i_8}{}_{,a_5]}&=-{5\over 4}
\epsilon_{i_1\dots i_7 [a_1\dots a_4|}A^{i_1i_2i_3}{}_{j}
Y^{ji_4\dots i_7}{}_{,|a_5]}\nn\\
&={5\over 4}\epsilon_{i_1\dots i_8 [a_1a_2a_3}A^{i_1\dots i_4}
Y^{i_5\dots i_8}{}_{a_4,a_5]}\nn\\
&={15\over 4}\epsilon_{i_1\dots i_7 [a_1\dots a_4}A_{a_5]}{}^{i_1i_2j}
Y_{j}{}^{i_3\dots i_6,i_7}\nn\\
&=-{3\over 2}\epsilon_{i_1\dots i_6 a_1\dots a_5}A^{i_1i_2jk}
Y_{jk}{}^{i_3i_4i_5, i_6}\nn\\
&={5\over 2}\epsilon_{i_1\dots i_7 [a_1\dots a_4|}A^{i_1i_2i_3j}
Y_{j|a_5]}{}^{i_4i_5i_6,i_7}\punkt
\label{blirp}
\end{align}
From eqs. (\ref{(00002)a}-\ref{(00002)g}) we get
\begin{align}
A_{a_1a_2a_3a_4a_5}&=
\frac{11}{160264}(10272A_{[a_1a_2a_3}{}^iX_{a_4a_5],i}
-29\epsilon_{[a_1a_2a_3|i_1\dots
i_8} A^{i_1i_2 i_3}{}_{|a_4|}Y^{i_4\dots
i_8}{}_{,|a_5]})\nn\\
&+\frac{120}{20033}D^eY_{a_1\dots a_5,e}
+\frac{6}{53935}(4151X^2_{a_1\dots
a_5}+8150Y^2_{a_1\dots a_5}+15325Y^{2\prime}_{a_1\dots a_5}) \nn\komma\\
A^{\prime}_{a_1a_2a_3a_4a_5}&
=-\frac{1}{2884752}(2569320A_{[a_1a_2a_3}{}^iX_{a_4a_5],i}
-7819\epsilon_{[a_1a_2a_3|i_1\dots
i_8} A^{i_1i_2 i_3}{}_{|a_4|}Y^{i_4\dots
i_8}{}_{,|a_5]})\nn\\
&-\frac{681}{80132}D^eY_{a_1\dots a_5,e}
+\frac{2}{10787}(4732X^2_{a_1\dots
a_5}-25(343Y^2_{a_1\dots a_5}+73Y^{2\prime}_{a_1\dots a_5})) \nn\komma\\
\frac{11}{8}D\Ga_{a_1\dots a_5}\tilde{S}&
=-4D^eY_{a_1\dots a_5,e}-120A_{[a_1a_2a_3}{}^iX_{a_4a_5],i}
+\frac{1}{3}\epsilon_{[a_1a_2a_3|i_1\dots
i_8} A^{i_1i_2 i_3}{}_{|a_4|}Y^{i_4\dots
i_8}{}_{,|a_5]}\nn\\
&
-40A_{a_1a_2a_3a_4a_5}+48A^{\prime}_{a_1a_2a_3a_4a_5}
\end{align}
and
\begin{align}
\epsilon_{[a_1a_2a_3|i_1\dots i_8} A^{i_1i_2 i_3}{}_{|a_4|}Y^{i_4\dots
i_8}{}_{,|a_5]}&=-2\epsilon_{[a_1a_2a_3|i_1\dots i_8}
A^{\prime i_1i_2 i_3}{}_{|a_4|}Y^{i_4\dots i_8}{}_{,|a_5]}\nn\komma\\
A_{[a_1a_2a_3}{}^iX_{a_4a_5],i}&=-2A^{\prime}_{[a_1a_2a_3}{}^iX_{a_4a_5],i}\punkt
\end{align}
Note that the last two equations are redundant, as they are implied by the
zeroth order relation
\begin{align}
A_{i_1\dots i_4}=-2A^{\prime}_{i_1\dots i_4}~.
\end{align}
\noindent{\it The $(20000)$'s.}

\noindent From the 1st SSBI we get
\begin{align}
\Pi\big[(\Ga_{a}{}^{c})^{\al\be}R_{\al\be bc}\big]&=
\frac{33792}{23}X^2_{a,b}+\frac{56320}{23}Y^2_{a,b}-1280 B_{a,b}\nn\\
&+\frac{992}{207}A^{\prime i_1\dots i_4}Y_{i_1\dots i_4 (a,b)}
+\frac{4544}{207}D^eX_{e(a,b)}
\komma
\end{align}
\begin{align}
\Pi\big[(\Ga_{a}{}^{c})^{\al\be}R_{\al\be cb}\big]&=
\frac{33792}{23}X^2_{a,b}+\frac{56320}{23}Y^2_{a,b}-128 B_{a,b}\nn\\
&-\frac{64}{3}A^{ i_1\dots i_4}Y_{i_1\dots i_4 (a,b)}
+\frac{6512}{207}A^{\prime i_1\dots i_4}Y_{i_1\dots i_4 (a,b)}\nn\\
&+\frac{17792}{207}D^eX_{e(a,b)} \komma
\end{align}
where we have denoted by $\Pi$ the projection onto the
hook-irreducible part. See app. C for a detailed discussion.

\noindent From the 2nd SSBI we get
\begin{align}
3\Pi\big[(\Ga_a)^{\al\be}(\Ga_{b})_\de{}^\ga
R_{(\al\be\ga)}{}^\de\big]&=
-2\Pi\big[(\Ga_{a}{}^{c})^{\al\be}R_{\al\be bc}\big]\nn\\
&=-\frac{55296}{23}X^2_{a,b}-\frac{92160}{23}Y^2_{a,b}+2304 B_{a,b}\nn\\
&+\frac{3616}{759}A^{\prime i_1\dots i_4}Y_{i_1\dots i_4 (a,b)}
-\frac{6144}{253}D^eX_{e(a,b)}~.
\end{align}
Implementing the zeroth-order relation
$A_{a_1\dots a_4}=-2A_{a_1\dots a_4}^\prime$ we get
\begin{align}
B_{a,b}&=
-\frac{1021}{36432}A^{ i_1\dots i_4}Y_{i_1\dots i_4 (a,b)}+\frac{349}{4554}D^eX_{e(a,b)}+\frac{48}{23}X^2_{a,b}+\frac{80}{23}Y^2_{a,b}\punkt
\end{align}

\noindent{\it The $(11000)$'s.}

\noindent From the 1st SSBI we get
\begin{align}
\Pi\big[(\Ga_{b}{}^{})^{\al\be}R_{\al\be a_1a_2}\big]
&=64 B_{a_1a_2,b}
+\frac{64}{3}A\circ X_{a_1a_2,b}
+\frac{2}{135}A^\prime \circ Y_{a_1a_2,b}
\komma
\end{align}
\begin{align}
\Pi\big[(\Ga_{[a_1|}{}^{})^{\al\be}R_{\al\be |a_2]b}\big]&=
-\frac{1}{2}\Pi\big[(\Ga_{b}{}^{})^{\al\be}R_{\al\be a_1a_2}\big]\nn\\
&=64 B_{a_1a_2,b}
-\frac{6400}{23}X^2_{a_1a_2,b} -\frac{256}{69}X^{2\prime}_{a_1a_2,b}\nn\\
&-\frac{14080}{69}Y^2_{a_1a_2,b} -\frac{89600}{23}Y^{2\prime}_{a_1a_2,b}
\nn\\
&-\frac{3008}{69}A\circ X_{a_1a_2,b}
-\frac{2}{69}A^\prime \circ Y_{a_1a_2,b}~,
\end{align}
where
\begin{align}
A\circ X_{a_1a_2,b}&:=A_{a_1a_2}{}^{ i_1i_2}X_{i_1  i_2 ,b}
-A_{b[a_1|}{}^{ i_1i_2}X_{i_1  i_2 ,|a_2]}\nn\\
A^\prime \circ Y_{a_1a_2,b}&:=\epsilon_{a_1a_2}{}^{i_1\dots i_9}
A^{\prime}_{ i_1\dots i_4}Y_{i_5\dots i_9 ,b}
-\epsilon_{b[a_1|}{}^{i_1\dots i_9}
A^{\prime}_{ i_1\dots i_4}Y_{i_5\dots i_9 ,|a_2]}~.
\end{align}

\noindent From the 2nd SSBI we get
\begin{align}
3\Pi\big[(\Ga_b)^{\al\be}(\Ga_{a_1a_2})_\de{}^\ga
R_{(\al\be\ga)}{}^\de\big]&=
-18\Pi\big[(\Ga_{b}{}^{})^{\al\be}R_{\al\be a_1a_2}\big]\nn\\
&=-2304 B_{a_1a_2,b}
+\frac{76800}{23}X^2_{a_1a_2,b} +\frac{1024}{23}X^{2\prime}_{a_1a_2,b}\nn\\
&+\frac{56320}{23}Y^2_{a_1a_2,b} +\frac{1075200}{23}Y^{2\prime}_{a_1a_2,b}
\nn\\
&+\frac{256}{23}A\circ X_{a_1a_2,b}
-\frac{8}{1035}A^\prime \circ Y_{a_1a_2,b}  ~,
\end{align}
\begin{align}
3\Pi\big[(\Ga_{b[a_1})^{\al\be}(\Ga_{a_2]})_\de{}^\ga
R_{(\al\be\ga)}{}^\de\big]&=
\Pi\big[(\Ga_{b}{}^{})^{\al\be}R_{\al\be a_1a_2}\big]\nn\\
&=128 B_{a_1a_2,b}
-\frac{217600}{3289}X^2_{a_1a_2,b}
-\frac{269824}{3289}X^{2\prime}_{a_1a_2,b}\nn\\
&+\frac{468480}{3289}Y^2_{a_1a_2,b}
-\frac{19532800}{3289}Y^{2\prime}_{a_1a_2,b}
\nn\\
&-\frac{172928}{9867}A\circ X_{a_1a_2,b}
+\frac{508}{40365}A^\prime \circ Y_{a_1a_2,b}~.
\end{align}
We get
\begin{align}
B_{a_1a_2,b}&=\frac{1327}{2815200}A^\prime \circ Y_{a_1a_2,b}
+\frac{2080}{391}X^2_{a_1a_2,b}
-\frac{616}{391}X^{2\prime}_{a_1a_2,b}\nn\\
&+\frac{3040}{391}Y^2_{a_1a_2,b}-\frac{10640}{391}Y^{2\prime}_{a_1a_2,b}\nn\\
A\circ X_{a_1a_2,b}
&=\frac{11}{15300}A^\prime \circ Y_{a_1a_2,b}+\frac{120}{17}X^2_{a_1a_2,b}
-\frac{80}{17}X^{2\prime}_{a_1a_2,b}\nn\\
&+\frac{280}{17}Y^2_{a_1a_2,b}
-\frac{3360}{17}Y^{2\prime}_{a_1a_2,b}~,
\end{align}
\noindent{\it The $(10100)$'s.}

\noindent From the 1st SSBI we get
\begin{align}
\Pi\big[(\Ga_{b[a_1|}{}^{})^{\al\be}R_{\al\be |a_2a_3]}\big]&=
\frac{128}{3} B_{a_1a_2a_3,b}
-32(D_{[a_1}X_{a_2a_3],b}-\frac{1}{9}\eta_{b[a_1}D^iX_{a_2a_3],i} )\nn\\
&+\frac{64}{3}A\circ Y^{(1)}_{a_1a_2a_3,b}
-\frac{16}{3}A\circ Y^{(2)}_{a_1a_2a_3,b}
+\frac{32}{3}A^{\prime}\circ Y^{(1)}_{a_1a_2a_3,b}
\komma
\end{align}
\begin{align}
\Pi\big[(\Ga_{[a_1a_2|}{}^{})^{\al\be}R_{\al\be |a_3]b}\big]&=
\Pi\big[(\Ga_{b[a_1|}{}^{})^{\al\be}R_{\al\be |a_2a_3]}\big]\nn\\
&=-\frac{128}{3} B_{a_1a_2a_3,b}
-\frac{4864}{115}X^2_{a_1a_2a_3,b}
-\frac{8192}{345}X^{2\prime}_{a_1a_2a_3,b}\nn\\
&-\frac{2048}{23}Y^2_{a_1a_2a_3,b}
-\frac{14848}{69}Y^{2\prime}_{a_1a_2a_3,b}
-\frac{35840}{23}Y^{2\prime\prime}_{a_1a_2a_3,b}\nn\\
&-\frac{1504}{23}
(D_{[a_1}X_{a_2a_3],b}-\frac{1}{9}\eta_{b[a_1}D^iX_{a_2a_3],i} )\nn\\
&+\frac{32}{3}A\circ Y^{(1)}_{a_1a_2a_3,b}
-\frac{2608}{621}A\circ Y^{(2)}_{a_1a_2a_3,b}\nn\\
&-\frac{1504}{69}A^{\prime}\circ Y^{(1)}_{a_1a_2a_3,b}
+\frac{5696}{621}A^{\prime}\circ Y^{(2)}_{a_1a_2a_3,b}~,
\end{align}
where we have defined
\begin{align}
A\circ Y^{(1)}_{a_1a_2a_3,b}&:=
A_{[a_1|}{}^{i_1i_2i_3}Y_{i_1i_2i_3|a_2a_3],b}
+A_{b}{}^{i_1i_2i_3}Y_{i_1i_2i_3[a_1a_2,a_3]}
+\frac{1}{9} \eta_{b[a_1|}
A^{i_1\dots i_4}Y_{i_1\dots i_4|a_2,a_3]}, \nn\\
A\circ Y^{(2)}_{a_1a_2a_3,b}&:=
5A_{[a_1|}{}^{i_1i_2i_3}Y_{i_1i_2i_3|a_2a_3],b}
-2A_{[a_1|}{}^{i_1i_2i_3}Y_{i_1i_2i_3b|a_2,a_3]}
+3A_{b}{}^{i_1i_2i_3}Y_{i_1i_2i_3[a_1a_2,a_3]}~.
\end{align}
Note that indeed there are two $(10100)$'s in the decomposition
of the tensor product
$A_{a_1\dots a_4}\otimes Y_{b_1\dots b_5,c}\sim (00010)\otimes(10002)$.

\noindent From the 2nd SSBI we get
\begin{align}
3\Pi\big[(\Ga_b)^{\al\be}(\Ga_{a_1a_2a_3})_\de{}^\ga
 & R_{(\al\be\ga)}{}^\de\big]=
-6\Pi\big[(\Ga_{b[a_1|}{}^{})^{\al\be}R_{\al\be |a_2a_3]}\big]\nn\\
&=-2304 B_{a_1a_2a_3,b}
+\frac{8036352}{16445}X^2_{a_1a_2a_3,b}
+\frac{15357952}{16445}X^{2\prime}_{a_1a_2a_3,b}\nn\\
&-\frac{2082816}{3289}Y^2_{a_1a_2a_3,b}
-\frac{14513152}{3289}Y^{2\prime}_{a_1a_2a_3,b}
-\frac{43868160}{3289}Y^{2\prime\prime}_{a_1a_2a_3,b}\nn\\
&-\frac{149568}{3289}
(D_{[a_1}X_{a_2a_3],b}-\frac{1}{9}\eta_{b[a_1}D^iX_{a_2a_3],i} )\nn\\
&+192A\circ Y^{(1)}_{a_1a_2a_3,b}
-\frac{404000}{9867}A\circ Y^{(2)}_{a_1a_2a_3,b}\nn\\
&+\frac{113472}{253}A^{\prime}\circ Y^{(1)}_{a_1a_2a_3,b}
-\frac{273664}{3289}A^{\prime}\circ Y^{(2)}_{a_1a_2a_3,b}~,
\end{align}
\begin{align}
3\Pi\big[(\Ga_{b[a_1})^{\al\be}(\Ga_{a_2a_3]})_\de{}^\ga
 & R_{(\al\be\ga)}{}^\de\big]=
-16\Pi\big[(\Ga_{b[a_1|}{}^{})^{\al\be}R_{\al\be |a_2a_3]}\big]\nn\\
&=-\frac{512}{3} B_{a_1a_2a_3,b}
-\frac{20556288}{16445} X^2_{a_1a_2a_3,b}
-\frac{22386688}{16445}X^{2\prime}_{a_1a_2a_3,b}\nn\\
&-\frac{3188736}{3289}Y^2_{a_1a_2a_3,b}
+\frac{1773568}{3289}Y^{2\prime}_{a_1a_2a_3,b}
-\frac{48384000}{3289}Y^{2\prime\prime}_{a_1a_2a_3,b}\nn\\
&+\frac{488192}{3289}
(D_{[a_1}X_{a_2a_3],b}-\frac{1}{9}\eta_{b[a_1}D^iX_{a_2a_3],i} )\nn\\
&-128A\circ Y^{(1)}_{a_1a_2a_3,b}
+\frac{289600}{9867}A\circ Y^{(2)}_{a_1a_2a_3,b}\nn\\
&+\frac{46336}{253}A^{\prime}\circ Y^{(1)}_{a_1a_2a_3,b}
-\frac{165632}{3289}A^{\prime}\circ Y^{(2)}_{a_1a_2a_3,b}~.
\end{align}
Implementing
the zeroth-order relation $A_{a_1\dots a_4}=-2A_{a_1\dots a_4}^\prime$ we get
\begin{align}
B_{a_1a_2a_3,b}&=\frac{495}{47656}A\circ Y^{(1)}_{a_1a_2a_3,b}+\frac{351}{6808}DX_{a_1a_2a_3,b}
+\frac{2916}{4255}X^{2}_{a_1a_2a_3,b}
+\frac{167}{185}X^{2\prime}_{a_1a_2a_3,b}\nn\\
&+\frac{117}{851}Y^{2}_{a_1a_2a_3,b}
-\frac{3287}{1702}Y^{2\prime}_{a_1a_2a_3,b}
+\frac{540}{851}Y^{2\prime\prime}_{a_1a_2a_3,b}\komma\nn\\
A\circ Y^{(2)}_{a_1a_2a_3,b}&=\frac{351}{259}A\circ Y^{(1)}_{a_1a_2a_3,b}
-\frac{405}{37}DX_{a_1a_2a_3,b}
-\frac{1080}{37}X^{2}_{a_1a_2a_3,b}
-\frac{1080}{37}X^{2\prime}_{a_1a_2a_3,b}\nn\\
&-\frac{1080}{37}Y^{2}_{a_1a_2a_3,b}
-\frac{540}{37}Y^{2\prime}_{a_1a_2a_3,b}
-\frac{17280}{37}Y^{2\prime\prime}_{a_1a_2a_3,b}\komma
\end{align}
where
\begin{equation}
DX_{a_1a_2a_3,b}:=D_{[a_1}X_{a_2a_3],b}-\frac{1}{9}\eta_{b[a_1}
D^iX_{a_2a_3],i}~.
\end{equation}

\noindent{\it The $(10010)$'s.}

\noindent From the 1st SSBI we get
\begin{align}
\Pi\big[(\Ga_{a_1\dots a_4}{}^{c})^{\al\be}R_{\al\be bc}\big]&=
896B_{a_1\dots a_4,b}
-\frac{43008}{115}X^{2}_{a_1\dots a_4,b}
-512X^{2\prime}_{a_1\dots a_4,b}\nn\\
&+\frac{87040}{161}Y^{2}_{a_1\dots a_4,b}
+\frac{122880}{161}Y^{2\prime}_{a_1\dots a_4,b}
+\frac{378880}{483}Y^{2\prime\prime}_{a_1\dots a_4,b}\nn\\
&+\frac{51200}{483}Y^{2\prime\prime\prime}_{a_1\dots a_4,b}
+\frac{25088}{69}A\circ X^{(1)}_{a_1\dots a_4,b}
-\frac{19456}{115}A\circ X^{(2)}_{a_1\dots a_4,b}\nn\\
&+\frac{39680}{69}A^{\prime}\circ X^{(1)}_{a_1\dots a_4,b}
-\frac{13568}{115}A^{\prime}\circ X^{(2)}_{a_1\dots a_4,b}\nn\\
&-\frac{1984}{1035}A\circ Y^{(1)}_{a_1\dots a_4,b}
+\frac{1312}{3105}A\circ Y^{(2)}_{a_1\dots a_4,b}\nn\\
&-\frac{560}{207}A^{\prime}\circ Y^{(1)}_{a_1\dots a_4,b}
+\frac{64}{115}A^{\prime}\circ Y^{(2)}_{a_1\dots a_4,b}\nn\\
&+\frac{9344}{1035}(
D^iY_{ia_1\dots a_4,b}-
D^iY_{ib[a_1\dots ,a_4]}
)
\komma
\label{plato}
\end{align}
\begin{align}
\Pi\big[(\Ga_{a_1\dots a_4}{}^{c})^{\al\be}R_{\al\be cb}\big]&=
128B_{a_1\dots a_4,b}
-\frac{43008}{115}X^{2}_{a_1\dots a_4,b}
-512X^{2\prime}_{a_1\dots a_4,b}\nn\\
&+\frac{87040}{161}Y^{2}_{a_1\dots a_4,b}
+\frac{122880}{161}Y^{2\prime}_{a_1\dots a_4,b}
+\frac{378880}{483}Y^{2\prime\prime}_{a_1\dots a_4,b}\nn\\
&+\frac{51200}{483}Y^{2\prime\prime\prime}_{a_1\dots a_4,b}
-\frac{10240}{69}A\circ X^{(1)}_{a_1\dots a_4,b}
+\frac{15872}{115}A\circ X^{(2)}_{a_1\dots a_4,b}\nn\\
&+\frac{4352}{69}A^{\prime}\circ X^{(1)}_{a_1\dots a_4,b}
+\frac{233728}{115}A^{\prime}\circ X^{(2)}_{a_1\dots a_4,b}\nn\\
&+\frac{224}{1035}A\circ Y^{(1)}_{a_1\dots a_4,b}
+\frac{5728}{3105}A\circ Y^{(2)}_{a_1\dots a_4,b}\nn\\
&-\frac{592}{1035}A^{\prime}\circ Y^{(1)}_{a_1\dots a_4,b}
-\frac{2368}{1035}A^{\prime}\circ Y^{(2)}_{a_1\dots a_4,b}\nn\\
&+\frac{62336}{1035}(
D^iY_{ia_1\dots a_4,b}-
D^iY_{ib[a_1\dots ,a_4]}
) ~,
\label{plotinus}
\end{align}
where we have defined
\begin{align}
A\circ X^{(1)}_{a_1\dots a_4,b}
&=A_{[a_1a_2a_3|}{}^iX_{i|a_4],b}+
A_{[a_1a_2a_3|}{}^iX_{ib,|a_4]}
+\frac{9}{16}
\eta_{b[a_1}A_{a_2a_3|}{}^{ij}X_{ij,|a_4]}
\nn\\
A\circ X^{(2)}_{a_1\dots a_4,b}&=
\frac{4}{3}A_{[a_1a_2a_3|}{}^iX_{i|a_4],b}
+\frac{1}{3}A_{[a_1a_2a_3|}{}^iX_{ib,|a_4]}
+\frac{1}{2}A_{b[a_1a_2}{}^iX_{a_3a_4],i}\nn\\
&+\frac{5}{16}\eta_{b[a_1}A_{a_2a_3|}{}^{ij}X_{ij,|a_4]}
\end{align}
and
\begin{align}
A\circ Y^{(1)}_{a_1\dots a_4,b}
&:=\Pi(\epsilon_{b[a_1a_2|}{}^{i_1\dots i_8}A_{i_1i_2i_3|a_3|}
Y_{i_4\dots i_8,|a_4]} )\nn\\
&=-\frac{2}{5}\epsilon_{b[a_1a_2}{}^{i_1\dots i_8}A_{a_3|i_1i_2i_3}
Y_{i_4\dots i_8,|a_4]}
+\frac{1}{5}\epsilon_{[a_1a_2a_3}{}^{i_1\dots i_8}A_{a_4]i_1i_2i_3}
Y_{i_4\dots i_8,b}\nn\\
&-\frac{1}{5}\epsilon_{[a_1a_2a_3|}{}^{i_1\dots i_8}A_{bi_1i_2i_3}
Y_{i_4\dots i_8,|a_4]}
+\frac{1}{8}\eta_{b[a_1}\epsilon_{a_2a_3|}{}^{i_1\dots i_9}
A_{i_1\dots i_4}Y_{i_5\dots i_9,|a_4]}
\nn\\
A\circ Y^{(2)}_{a_1\dots a_4,b}
&:=\Pi(\epsilon_{[a_1a_2a_3}{}^{i_1\dots i_8}A_{a_4]i_1i_2i_3}
Y_{i_4\dots i_8,b})\nn\\
&=-\frac{3}{5}
\epsilon_{b[a_1a_2}{}^{i_1\dots i_8}A_{a_3|i_1i_2i_3}
Y_{i_4\dots i_8,|a_4]}
+\frac{4}{5}\epsilon_{[a_1a_2a_3}{}^{i_1\dots i_8}A_{a_4]i_1i_2i_3}
Y_{i_4\dots i_8,b}\nn\\
&+\frac{1}{5}\epsilon_{[a_1a_2a_3|}{}^{i_1\dots i_8}A_{bi_1i_2i_3}
Y_{i_4\dots i_8,|a_4]}
+\frac{3}{32}\eta_{b[a_1}\epsilon_{a_2a_3|}{}^{i_1\dots i_9}
A_{i_1\dots i_4}Y_{i_5\dots i_9,|a_4]}~.
\end{align}
In deriving (\ref{plato}, \ref{plotinus}) we have used the relations
\begin{align}
\epsilon_{[a_1a_2|}{}^{i_1\dots i_9}A_{i_1\dots i_4}Y_{|a_3]i_5\dots ,i_9}
&=\frac{1}{5}
\epsilon_{[a_1a_2|}{}^{i_1\dots i_9}A_{i_1\dots i_4}Y_{i_5\dots i_9,|a_3]}\nn\\
&=-\frac{4}{3}
\epsilon_{a_1a_2a_3}{}^{i_1\dots i_8}A_{i_1i_2 i_3}{}^j
Y_{ji_4\dots ,i_8}
\end{align}
and
\begin{align}
A\circ Y^{(3)}&=-{1\over 5}A\circ Y^{(1)}+{1\over15}A\circ Y^{(2)}  \nn\\
A\circ Y^{(4)}&=-{3}A\circ Y^{(1)}+{}A\circ Y^{(2)}   \nn\\
A\circ Y^{(5)}&=-{4\over5}A\circ Y^{(1)}+{4\over15}A\circ Y^{(2)}   \nn\\
A\circ Y^{(6)}&={4\over5}A\circ Y^{(1)}   \nn\\
A\circ Y^{(7)}&=-{4\over5}A\circ Y^{(2)}   \nn\\
A\circ Y^{(8)}&={12\over5}A\circ Y^{(1)}-{4\over5}A\circ Y^{(2)}   \nn\\
A\circ Y^{(9)}&= {3\over20}A\circ Y^{(1)}  \nn\\
A\circ Y^{(10)}&={3\over5}A\circ Y^{(1)}  \nn\\
A\circ Y^{(11)}&={1\over5}A\circ Y^{(2)}   \nn\\
A\circ Y^{(12)}&={4\over5}A\circ Y^{(2)} ~,
\label{wittgenstein}
\end{align}
where
\begin{align}
A\circ Y_{a_1\dots a_4,b}^{(3)}&:=\Pi(\epsilon_{b[a_1a_2a_3}{}^{i_1\dots i_7}
A^j{}_{a_4]i_1i_2}
Y_{ji_3\dots ,i_7})\nn\\
A\circ Y_{a_1\dots a_4,b}^{(4)}&:=\Pi(\epsilon_{[a_1a_2a_3|}{}^{i_1\dots i_8}A_{bi_1i_2i_3}
Y_{i_4\dots i_8,|a_4]})\nn\\
A\circ Y_{a_1\dots a_4,b}^{(5)}&:=\Pi(\epsilon_{a_1\dots a_4}{}^{i_1\dots i_7}A_{bi_1i_2}{}^j
Y_{ji_3\dots ,i_7})\nn\\
A\circ Y_{a_1\dots a_4,b}^{(6)}&:=\Pi(\epsilon_{b[a_1a_2|}{}^{i_1\dots i_8}A_{i_1\dots i_4}
Y_{i_5\dots i_8|a_3,a_4]})\nn\\
A\circ Y_{a_1\dots a_4,b}^{(7)}&:=\Pi(\epsilon_{[a_1a_2a_3|}{}^{i_1\dots i_8}A_{i_1\dots i_4}
Y_{i_5\dots i_8|a_4],b})\nn\\
A\circ Y_{a_1\dots a_4,b}^{(8)}&:=\Pi(\epsilon_{[a_1a_2a_3|}{}^{i_1\dots i_8}A_{i_1\dots i_4}
Y_{i_5\dots i_8b,|a_4]})\nn\\
A\circ Y_{a_1\dots a_4,b}^{(9)}&:=\Pi(\epsilon_{b[a_1a_2a_3|}{}^{i_1\dots i_7}A_{i_1i_2i_3}{}^j
Y_{j|a_4]i_4\dots ,i_7})\nn\\
A\circ Y_{a_1\dots a_4,b}^{(10)}&:=\Pi(\epsilon_{a_1\dots a_4}{}^{i_1\dots i_7}A^j{}_{i_1i_2i_3}
Y_{jbi_4\dots ,i_7})\nn\\
A\circ Y_{a_1\dots a_4,b}^{(11)}&:=\Pi(\epsilon_{b[a_1a_2a_3|}{}^{i_1\dots i_7}A^j{}_{i_1i_2i_3}
Y_{ji_4\dots i_7,|a_4]})\nn\\
A\circ Y_{a_1\dots a_4,b}^{(12)}&:=\Pi(\epsilon_{a_1\dots a_4}{}^{i_1\dots i_7}A_{i_1i_2i_3}{}^j
Y_{ji_4\dots i_7,b})~.
\end{align}
Note that there are only two independent $A\circ Y$ structures,
as can be seen from the fact that
$$
A_{a_1\dots a_4}\otimes Y_{b_1\dots b_5,c}\sim (00010)\otimes(10002)
= 2(10010)\oplus\dots ~.
$$
\noindent From the 2nd SSBI we get
\begin{align}
3\Pi\big[(\Ga_{b})^{\al\be}(\Ga_{a_1\dots a_4})_\de{}^\ga
R_{(\al\be\ga)}{}^\de\big]
&=2\Pi\big[(\Ga_{a_1\dots a_4}{}^{c})^{\al\be}R_{\al\be bc}\big]\nn\\
&=2304B_{a_1\dots a_4,b}
-\frac{786432}{16445}X^{2}_{a_1\dots a_4,b}
-\frac{66048}{143}X^{2\prime}_{a_1\dots a_4,b}\nn\\
&+\frac{24975360}{23023}Y^{2}_{a_1\dots a_4,b}
+\frac{27985920}{23023}Y^{2\prime}_{a_1\dots a_4,b}\nn\\
&+\frac{62423040}{23023}Y^{2\prime\prime}_{a_1\dots a_4,b}
-\frac{25651200}{23023}Y^{2\prime\prime\prime}_{a_1\dots a_4,b}\nn\\
&-\frac{3348480}{3289}A\circ X^{(1)}_{a_1\dots a_4,b}
+\frac{19915776}{16445}A\circ X^{(2)}_{a_1\dots a_4,b}\nn\\
&-\frac{2998784}{3289}A^{\prime}\circ X^{(1)}_{a_1\dots a_4,b}
+\frac{15641088}{16445}A^{\prime}\circ X^{(2)}_{a_1\dots a_4,b}\nn\\
&+\frac{39232}{9867}A\circ Y^{(1)}_{a_1\dots a_4,b}
-\frac{153152}{148005}A\circ Y^{(2)}_{a_1\dots a_4,b}\nn\\
&+\frac{139712}{16445}A^{\prime}\circ Y^{(1)}_{a_1\dots a_4,b}
-\frac{24064}{16445}A^{\prime}\circ Y^{(2)}_{a_1\dots a_4,b}\nn\\
&+\frac{179456}{16445}(
D^iY_{ia_1\dots a_4,b}-
D^iY_{ib[a_1\dots ,a_4]}
)
\komma
\end{align}
\begin{align}
3\Pi\big[(\Ga_{b[a_1})^{\al\be}(\Ga_{a_2a_3a_4]})_\de{}^\ga
R_{(\al\be\ga)}{}^\de\big]&=
\frac{3}{2}\Pi\big[(\Ga_{a_1\dots a_4}{}^{c})^{\al\be}R_{\al\be bc}\big]\nn\\
&=-192B_{a_1\dots a_4,b}
-\frac{26102784}{16445}X^{2}_{a_1\dots a_4,b}
-\frac{197376}{143}X^{2\prime}_{a_1\dots a_4,b}\nn\\
&+\frac{6059520}{23023}Y^{2}_{a_1\dots a_4,b}
+\frac{23101440}{23023}Y^{2\prime}_{a_1\dots a_4,b}\nn\\
&-\frac{43576320}{23023}Y^{2\prime\prime}_{a_1\dots a_4,b}
+\frac{62284800}{23023}Y^{2\prime\prime\prime}_{a_1\dots a_4,b}\nn\\
&+\frac{4500480}{3289}A\circ X^{(1)}_{a_1\dots a_4,b}
-\frac{14462208}{16445}A\circ X^{(2)}_{a_1\dots a_4,b}\nn\\
&-\frac{1488768}{3289}A^{\prime}\circ X^{(1)}_{a_1\dots a_4,b}
+\frac{23135616}{16445}A^{\prime}\circ X^{(2)}_{a_1\dots a_4,b}\nn\\
&-\frac{44592}{16445}A\circ Y^{(1)}_{a_1\dots a_4,b}
+\frac{6832}{16445}A\circ Y^{(2)}_{a_1\dots a_4,b}\nn\\
&-\frac{6008}{16445}A^{\prime}\circ Y^{(1)}_{a_1\dots a_4,b}
+\frac{11488}{9867}A^{\prime}\circ Y^{(2)}_{a_1\dots a_4,b}\nn\\
&-\frac{111808}{16445}(
D^iY_{ia_1\dots a_4,b}-
D^iY_{ib[a_1\dots ,a_4]}
)~.
\end{align}
Implementing the zeroth-order relation
$A_{a_1\dots a_4}=-2A_{a_1\dots a_4}^\prime$ we get
\begin{align}
B_{a_1\dots a_4,b}&=\frac{5052}{4255}A\circ X^{(2)}_{a_1\dots a_4,b}
+\frac{401}{2042400}A\circ Y^{(1)}_{a_1\dots a_4,b}
-\frac{577}{170200}A\circ Y^{(2)}_{a_1\dots a_4,b}\nn\\
&-\frac{63}{851}DY_{a_1\dots a_4,b}
+\frac{3808}{4255}X^{2}_{a_1\dots a_4,b}
+\frac{991}{851}X^{2\prime}_{a_1\dots a_4,b}
-\frac{6780}{5957}Y^{2}_{a_1\dots a_4,b}\nn\\
&-\frac{9860}{5957}Y^{2\prime}_{a_1\dots a_4,b}
-\frac{8800}{5957}Y^{2\prime\prime}_{a_1\dots a_4,b}
-\frac{2540}{5957}Y^{2\prime\prime\prime}_{a_1\dots a_4,b}
\komma\nn\\
A\circ X^{(1)}_{a_1\dots a_4,b}&=\frac{81}{37}A\circ X^{(2)}_{a_1\dots a_4,b}
+\frac{119}{88800}A\circ Y^{(1)}_{a_1\dots a_4,b}
-\frac{217}{66600}A\circ Y^{(2)}_{a_1\dots a_4,b}
-\frac{7}{111}DY_{a_1\dots a_4,b}\nn\\
&+\frac{60}{37}X^{2}_{a_1\dots a_4,b}
+\frac{60}{37}X^{2\prime}_{a_1\dots a_4,b}
-\frac{30}{37}Y^{2}_{a_1\dots a_4,b}
-\frac{60}{37}Y^{2\prime}_{a_1\dots a_4,b}
+\frac{20}{37}Y^{2\prime\prime}_{a_1\dots a_4,b}\nn\\
&-\frac{80}{37}Y^{2\prime\prime\prime}_{a_1\dots a_4,b}\komma
\end{align}
where
\begin{equation}
DY_{a_1\dots a_4,b}:=D^iY_{ia_1\dots a_4,b}-
D^iY_{ib[a_1\dots ,a_4]}\punkt
\end{equation}

\noindent{\it The $(10002)$'s.}

\noindent From the 1st SSBI we get
\begin{align}
\Pi\big[(\Ga_{a_1\dots a_5}{}^{c})^{\al\be}R_{\al\be bc}\big]&=
-768B_{a_1\dots a_5,b}
-\frac{8960}{23}X^{2}_{a_1\dots a_5,b}
+\frac{4480}{23}X^{2\prime}_{a_1\dots a_5,b}\nn\\
&-\frac{13440}{23}Y^{2}_{a_1\dots a_5,b}
-\frac{21760}{23}Y^{2\prime}_{a_1\dots a_5,b}
-\frac{17920}{23}Y^{2\prime\prime}_{a_1\dots a_5,b}\nn\\
&-\frac{10240}{23}Y^{2\prime\prime\prime}_{a_1\dots a_5,b}
+\frac{8960}{69}Y^{2\prime\prime\prime\prime}_{a_1\dots a_5,b}\nn\\
&-\frac{8320}{69}A\circ Y^{(1)}_{a_1\dots a_5,b}
-\frac{53120}{69}A\circ Y^{(2)}_{a_1\dots a_5,b}\nn\\
&-\frac{20480}{69}A^{\prime}\circ Y^{(1)}_{a_1\dots a_5,b}
-\frac{81920}{69}A^{\prime}\circ Y^{(2)}_{a_1\dots a_5,b}\nn\\
&-\frac{4}{69}(\epsilon_{a_1\dots a_5}{}^{i_1\dots i_6}
D_{i_1}Y_{i_2\dots i_6,b}+\epsilon_{b[a_1\dots a_4|}{}^{i_1\dots i_6}
D_{i_1}Y_{i_2\dots i_6,|a_5]}
)\nn\\
&+\frac{440}{207}(\epsilon_{a_1\dots a_5}{}^{i_1\dots i_6}
A^{\prime}_{i_1\dots i_4}X_{i_5i_6,b}
+\epsilon_{b[a_1\dots a_4|}{}^{i_1\dots i_6}
A^{\prime}_{i_1\dots i_4}X_{i_5i_6,|a_5]})
\komma
\end{align}
\begin{align}
\Pi\big[(\Ga_{a_1\dots a_5}{}^{c})^{\al\be}R_{\al\be cb}\big]&=
-128B_{a_1\dots a_5,b}
-\frac{8960}{23}X^{2}_{a_1\dots a_5,b}
+\frac{4480}{23}X^{2\prime}_{a_1\dots a_5,b}\nn\\
&-\frac{13440}{23}Y^{2}_{a_1\dots a_5,b}
-\frac{21760}{23}Y^{2\prime}_{a_1\dots a_5,b}
-\frac{17920}{23}Y^{2\prime\prime}_{a_1\dots a_5,b}\nn\\
&-\frac{10240}{23}Y^{2\prime\prime\prime}_{a_1\dots a_5,b}
+\frac{8960}{69}Y^{2\prime\prime\prime\prime}_{a_1\dots a_5,b}\nn\\
&+\frac{50560}{69}A\circ Y^{(1)}_{a_1\dots a_5,b}
+\frac{35200}{69}A\circ Y^{(2)}_{a_1\dots a_5,b}\nn\\
&-\frac{145600}{69}A^{\prime}\circ Y^{(1)}_{a_1\dots a_5,b}
-\frac{104000}{69}A^{\prime}\circ Y^{(2)}_{a_1\dots a_5,b}\nn\\
&-\frac{104}{207}(\epsilon_{a_1\dots a_5}{}^{i_1\dots i_6}
D_{i_1}Y_{i_2\dots i_6,b}+\epsilon_{b[a_1\dots a_4|}{}^{i_1\dots i_6}
D_{i_1}Y_{i_2\dots i_6,|a_5]}
)\nn\\
&+\frac{80}{9}(\epsilon_{a_1\dots a_5}{}^{i_1\dots i_6}
A_{i_1\dots i_4}X_{i_5i_6,b}+\epsilon_{b[a_1\dots a_4|}{}^{i_1\dots i_6}
A_{i_1\dots i_4}X_{i_5i_6,|a_5]})\nn\\
&-\frac{20}{207}(\epsilon_{a_1\dots a_5}{}^{i_1\dots i_6}
A^{\prime}_{i_1\dots i_4}X_{i_5i_6,b}
+\epsilon_{b[a_1\dots a_4|}{}^{i_1\dots i_6}
A^{\prime}_{i_1\dots i_4}X_{i_5i_6,|a_5]}) ~,
\end{align}
where we have defined the two irreducible-hook combinations
\begin{align}
A\circ Y^{(1)}_{a_1\dots a_5,b}&:=
A_{[a_1a_2|}{}^{i_1i_2}Y_{i_1i_2|a_3a_4a_5],b}
+A_{b[a_1|}{}^{i_1i_2}Y_{i_1i_2|a_2\dots ,a_5]},\nn\\
A\circ Y^{(2)}_{a_1\dots a_5,b}&:=
A_{[a_1a_2|}{}^{i_1i_2}Y_{i_1i_2b|a_3a_4,a_5]}
-A_{b[a_1|}{}^{i_1i_2}Y_{i_1i_2|a_2\dots ,a_5]}\nn\\
&+\frac{4}{7} \eta_{b[a_1}A_{a_2|}{}^{i_1i_2 i_3}Y_{i_1i_2 i_3|a_3a_4,a_5]}~.
\end{align}
Note that indeed there are two $(10002)$'s in the decomposition
of the tensor product
$A_{a_1\dots a_4}\otimes Y_{b_1\dots b_5,c}\sim (00010)\otimes(10002)$.

\noindent From the 2nd SSBI we get
\begin{align}
3\Pi\big[(\Ga_{b})^{\al\be}(\Ga_{a_1\dots a_5})_\de{}^\ga
& R_{(\al\be\ga)}{}^\de \big]=
-2\Pi\big[(\Ga_{a_1\dots a_5}{}^{c})^{\al\be}R_{\al\be bc}\big]\nn\\
&=2304B_{a_1\dots a_5,b}
+\frac{990720}{3289}X^{2}_{a_1\dots a_5,b}\nn\\
&+\frac{1388800}{3289}X^{2\prime}_{a_1\dots a_5,b}
-\frac{162560}{3289}Y^{2}_{a_1\dots a_5,b}\nn\\
&+\frac{42984960}{23023}Y^{2\prime}_{a_1\dots a_5,b}
+\frac{7398400}{3289}Y^{2\prime\prime}_{a_1\dots a_5,b}\nn\\
&+\frac{49612800}{23023}Y^{2\prime\prime\prime}_{a_1\dots a_5,b}
-\frac{212480}{3289}Y^{2\prime\prime\prime\prime}_{a_1\dots a_5,b}\nn\\
&+\frac{51200}{299}A\circ Y^{(1)}_{a_1\dots a_5,b}
-\frac{4968960}{3289}A\circ Y^{(2)}_{a_1\dots a_5,b}\nn\\
&-\frac{197760}{3289}A^{\prime}\circ Y^{(1)}_{a_1\dots a_5,b}
-\frac{836480}{253}A^{\prime}\circ Y^{(2)}_{a_1\dots a_5,b}\nn\\
&+\frac{2720}{29601}(\epsilon_{a_1\dots a_5}{}^{i_1\dots i_6}
D_{i_1}Y_{i_2\dots i_6,b}+\epsilon_{b[a_1\dots a_4|}{}^{i_1\dots i_6}
D_{i_1}Y_{i_2\dots i_6,|a_5]}
)\nn\\
&-\frac{2920}{897}(\epsilon_{a_1\dots a_5}{}^{i_1\dots i_6}
A^{\prime}_{i_1\dots i_4}X_{i_5i_6,b}
+\epsilon_{b[a_1\dots a_4|}{}^{i_1\dots i_6}
A^{\prime}_{i_1\dots i_4}X_{i_5i_6,|a_5]})
\komma
\end{align}
\begin{align}
3\Pi\big[(\Ga_{b[a_1})^{\al\be}(\Ga_{a_2\dots a_5]})_\de{}^\ga
& R_{(\al\be\ga)}{}^\de\big]=
-\frac{8}{5}\Pi\big[(\Ga_{a_1\dots a_5}{}^{c})^{\al\be}R_{\al\be bc}\big]\nn\\
&=\frac{1024}{5}B_{a_1\dots a_5,b}
+\frac{2936832}{3289}X^{2}_{a_1\dots a_5,b}\nn\\
&-\frac{3376128}{3289}X^{2\prime}_{a_1\dots a_5,b}
+\frac{5394432}{3289}Y^{2}_{a_1\dots a_5,b}\nn\\
&+\frac{18696192}{23023}Y^{2\prime}_{a_1\dots a_5,b}
+\frac{1069056}{3289}Y^{2\prime\prime}_{a_1\dots a_5,b}\nn\\
&-\frac{13953024}{23023}Y^{2\prime\prime\prime}_{a_1\dots a_5,b}
-\frac{531456}{3289}Y^{2\prime\prime\prime\prime}_{a_1\dots a_5,b}\nn\\
&+\frac{8192}{23}A\circ Y^{(1)}_{a_1\dots a_5,b}
+\frac{2085888}{3289}A\circ Y^{(2)}_{a_1\dots a_5,b}\nn\\
&+\frac{657920}{3289}A^{\prime}\circ Y^{(1)}_{a_1\dots a_5,b}
-\frac{936448}{3289}A^{\prime}\circ Y^{(2)}_{a_1\dots a_5,b}\nn\\
&-\frac{8512}{148005}(\epsilon_{a_1\dots a_5}{}^{i_1\dots i_6}
D_{i_1}Y_{i_2\dots i_6,b}+\epsilon_{b[a_1\dots a_4|}{}^{i_1\dots i_6}
D_{i_1}Y_{i_2\dots i_6,|a_5]}
)\nn\\
&-\frac{7840}{897}(\epsilon_{a_1\dots a_5}{}^{i_1\dots i_6}
A^{\prime}_{i_1\dots i_4}X_{i_5i_6,b}
+\epsilon_{b[a_1\dots a_4|}{}^{i_1\dots i_6}
A^{\prime}_{i_1\dots i_4}X_{i_5i_6,|a_5]})
\komma
\end{align}
\begin{align}
3\Pi\big[(\Ga_{a_1\dots a_5})^{\al\be}(\Ga_{b})_\de{}^\ga
 & R_{(\al\be\ga)}{}^\de\big]=
-2\Pi\big[(\Ga_{a_1\dots a_5}{}^{c})^{\al\be}R_{\al\be bc}\big]\nn\\
&=256B_{a_1\dots a_5,b}
+\frac{2434560}{3289}X^{2}_{a_1\dots a_5,b}\nn\\
&-\frac{2983680}{3289}X^{2\prime}_{a_1\dots a_5,b}
+\frac{8597760}{3289}Y^{2}_{a_1\dots a_5,b}\nn\\
&+\frac{40680960}{23023}Y^{2\prime}_{a_1\dots a_5,b}
-\frac{3609600}{3289}Y^{2\prime\prime}_{a_1\dots a_5,b}\nn\\
&-\frac{43407360}{23023}Y^{2\prime\prime\prime}_{a_1\dots a_5,b}
-\frac{3755520}{3289}Y^{2\prime\prime\prime\prime}_{a_1\dots a_5,b}\nn\\
&+\frac{192000}{299}A\circ Y^{(1)}_{a_1\dots a_5,b}
+\frac{1812480}{3289}A\circ Y^{(2)}_{a_1\dots a_5,b}\nn\\
&+\frac{734080}{3289}A^{\prime}\circ Y^{(1)}_{a_1\dots a_5,b}
-\frac{5027200}{3289}A^{\prime}\circ Y^{(2)}_{a_1\dots a_5,b}\nn\\
&-\frac{3232}{29601}(\epsilon_{a_1\dots a_5}{}^{i_1\dots i_6}
D_{i_1}Y_{i_2\dots i_6,b}+\epsilon_{b[a_1\dots a_4|}{}^{i_1\dots i_6}
D_{i_1}Y_{i_2\dots i_6,|a_5]}
)\nn\\
&-\frac{13480}{897}(\epsilon_{a_1\dots a_5}{}^{i_1\dots i_6}
A^{\prime}_{i_1\dots i_4}X_{i_5i_6,b}
+\epsilon_{b[a_1\dots a_4|}{}^{i_1\dots i_6}
A^{\prime}_{i_1\dots i_4}X_{i_5i_6,|a_5]})~.
\end{align}
Implementing
the zeroth-order relation
$A_{a_1\dots a_4}=-2A^\prime_{a_1\dots a_4}$ we get
\begin{align}
B_{a_1\dots a_5,b}&=-\frac{65}{92736}A\circ X_{a_1\dots a5,b}
-\frac{295}{966}A\circ Y^{(1)}_{a_1\dots a_5,b}
+\frac{20}{23}X^{2}_{a_1\dots a_5,b}
-\frac{30}{23}X^{2\prime}_{a_1\dots a_5,b}\nn\\
&+\frac{15}{23}Y^{2}_{a_1\dots a_5,b}
-\frac{165}{161}Y^{2\prime}_{a_1\dots a_5,b}
+\frac{15}{23}Y^{2\prime\prime}_{a_1\dots a_5,b}
-\frac{55}{161}Y^{2\prime\prime\prime}_{a_1\dots a_5,b}
+\frac{5}{6}Y^{2\prime\prime\prime\prime}_{a_1\dots a_5,b}
\komma\nn\\
A\circ Y^{(2)}_{a_1\dots a_5,b}&=-\frac{11}{2016}A\circ X_{a_1\dots a5,b}
-\frac{11}{42}A\circ Y^{(1)}_{a_1\dots a_5,b}
+X^{2}_{a_1\dots a_5,b}
-X^{2\prime}_{a_1\dots a_5,b}\nn\\
&-\frac{5}{2}Y^{2}_{a_1\dots a_5,b}
-3Y^{2\prime}_{a_1\dots a_5,b}
+5Y^{2\prime\prime}_{a_1\dots a_5,b}
+4Y^{2\prime\prime\prime}_{a_1\dots a_5,b}
+\frac{10}{3}Y^{2\prime\prime\prime\prime}_{a_1\dots a_5,b}\komma\nn\\
DY_{a_1\dots a_5,b}&=\frac{55}{12}A\circ X_{a_1\dots a5,b}
+3220 A\circ Y^{(1)}_{a_1\dots a_5,b}
-840 X^{2}_{a_1\dots a_5,b}
+840 X^{2\prime}_{a_1\dots a_5,b}\nn\\
&-7980 Y^{2}_{a_1\dots a_5,b}
-7560 Y^{2\prime}_{a_1\dots a_5,b}
+5880 Y^{2\prime\prime}_{a_1\dots a_5,b}
+6720 Y^{2\prime\prime\prime}_{a_1\dots a_5,b}
+5600 Y^{2\prime\prime\prime\prime}_{a_1\dots a_5,b}\komma
\end{align}
where
\begin{align}
DY_{a_1\dots a_5,b}&:=\epsilon_{a_1\dots a_5}{}^{i_1\dots i_6}
D_{i_1}Y_{i_2\dots i_6,b}+\epsilon_{b[a_1\dots a_4|}{}^{i_1\dots i_6}
D_{i_1}Y_{i_2\dots i_6,|a_5]}\komma\nn\\
A\circ X_{a_1\dots a5,b}&:=\epsilon_{a_1\dots a_5}{}^{i_1\dots i_6}
A_{i_1\dots i_4}X_{i_5i_6,b}
+\epsilon_{b[a_1\dots a_4|}{}^{i_1\dots i_6}
A_{i_1\dots i_4}X_{i_5i_6,|a_5]}\punkt
\end{align}

\subsection{The dimension-$3\over2$ SSBI's}

We consider now the SSBI's at dimension $\frac{3}{2}$, which read
\begin{align}
\begin{split}
&2R_{\al[bc]}{}^d=D_\al T_{bc}{}^d+2D_{[b}T_{c]\al}{}^d
    +2T_{\al[b}{}^ET_{|E|c]}{}^d+T_{bc}{}^ET_{E\al}{}^d\\
&2R_{a(\be\ga)}{}^\de=D_a T_{\be\ga}{}^\de+2D_{(\be}T_{\ga)a}{}^\de
    +2T_{a(\be}{}^ET_{|E|\ga)}{}^\de+T_{\be\ga}{}^ET_{Ea}{}^\de
\label{DimThreeHalvesBI}
\end{split}
\end{align}
In an unconstrained superfield in the representation 4290, there are
two spinors at level $\theta^3$. The index structure of the SSBI's at this
level also contains two spinor equations.
By contracting the first of the SSBI's with $\delta^c_d\Ga^b$ we find
\begin{align}
-( \Ga^b )_\al{}^\be R_{\be c b}{}^c= &-{13\over 6}( \Ga^{i_1i_2i_3}
\tilde{S}^{i_4})_\al A_{i_1\dots i_4}
-{13\over 6}( \Ga^{i_1i_2i_3} \tilde{S}^{i_4})_\al A^{\prime}_{i_1\dots i_4}\nn\\
&+{5\over 3}( \Ga^{i_1\dots i_4} \tilde{S})_\al A_{i_1\dots i_4}
-{35\over 12}( \Ga^{i_1\dots i_4} \tilde{S})_\al A^{\prime}_{i_1\dots i_4}\nn\\
&+13D^i\tilde{S}_{i\al}-10\Dslash\tilde{S}_{\al}\nn\\
&-220\tilde{t}_{\al}+2( \Ga^{k} \tilde{t}^{i_1i_2})_\al X_{i_1i_2,k}
+{1\over 6}( \Ga^{i_1\dots i_4} \tilde{t}^{i_5i_6})_\al Y_{i_1\dots
i_5,i_6}. \label{eva}
\end{align}
Contracting with $\Ga^{bcd}$ we get
\begin{equation}
2( \Ga^{bcd} )_\al{}^\be R_{\be bcd} =-1980\tilde{t}_{\al}-4(
\Ga^{k} \tilde{t}^{i_1i_2})_\al X_{i_1i_2,k} -{1\over 3}(
\Ga^{i_1\dots i_4} \tilde{t}^{i_5i_6})_\al Y_{i_1\dots i_5,i_6}.
\label{duo}
\end{equation}
Contracting the second of the SSBI's with
$C_{\al\de}(\Ga^a)^{\be\ga}$ and taking the Lorentz condition into
account, we get
\begin{align}
{1\over 2}( \Ga^{bcd} )_\al{}^\be R_{\be bcd}
+( \Ga^b )_\al{}^\be &R_{\be c b}{}^c=\nn\\
&+22  D_{\al}A-20(\Ga^iD)_{\al}A_i+2 (\Ga^{i_1i_2}D)_{\al}A_{i_1i_2} \nn\\
&+(\Ga^{i_1i_2i_3}D)_{\al}A_{i_1i_2i_3}-{1\over 3}(\Ga^{i_1\dots
i_4}D)_{\al}A_{i_1\dots i_4}
-{1\over 12}(\Ga^{i_1\dots i_5}D)_{\al}A_{i_1\dots i_5}\nn\\
&-2 (\Ga^{i}D)_{\al}A^{\prime}_{i} -9
(\Ga^{i_1i_2}D)_{\al}A^{\prime}_{i_1i_2}
+{8\over 3}(\Ga^{i_1i_2i_3}D)_{\al}A^{\prime}_{i_1i_2i_3}\nn\\
&+{7\over 12}(\Ga^{i_1\dots i_4}D)_{\al}A^{\prime}_{i_1 \dots i_4}
-{1\over 10}(\Ga^{i_1\dots i_5}D)_{\al}A^{\prime}_{i_1\dots i_5}\nn\\
&-{13\over 3}( \Ga^{i_1i_2i_3} \tilde{S}^{i_4})_\al A_{i_1\dots i_4}
-{13\over 3}( \Ga^{i_1i_2i_3} \tilde{S}^{i_4})_\al A^{\prime}_{i_1\dots i_4}\nn\\
&-{1\over 3}( \Ga^{i_1\dots i_4} \tilde{S})_\al A_{i_1\dots i_4}
+{7\over 12}( \Ga^{i_1\dots i_4} \tilde{S})_\al A^{\prime}_{i_1\dots
i_4}. \label{tria}
\end{align}
Contracting  with $(\Ga^{ai})_{\al\de}(\Ga_i)^{\be\ga}$  we get
\begin{align}
3( \Ga^{bcd} )_\al{}^\be R_{\be bcd}
-8( \Ga^b )_\al{}^\be &R_{\be c b}{}^c=\nn\\
&-220  D_{\al}A+160(\Ga^iD)_{\al}A_i+16 (\Ga^{i_1i_2}D)_{\al}A_{i_1i_2} \nn\\
&+6(\Ga^{i_1i_2i_3}D)_{\al}A_{i_1i_2i_3}-{4\over 3}(\Ga^{i_1\dots
i_4}D)_{\al}A_{i_1\dots i_4}\nn\\
&-{1\over 6}(\Ga^{i_1\dots i_5}D)_{\al}A_{i_1\dots i_5} -20
(\Ga^{i}D)_{\al}A^{\prime}_{i}
+54 (\Ga^{i_1i_2}D)_{\al}A^{\prime}_{i_1i_2}\nn\\
&-{32\over 3}(\Ga^{i_1i_2i_3}D)_{\al}A^{\prime}_{i_1i_2i_3}
-{7\over 6}(\Ga^{i_1\dots i_4}D)_{\al}A^{\prime}_{i_1\dots i_4}\nn\\
&-{52\over 3}( \Ga^{i_1i_2i_3} \tilde{S}^{i_4})_\al A_{i_1\dots i_4}
+{26\over 3}( \Ga^{i_1i_2i_3} \tilde{S}^{i_4})_\al A^{\prime}_{i_1\dots i_4}\nn\\
&-{4\over 3}( \Ga^{i_1\dots i_4} \tilde{S})_\al A_{i_1\dots i_4}
-{7\over 6}( \Ga^{i_1\dots i_4} \tilde{S})_\al A^{\prime}_{i_1\dots i_4}\nn\\
&+32( \Ga^{i_1\dots i_4} \tilde{Z}^{\prime})_\al A_{i_1\dots i_4}
+64( \Ga^{i_1i_2i_3} \tilde{Z}^{\prime i_4})_\al A_{i_1\dots
i_4}\nn\\
&+32( \Ga^{i_1i_2} \tilde{Z}^{\prime i_3i_4})_\al A_{i_1\dots i_4}
+112( \Ga^{i_1\dots i_4} \tilde{Z})_\al A^{\prime}_{i_1\dots
i_4}\nn\\
&-320( \Ga^{i_1i_2i_3} \tilde{Z}^{i_4})_\al A^{\prime}_{i_1\dots
i_4}+320( \Ga^{i_1i_2} \tilde{Z}^{i_3i_4})_\al A^{\prime}_{i_1\dots
i_4}\nn\\
&-128( \Ga^{i_1} \tilde{Z}^{i_2i_3i_4})_\al A^{\prime}_{i_1\dots
i_4}
+16 \tilde{Z}^{i_1\dots i_4}_\al A^{\prime}_{i_1\dots i_4}\nn\\
&+7040\tilde{t}_\al\komma \label{tessera}
\end{align}
where we have used the conventions $\G_{012345678910}=-1$ and
$\eps_{012345678910}=1$ and also the relation
\begin{equation}
\Ga_{a_1\ldots
a_p}=-(-1)^{(p+1)(p-2)/2}\frac{1}{(11-p)!}\eps_{a_1\ldots
a_p}{}^{a_{p+1}\ldots a_{11}}\Ga_{a_{p+1}\ldots a_{11}}\punkt
\end{equation}
Combining the above equations, eliminating $\tilde{t}_{ij}{}^\al$,
we finally get
\begin{align}
-38060\tilde{t}_\al=
&-440  D_{\al}A+280(\Ga^iD)_{\al}A_i+68 (\Ga^{i_1i_2}D)_{\al}A_{i_1i_2} \nn\\
&+28(\Ga^{i_1i_2i_3}D)_{\al}A_{i_1i_2i_3}-{22\over 3}(\Ga^{i_1\dots
i_4}D)_{\al}A_{i_1\dots i_4}
-{4\over 3}(\Ga^{i_1\dots i_5}D)_{\al}A_{i_1\dots i_5}\nn\\
&-80 (\Ga^{i}D)_{\al}A^{\prime}_{i} +72
(\Ga^{i_1i_2}D)_{\al}A^{\prime}_{i_1i_2}
-{16\over 3}(\Ga^{i_1i_2i_3}D)_{\al}A^{\prime}_{i_1i_2i_3}\nn\\
&+{7\over 3}(\Ga^{i_1\dots i_4}D)_{\al}A^{\prime}_{i_1\dots i_4}
-(\Ga^{i_1\dots i_5}D)_{\al}A^{\prime}_{i_1\dots i_5}-65(
\Ga^{i_1i_2i_3} \tilde{S}^{i_4})_\al A_{i_1\dots i_4}\nn\\
&+13( \Ga^{i_1i_2i_3} \tilde{S}^{i_4})_\al A^{\prime}_{i_1\dots i_4}
-{92\over 3}( \Ga^{i_1\dots i_4} \tilde{S})_\al A_{i_1\dots
i_4}+{259\over 6}( \Ga^{i_1\dots i_4} \tilde{S})_\al
A^{\prime}_{i_1\dots i_4}\nn\\
& +96( \Ga^{i_1\dots i_4} \tilde{Z}^{\prime})_\al A_{i_1\dots i_4}
 +192( \Ga^{i_1i_2i_3} \tilde{Z}^{\prime i_4})_\al A_{i_1\dots i_4}
+96( \Ga^{i_1i_2} \tilde{Z}^{\prime i_3i_4})_\al A_{i_1\dots i_4}\nn\\
&+336( \Ga^{i_1\dots i_4} \tilde{Z})_\al A^{\prime}_{i_1\dots i_4}
-960( \Ga^{i_1i_2i_3} \tilde{Z}^{i_4})_\al A^{\prime}_{i_1\dots i_4}
+960( \Ga^{i_1i_2} \tilde{Z}^{i_3i_4})_\al A^{\prime}_{i_1\dots
i_4j}\nn\\
& -384( \Ga^{i_1} \tilde{Z}^{i_2i_3i_4})_\al A^{\prime}_{i_1\dots
i_4} +48\tilde{Z}^{i_1\dots i_4}_\al A^{\prime}_{i_1\dots
i_4}-182D^i\tilde{S}_{i\al}+140\Dslash\tilde{S}_{\al}. \label{pevte}
\end{align}

\subsection{The dimension-2 SSBI's}

We consider now the SSBI's at dimension 2, which read
\begin{align}
\begin{split}
&R_{[abc]}{}^d=D_{[a}T_{bc]}{}^d
    +T_{[ab}{}^ET_{|E|c]}{}^d\komma\\
&R_{ab\ga}{}^\de=2D_{[a} T_{b]\ga}{}^\de+D_{\ga}T_{ab}{}^\de
    +T_{ab}{}^ET_{|E|\ga}{}^\de+2T_{\ga[a}{}^ET_{|E|b]}{}^\de\punkt
\label{DimTwoBI}
\end{split}
\end{align}
We will focus on the representations associated with the Einstein
equations, (00000) and (20000), the 4-form equation of motion,
(00100), and the 4-form BI, (00002). Only the second SSBI in
\ref{DimTwoBI} contributes to these representations and the first
SSBI will therefore not be analysed below.

\bigskip
\noindent{\it The $(00000)$ and $(20000)$.}

\noindent The second SSBI contains one $(00000)$ and one $(20000)$.
They are obtained by contracting with $(\Ga_{bc})_\de{}^\ga$ and
symmetrising in $ac$,
\begin{align}
16R_{(a}{}^b{}_{c)b}&= -288D_{(a}A_{c)}-32\eta_{ac}D^iA_i+32D^i
B_{i(a,c)}+10\eta_{ac}D^\al\tilde{t}_\al-9D\Ga_{(a}\tilde{t}_{c)}\nn\\
&+2\tilde{S}_{(a,}{}^{i\al}\tilde{t}_{c)i\al}-11\tilde{S}_i\Ga_{(a}\tilde{t}_{c)}{}^i
+140\tilde{Z}_{(a}{}^{i\al}\tilde{t}_{c)i\al}+28\tilde{Z}^i\Ga_{(a}\tilde{t}_{c)i}-4\tilde{Z}^\prime_{(a}{}^{i\al}\tilde{t}_{c)i\al}\nn\\
& -14\tilde{Z}^{\prime
i}\Ga_{(a}\tilde{t}_{c)i}-32A_{(a}{}^{i_1i_2i_3}A_{c)i_1i_2i_3}-64A^\prime_{(a}{}^{i_1i_2i_3}A^\prime_{c)i_1i_2i_3}
+16\eta_{ac}A^\prime_{i_1\dots i_4}A^{\prime i_1\dots i_4}\nn\\
&+\frac{64}{3}A_{(a|i_1i_2i_3|}B^{i_1i_2i_3}{}_{,c)}+\frac{40}{3}A^\prime_{i_1\dots
i_4}B^{i_1\dots i_4}{}_{(a,c)} \punkt
\end{align}

\bigskip
\noindent{\it The $(00100)$'s.}

\noindent The second SSBI contains three $(00100)$'s, which are
obtained by contracting with $(\Ga_{c})_\de{}^\ga$ and
antisymmetrising in $abc$,
\begin{align}
0&=-64D_{[a}A_{bc]}-D\Ga_{[a}\tilde{t}_{bc]}-2D\Ga_{[ab}\tilde{t}_{c]}-D\Ga_{abc}\tilde{t}+4\tilde{S}_{[a}{}^\al\tilde{t}_{bc]\al}
-2\tilde{S}\Ga_{[a}\tilde{t}_{bc]}\nn\\
&-2\tilde{S}^i\Ga_{[ab}\tilde{t}_{c]i}-2\tilde{S}^i{}_{,[a}\Ga_b\tilde{t}_{c]i}-252\tilde{Z}_{[a}{}^{\al}\tilde{t}_{bc]\al}-42\tilde{Z}\Ga_{[a}\tilde{t}_{bc]}
-18\tilde{Z}^\prime_{[a}{}^{\al}\tilde{t}_{bc]\al}-35\tilde{Z}^\prime\Ga_{[a}\tilde{t}_{bc]}\nn\\
&-\frac{64}{3}A_{[a}{}^{i_1i_2i_3}A_{bc]i_1i_2i_3}+64A_{[ab}{}^{i_1i_2}A^\prime_{c]i_1i_2}+\frac{64}{3}A_{[a|i_1i_2i_3}B^{i_1i_2i_3}{}_{|b,c]}
-\frac{64}{3}A^\prime_{[a}{}^{i_1i_2i_3}A^\prime_{bc]i_1i_2i_3}\nn\\
&-\frac{1}{9}\eps_{abci_1\dots i_8}A^{\prime i_1\dots i_4}A^{\prime
i_5\dots i_8}-\frac{2}{45}\eps_{[ab|i_1\dots i_9}A^{\prime i_1\dots
i_4}B^{i_5\dots i_9}{}_{,|c]}\komma
\end{align}
$(\Ga^{b}{}_{cd})_\de{}^\ga$ and antisymmetrising in $acd$,
\begin{align}
0&=-256D_{[a} A^\prime_{cd]}+32D^iA_{iacd}+\frac{32}{3}D^iB_{acd,i}
+2D\Ga_{[a}\tilde{t}_{cd]}-5D\Ga_{[ac}\tilde{t}_{d]}-8D\Ga_{acd}\tilde{t}\nn\\
&-8\tilde{S}_{[a}{}^\al\tilde{t}_{cd]\al}
+4\tilde{S}\Ga_{[a}\tilde{t}_{cd]}
+13\tilde{S}^i\Ga_{[ac}\tilde{t}_{d]i}-56\tilde{Z}_{[a}{}^{\al}\tilde{t}_{cd]\al}+80\tilde{Z}_{[ac}{}^{i\al}\tilde{t}_{d]i\al}
-28\tilde{Z}\Ga_{[a}\tilde{t}_{cd]}\nn\\
&+40\tilde{Z}_{[a}{}^i\Ga_c\tilde{t}_{d]i}+28\tilde{Z}^i\Ga_{[ac}\tilde{t}_{d]i}
+20\tilde{Z}^\prime_{[a}{}^{\al}\tilde{t}_{cd]\al}+14\tilde{Z}^\prime\Ga_{[a}\tilde{t}_{cd]}+8\tilde{Z}^\prime_{[a}{}^i\Ga_c\tilde{t}_{d]i}
-10\tilde{Z}^{\prime i}\Ga_{[ac}\tilde{t}_{d]i}\nn\\
&-384A^i
A_{acdi}-64A_{[a}{}^{i_1i_2}A_{cd]i_1i_2}+\frac{64}{3}A_{[a}{}^{i_1i_2i_3}A^\prime_{cd]i_1i_2i_3}-\frac{128}{3}A_{[ac}{}^{i_1i_2i_3}
A^\prime_{d]i_1i_2i_3}\nn\\
&+192A^\prime_{[a}{}^{i_1i_2}A^\prime_{cd]i_1i_2}
-\frac{448}{3}A^\prime_{[a}{}^{i_1i_2i_3}B_{|i_1i_2i_3|c,d]}-\frac{4}{9}\eps_{acdi_1\dots
i_8}A^{i_1\dots i_4}A^{\prime i_5\dots
i_8}\nn\\
&-\frac{4}{45}\eps_{[ac|i_1\dots i_9|}A^{i_1\dots i_4}B^{i_5\dots
i_9}{}_{,d]} \komma
\end{align}
and with $(\Ga^{ab}{}_{cde})_\de{}^\ga$,
\begin{align}
0&=-448D^iA^\prime_{icde}
+6D\Ga_{[c}\tilde{t}_{de]}-42D\Ga_{[cd}\tilde{t}_{e]}+56D\Ga_{cde}\tilde{t}
-72\tilde{Z}_{[c}{}^{\al}\tilde{t}_{de]\al}-96\tilde{Z}_{[cd}{}^{i\al}\tilde{t}_{e]i\al}\nn\\
&-32\tilde{Z}_{cde}{}^{i_1i_2\al}\tilde{t}_{i_1i_2\al}-60\tilde{Z}\Ga_{[c}\tilde{t}_{de]}
-144\tilde{Z}_{[c}{}^i\Ga_d\tilde{t}_{e]i}-48\tilde{Z}_{[cd}{}^{i_1i_2}\Ga_{e]}\tilde{t}_{i_1i_2}+72\tilde{Z}^i\Ga_{[cd}\tilde{t}_{e]i}\nn\\
&-12\tilde{Z}^{i_1i_2}\Ga_{cde}\tilde{t}_{i_1i_2}-48\tilde{Z}_{[c}{}^{i_1i_2}\Ga_{de]}\tilde{t}_{i_1i_2}-12\tilde{Z}^\prime_{[c}{}^{\al}\tilde{t}_{de]\al}
+30\tilde{Z}^\prime\Ga_{[c}\tilde{t}_{de]}+48\tilde{Z}^\prime_{[c}{}^i\Ga_d\tilde{t}_{e]i}\nn\\
&+12\tilde{Z}^{\prime i}\Ga_{[cd}\tilde{t}_{e]i}+4\tilde{Z}^{\prime
i_1i_2}\Ga_{cde}\tilde{t}_{i_1i_2}
-384A_{[c}{}^{i_1i_2i_3}A_{de]i_1i_2i_3}+5376A^iA^\prime_{cdei}+2304A_{[c}{}^{i_1i_2}A^\prime_{de]i_1i_2}\nn\\
&+2304A_{[cd}{}^{i_1i_2}A^\prime_{e]i_1i_2}
+256A_{[c}{}^{i_1i_2i_3}B_{|i_1i_2i_3|d,e]}+\frac{16}{9}\eps_{cdei_1\dots
i_8}A^{i_1\dots i_4}A^{\prime i_5\dots i_8}\nn\\
&-\frac{16}{9}\eps_{cdei_1\dots i_8}A^{\prime i_1\dots i_4}A^{\prime
i_5\dots i_8}-\frac{32}{45}\eps_{cdei_1\dots i_8}A^{\prime i_1i_2i_3
j}B^{i_4\dots
i_8}{}_{,j}-1280A^\prime_{[c}{}^{i_1i_2i_3}A^\prime_{de]i_1i_2i_3}\nn\\
&+1152A^\prime_{[cd}{}^{i_1i_2}B_{|i_1i_2|,e]}\punkt
\end{align}

\bigskip
\noindent{\it The $(00002)$'s.}

\noindent The second SSBI contains three $(00002)$, which are
obtained by contracting with $(\Ga_{cde})_\de{}^\ga$ and
antisymmetrising in $abcde$,
\begin{align}
0&=64D_{[a}A_{bcde]}+D\Ga_{[abc}\tilde{t}_{de]}+2D\Ga_{[abcd}\tilde{t}_{e]}+D\Ga_{abcde}\tilde{t}
-26\tilde{S}_{[a}\Ga_{bc}\tilde{t}_{de]}-2\tilde{S}\Ga_{[abc}\tilde{t}_{de]}\nn\\
&+80\tilde{Z}_{[abc}{}^\al \tilde{t}_{de]\al}+60\tilde{Z}_{[ab}\Ga_c
\tilde{t}_{de]}+84\tilde{Z}_{[a}\Ga_{bc}\tilde{t}_{de]}-14\tilde{Z}\Ga_{[abc}\tilde{t}_{de]}+12\tilde{Z}^\prime_{[ab}\Ga_c
\tilde{t}_{de]}\nn\\
&-30\tilde{Z}^\prime_{[a}\Ga_{bc}\tilde{t}_{de]}+7\tilde{Z}^\prime
\Ga_{[abc}\tilde{t}_{de]} +384A_{[a}A_{bcde]}-384A_{[ab}{}^i
A_{cde]i}-192A_{[abc}{}^{i_1i_2}A^\prime_{de]i_1i_2}\nn\\
&-64A_{[ab}{}^{i_1i_2}A^\prime_{cde]i_1i_2}+192A_{[abc}{}^iB_{de],i}-384A^\prime_{[ab}{}^i
A^\prime_{cde]i}+128A^\prime_{[ab}{}^{i_1i_2}B_{cde]i_1,i_2}\nn\\
&-\frac{8}{9}\eps_{[abcd|i_1\dots i_7|}A^{i_1i_2i_3}{}_{e]}A^{\prime
i_4\dots i_7}-\frac{8}{45}\eps_{[abc|i_1\dots i_8|}A^{i_1i_2i_3}{}_d
B^{i_4\dots i_8}{}_{,e]}\nn\\
&-\frac{8}{3}\eps_{abcde i_1\dots i_6}A^{\prime i_1i_2}A^{\prime
i_3\dots i_6} -\frac{8}{9}\eps_{[abcd|i_1\dots i_7|}A^{\prime
i_1\dots i_4}B^{i_5i_6i_7}{}_{,e]}\komma
\end{align}
$(\Ga_{cdef})_\de{}^\ga$ and antisymmetrising in $abcdef$,
\begin{align}
0&=64D_{[a}A_{bcdef]}+D\Ga_{[abcd}\tilde{t}_{ef]}+2D\Ga_{[abcde}\tilde{t}_{f]}+D\Ga_{abcdef}\tilde{t}-26\tilde{S}_{[a}\Ga_{bcd}\tilde{t}_{ef]}\nn\\
&-2\tilde{S}\Ga_{[abcd}\tilde{t}_{ef]}+48\tilde{Z}_{[abcd}{}^\al
\tilde{t}_{ef]\al}-64\tilde{Z}_{[abc}\Ga_d
\tilde{t}_{ef]}+120\tilde{Z}_{[ab}\Ga_{cd}\tilde{t}_{ef]}-48\tilde{Z}_{[a}\Ga_{bcd}\tilde{t}_{ef]}\nn\\
&-14\tilde{Z}\Ga_{[abcd}\tilde{t}_{ef]}
+24\tilde{Z}^\prime_{[ab}\Ga_{cd}\tilde{t}_{ef]}-24\tilde{Z}^\prime_{[a}\Ga_{bcd}\tilde{t}_{ef]}-\tilde{Z}^\prime\Ga_{[abcd}\tilde{t}_{ef]}
-512A_{[ab}A_{cdef]}\nn\\
&+512A_{[abc}{}^i
A^\prime_{def]i}-128A_{[ab}{}^{i_1i_2}B_{cdef]i_1,i_2}+512A^\prime_{[ab}A^\prime_{cdef]}+\frac{512}{3}A^\prime_
{[abc}{}^i B_{def],i}\nn\\
& -\frac{32}{15}\eps_{[abcd|i_1\dots
i_7|}A^{i_1i_2}{}_{ef]}A^{\prime i_3\dots i_7}
-\frac{32}{9}\eps_{[abcd|i_1\dots
i_7|}A^{i_1i_2i_3}{}_{ef]}A^{\prime i_4\dots
i_7}\nn\\
&-\frac{64}{9}\eps_{abcdefi_1\dots i_5}A^{\prime i_1i_2j}A^{\prime
i_3i_4i_5}{}_j-\frac{32}{9}\eps_{[abcd|i_1\dots i_7|}A^{\prime
i_1\dots i_4}B^{i_5i_6i_7}{}_{e,f]}\komma
\end{align}
and with $(\Ga^{b}{}_{cdef})_\de{}^\ga$ and antisymmetrising in
$acdef$,
\begin{align}
0&=192D_{[a}A^\prime_{cdef]}-\frac{32}{5}D^iB_{acdef,i}-\frac{4}{15}\eps_{acdef}{}^{i_1\dots
i_6}D_{i_1}A^\prime_{i_2\dots
i_6}-4D\Ga_{[acd}\tilde{t}_{ef]}+D\Ga_{[acde}\tilde{t}_{f]}\nn\\
&+6D\Ga_{acdef}\tilde{t}+33\tilde{S}_{[a}\Ga_{cd}\tilde{t}_{ef]}
+4\tilde{S}\Ga_{[acd}\tilde{t}_{ef]}+\tilde{S}_{[a,}{}^i\Ga_{cde}\tilde{t}_{f]i}+\tilde{S}^i\Ga_{[acde}\tilde{t}_{f]i}
+64\tilde{Z}_{[acd}{}^\al\tilde{t}_{ef]\al}\nn\\
&-32\tilde{Z}_{[acde}{}^{i\al}\tilde{t}_{f]\al}-144\tilde{Z}_{[ac}\Ga_d\tilde{t}_{ef]}
+64\tilde{Z}_{[acd}{}^i\Ga_e\tilde{t}_{f]i}-144\tilde{Z}_{[a}\Ga_{cd}\tilde{t}_{ef]}-96\tilde{Z}_{[ac}{}^i\Ga_{de}\tilde{t}_{f]i}\nn\\
&+40\tilde{Z}\Ga_{[acd}\tilde{t}_{ef]}+48\tilde{Z}_{[a}{}^i
\Ga_{cde}\tilde{t}_{f]i}-12\tilde{Z}^i\Ga_{[acde}\tilde{t}_{f]i}+48\tilde{Z}^\prime_{[ac}\Ga_d\tilde{t}_{ef]}
-24\tilde{Z}^\prime_{[a}\Ga_{cd}\tilde{t}_{ef]}\nn\\
&-20\tilde{Z}^\prime\Ga_{[acd}\tilde{t}_{ef]}-16\tilde{Z}^\prime_{[a}{}^i\Ga_{cde}\tilde{t}_{f]i}
-2\tilde{Z}^{\prime
i}\Ga_{[acde}\tilde{t}_{f]i}+1152A_{[a}A^\prime_{cdef]}-1024A_{[ac}{}^iA^\prime_{def]i}\nn\\
&-128A_{[ac}{}^{i_1i_2}A_{def]i_1i_2}+1536A_{[acd}{}^iA^\prime_{ef]i}+256A^\prime_{[ac}{}^{i_1i_2}A^\prime_{def]i_1i_2}
+896A^\prime_{[acd}{}^iB_{ef],i}\nn\\
&-\frac{8}{3}\eps_{acdefi_1\dots i_6}A^{i_1i_2}A^{\prime i_3\dots
i_6}-\frac{16}{3}\eps_{acdefi_1\dots i_6}A^{i_1\dots i_4}A^{\prime
i_5i_6}\nn\\
&-\frac{16}{9}\eps_{[acde|i_1\dots i_7|}A^{i_1\dots
i_4}B^{i_5i_6i_7}{}_{,f]}-\frac{112}{45}\eps_{[acd|i_1\dots
i_8|}A^{\prime i_1i_2i_3}{}_eB^{i_4\dots i_8}{}_{,f]} \komma
\end{align}
where we have used \eqref{blirp} and the following identities
\begin{align}
\eps_{abcde i_1\dots i_6}A^{i_1i_2i_3j}A^{\prime
i_4i_5i_6}{}_j&=-\frac{5}{4}\eps_{[abcd|i_1\dots i_7|}A^{i_1\dots
i_4}A^{\prime i_5i_6i_7}{}_{e]}\komma\nn\\
\eps_{abcde i_1\dots i_6}A^{i_1i_2i_3j}A^{
i_4i_5i_6}{}_j&=-\frac{5}{4}\eps_{[abcd|i_1\dots i_7|}A^{i_1\dots
i_4}A^{i_5i_6i_7}{}_{e]}=0\komma\nn\\
\eps_{abcdei_1\dots i_6}A^{\prime
i_1i_2i_3j}B^{i_4i_5i_6}{}_{,j}&=-\frac{5}{4}\eps_{[abcd|i_1\dots
i_7|}A^{\prime i_1\dots i_4}B^{i_5i_6i_7}{}_{,e]}\komma\nn\\
\eps_{[abcd|i_1\dots i_7|}A^{i_1i_2i_3}{}_{ef]}A^{\prime i_4\dots
i_7}&=\frac{4}{5}\eps_{[abcde|i_1\dots
i_6|}A^{i_1i_2i_3j}{}_{f]}A^{\prime i_4i_5i_6}{}_j\komma\nn\\
\eps_{[abcd|i_1\dots i_7|}A^{i_1i_2}{}_{ef]}A^{\prime i_3\dots
i_7}&=\eps_{[abcde|i_1\dots i_6|}A^{i_1i_2j}{}_{f]}A^{\prime
i_3\dots i_6}{}_j\komma\nn\\
\eps_{[abcd|i_1\dots i_7|}A^{\prime i_1\dots
i_4}B^{i_5i_6i_7}{}_{e,f]}&=-\frac{4}{5}\eps_{[abcde|i_1\dots
i_6}A^{\prime i_1i_2i_3j}B^{i_4i_5i_6}{}_{j,|f]}\komma\nn\\
\eps_{abcdefi_1\dots i_5}A^{\prime i_1i_2j}A^{\prime i_3i_4i_5}{}_j&=-\frac{3}{2}\eps_{[abcde|i_1\dots i_6|}A^{\prime i_1i_2}{}_{f]}A^{\prime i_3\dots i_6}\nn\\
&=2\eps_{[abcde|i_1\dots i_6|}A^{\prime i_1i_2i_3}A^{\prime
i_4i_5i_6}{}_{f]}
\end{align}
and
\begin{align}
\eps_{[abcd|i_1\dots i_7|}A^{\prime i_1\dots i_4}A^{\prime
i_5i_6i_7}{}_{ef]}&=-\frac{4}{5}\eps_{[abcde|i_1\dots i_6}A^{\prime
i_1i_2i_3j}A^{\prime i_4i_5i_6}{}_{j|f]} \nn\\
\eps_{[abcde|i_1\dots i_6}A^{\prime i_1i_2i_3j}A^{\prime
i_4i_5i_6}{}_{j|f]}&=-\frac{1}{2}\eps_{abcdefi_1\dots i_5}A^{\prime
i_1i_2jk}A^{\prime i_3i_4i_5}{}_{jk} \nn\\
\eps_{[abcd|i_1\dots i_7|}A^{\prime i_1i_2}{}_{ef]}A^{\prime
i_3\dots i_7}&=\eps_{[abcde|i_1\dots i_6|}A^{\prime
i_1i_2j}{}_{f]}A^{\prime i_3\dots i_6}{}_{j} \nn\\
\eps_{[abcde|i_1\dots i_6|}A^{\prime i_1i_2j}{}_{f]}A^{\prime
i_3\dots i_6}{}_j&=\frac{2}{3}\eps_{abcdefi_1\dots i_5}A^{\prime
i_1i_2jk}A^{\prime i_3i_4i_5}{}_{jk} \punkt
\end{align}

\section{Decomposition of tensor-spinors}\label{appenixb}

{\bf Decomposition of tensor-spinors of the types
$\quad\begin{array}{c}{\tiny\yng(1,1,1,1)}\\ \vdots\\ {\tiny\yng(1)}\end{array}\quad$ and
$\quad\begin{array}{l}{\tiny\yng(2,1,1,1)}\\ \vdots\\ {\tiny\yng(1)}\end{array}\quad$}

{\bf 1.} Consider a general (=reducible) rank-$n$ antisymmetric tensor-spinor
in D dimensions, $V^\al_{a_1\ldots a_n}$. We want to decompose it into
irreducible ($\Ga$-traceless) representations. An irreducible rank-$n$ tensor
spinor is obtained from a reducible one as
\begin{equation}
V'{}_{a_1\ldots a_n}= \sum_{p=0}^nN_{n,p}\Ga{}_{[a_1\ldots
a_p}\Ga{}^{b_1\ldots b_p}
        V_{|b_1\ldots b_p|a_{p+1}\ldots a_n]}\komma
\label{One}
\end{equation}
where $N_{n,p}=\frac{(-1)^{\frac{p(p+1)}{2}}}{p!}
\frac{\binom{n}{p}}{\binom{D-2n+p+1}{p}}$.

If we define the expansion of $V$ in irreducible representations as
\begin{equation}
V{}_{a_1\ldots a_n}=\sum_{p=0}^n{n\choose p}
    \Ga{}_{[a_1\ldots a_p}\tilde V{}_{a_{p+1}\ldots a_n]}\komma
\end{equation}
the $\Ga$-traces of $V$ are
\begin{align}
\begin{split}
v{}_{a_{p+1}\ldots a_n}&\equiv
    {1\over p!}\Ga{}^{a_1\ldots a_p}V_{a_1\ldots a_pa_{p+1}\ldots a_n}\cr
&=(-1)^{p(p-1)\over2}\sum_{r=0}^{n-p}{n-p\choose r}{D-2n+2p+r\choose p}
    \Ga{}_{[a_{p+1}\ldots a_{p+r}}\tilde V{}_{a_{p+r+1}\ldots a_n]}
\punkt
\end{split}
\end{align}
Subtracting the $\Ga$-trace leaves only the first term in the sum:
\begin{equation}
v'{}_{a_{p+1}\ldots a_n}
    =(-1)^{p(p-1)\over2}{D-2n+2p\choose p}\tilde V{}_{a_{p+1}\ldots a_n}
    \komma\label{Two}
\end{equation}
or, explicitly, using (\ref{One}):
\begin{equation}
\tilde V{}_{a_{p+1}\ldots a_n}= {(-1)^p\over{D-2n+2p\choose
p}}\sum_{r=p}^n{(-1)^{r(r+1)\over2}\over r!} {{r\choose p}{n-p\choose
r-p}\over{D-2n+p+r+1\choose r-p}} \Ga_{[a_{p+1}\ldots a_r}\Ga^{b_1\ldots b_r}
V{}_{|b_1\ldots b_r|a_{r+1}\ldots a_n]} \komma
\end{equation}
which of course coincides with (\ref{One}) for $p=0$.

\phantom{}

{\bf 2.} Consider the tensor product of an irreducible hook tensor, \ie, a
tensor of the type
\begin{equation}
U_{a_{1}\ldots a_{n},a}:\quad U_{[a_{1}\ldots a_{n},a]}=0, \quad
U_{a_{1}\ldots a_{n-1}a}{}^a=0\komma
\end{equation}
with a spinor. The $\Ga$-traceless part is
\begin{align}
\begin{split}
U'_{a_{1}\ldots a_{n},a}=&\sum_{p=1}^nk_{n,p}\Ga_{[a_1\ldots
a_p}\Ga^{b_1\ldots b_p}
     U_{\vert b_1\ldots b_p\vert a_{p+1}\ldots a_n],a}\cr
+&\sum_{p=1}^nl_{n,p}\Ga_{a[a_1\ldots a_{p-1}}\Ga^{b_1\ldots b_p}
     U_{\vert b_1\ldots b_p\vert a_p\ldots a_{n-1},a_n]}\cr
+&\sum_{p=2}^nm_{n,p}\eta_{a[a_1}\Ga_{a_2\ldots a_{p-1}}\Ga^{b_1\ldots b_p}
     U_{\vert b_1\ldots b_p\vert a_p\ldots a_{n-1},a_n]}\cr
-&[a_1\ldots a_na]\komma\label{Three}
\end{split}
\end{align}
where
\begin{align}
k_{n,p}&=N_{n,p}\komma\cr
l_{n,p}&=(-1)^{n+1}{(D-n+1)(D-2n+1)-(n+1)p\over(D+2)(D-n+2)}N_{n,p}\komma\cr
m_{n,p}&=(-1)^n{(D-2n+p+1)(n+1)(p-1)\over(D+2)(D-n+2)}N_{n,p}
\end{align}
(at $p=n$, only the combination ${1\over n+1}(nk_{n,n}+(-1)^{n+1}l_{n,n})
={(D+1)(D-n+1)\over(D+2)(D-n+2)}N_{n,n}$ enters).

The vanishing of the completely antisymmetric part implies that $\Ga$-traces
only have to be taken on $a_{1}\ldots a_{n}$, and $\Ga$-tracelessness in these
indices implies $\Ga$-tracelessness in $a$. The tracelessness wrt the vector
indices survives after a multiple $\Ga$-trace, but the vanishing of the
antisymmetrised tensor does not, so in order to get an irreducible
representation out of a $\Ga$-trace on $U$ one has to divide it into an
antisymmetric part and an irreducible hook, and subtract the $\Ga$-traces of
both. Explicitly:
\begin{equation}
u_{a_{p+1}\ldots a_{n},a}\equiv{1\over p!}\Ga^{a_{1}\ldots a_{p}}
U_{a_{1}\ldots a_{n},a}
\end{equation}
splits into
\begin{equation}
u_{a_{p+1}\ldots a_{n},a} =v_{a_{p+1}\ldots a_{n}a}+w_{a_{p+1}\ldots
a_{n},a}\komma
\end{equation}
where
\begin{align}
v_{a_{p+1}\ldots a_{n}a}&\equiv u_{[a_{p+1}\ldots a_{n},a]}\komma\cr
w_{a_{p+1}\ldots a_{n},a}&\equiv u_{a_{p+1}\ldots a_{n},a}
    -u_{[a_{p+1}\ldots a_{n},a]}\cr
&={n-p\over n-p+1} \left(u_{a_{p+1}\ldots a_{n},a}
+(-1)^{n-p+1}u_{a[a_{p+1}\ldots a_{n-1},a_{n}]}\right) \punkt
\end{align}
The tensor $u'_{a_{p+1}\ldots a_{n},a}$ is defined using the subtraction of
$\Ga$-traces according to (\ref{One}) and (\ref{Three}), and consists of two
irreducible tensors $v'$ and $w'$. One obtains:
\begin{equation}
v'_{a_{p+1}\ldots a_na}=\sum_{r=p}^n(-1)^{{p(p+1)\over2}+{r(r+1)\over2}}
{1\over r!}{{r-1\choose p-1}{n-p\choose r-p}\over{D-2n+p+r-1\choose r-p}}
\Ga_{[a_{p+1}\ldots a_r}\Ga^{b_1\ldots b_r} U_{\vert b_1\ldots b_r\vert
a_{r+1}\ldots a_n,a]}
\end{equation}

On the other hand, an expansion of $U$ in irreducible tensors is defined by
\begin{align}
\begin{split}
U_{a_{1}\ldots a_{n},a}&=\sum_{p=0}^{n-1}{n\choose p} \Ga{}_{[a_1\ldots
a_p}\tilde W{}_{a_{p+1}\ldots a_n],a}\cr &+\sum_{p=1}^n{n\choose p}\Bigl(
\Ga{}_{a[a_1\ldots a_{p-1}}\tilde V{}_{a_{p}\ldots a_n]}
+(-1)^{n-1}\Ga{}_{[a_1\ldots a_{p}}\tilde V{}_{a_{p+1}\ldots a_n]a}\cr
&\qquad\qquad\qquad+{(n+1)(p-1)\over D-n+1} \eta_{a[a_1}\Ga{}_{a_2\ldots
a_{p-1}}\tilde V{}_{a_{p}\ldots a_n]}\Bigr)
\end{split}
\end{align}

Performing the $\Ga$-traces on this expansion
yields (the second eq. directly from (\ref{Two}), the first after some
computing)
\begin{align}
v'_{a_{p+1}\ldots a_{n}a}&
=(-1)^{n+1+{p(p-1)\over2}}{(D+2)(D-n+p)\over(D-n+1)(D-2n+p-1)}
{D-2n+2p-2\choose p}\tilde V{}_{a_{p+1}\ldots a_na}\komma\cr w'_{a_{p+1}\ldots
a_{n},a}& =(-1)^{p(p-1)\over2}{D-2n+2p\choose p}\tilde W{}_{a_{p+1}\ldots
a_n,a}\punkt
\end{align}

%
%
\cleardoublepage
\bibliography{11dref}
\bibliographystyle{utphysmod}
\end{document}